\documentclass[12pt]{unibz_ase_thesis}
\usepackage[hyphens]{url}
\usepackage{lipsum}
\usepackage{graphicx}
\usepackage{pdfpages}
\usepackage{enumitem}
\usepackage{hyperref}

\usepackage{todonotes}

\usepackage[utf8]{inputenc}
\usepackage[T1]{fontenc}
\usepackage{newtxtext}
\usepackage{tcolorbox}
\usepackage{titlesec}
\usepackage{hyperref}
\usepackage{caption} 
\usepackage[a-2b]{pdfx}

\usepackage{blkarray}
\usepackage{longtable}
\usepackage{rotating}
\usepackage{algorithm}
\usepackage{algpseudocode}
\usepackage{textcomp}
\usepackage[ngerman,italian,english]{babel}
\usepackage{csquotes}
\usepackage{tikz}
\usetikzlibrary{trees}

\usepackage{amsmath, amsthm, amssymb}
\usepackage[
    backend=biber,natbib,
    bibstyle=ieee,
    citestyle=numeric-comp
]{biblatex} 
\usepackage{chancery} 
\usepackage[T1]{fontenc} 

\newcommand{\myPart}[1]{
  \cleardoublepage
  \stepcounter{part}
  \phantomsection
  \addcontentsline{toc}{part}{\thepart\hspace{1em} \MakeUppercase{#1}} 
  \begin{center}
    \Huge\bfseries PART \thepart\\ \MakeUppercase{#1} 
  \end{center}
  \vspace{1cm}
  \setcounter{figure}{0}
}
\algnewcommand\algorithmicswitch{\textbf{switch}}
\algnewcommand\algorithmiccase{\textbf{case}}
\algdef{SE}[SWITCH]{Switch}{EndSwitch}[1]{\algorithmicswitch\ #1\ \algorithmicdo}{\algorithmicend\ \algorithmicswitch}%
\algdef{SE}[CASE]{Case}{EndCase}[1]{\algorithmiccase\ #1}{\algorithmicend\ \algorithmiccase}%
\algtext*{EndSwitch}%
\algtext*{EndCase}%
\newcommand{\BlankLine}{\vspace{\baselineskip}}

\begin{document}
\pagenumbering{gobble}

\title{Green Resilience of Cyber-Physical Systems}  

\author{Diaeddin Rimawi} 

\unilogo{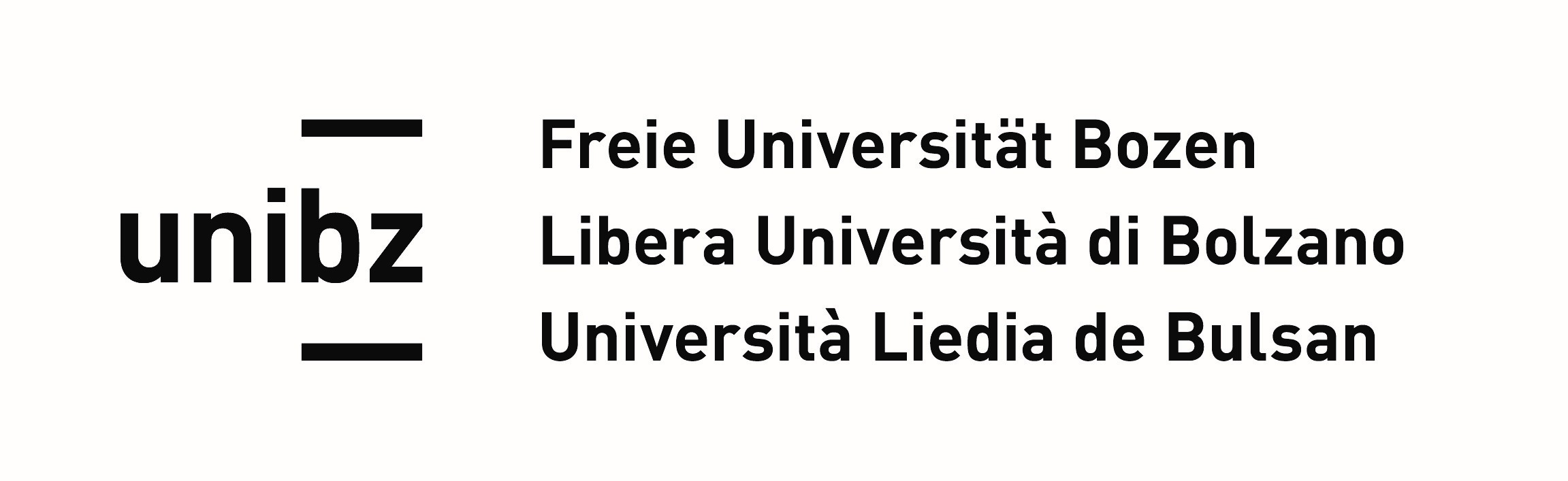}                                 
\degreeaward{Ph.D.\ in Advanced-Systems Engineering}       
\PhDcycle{$37^\mathrm{th}$ Cycle}          
\copyyear{2025}  
\defenddate{15.05.2025}          

\birthday{[28.09.1992]}				
\birthplace{[Ramallah, Palestine]}			
\signdate{[15.05.2025]}				
\signplace{[Bolzano, Italy]}			

\supervisor{Professor Antonio Liotta}						
\secondsupervisor{Professor Barbara Russo}		

\orcid{0000-0003-3791-399X}

\rightsstatement{All rights reserved}

\maketitle[logo]


\begin{acknowledgements}

I am thankful to the Italian Ministry of Universities and Research for supporting this Ph.D. through a grant under the Programma Operativo Nazionale Ricerca e Innovazione 2014-2020 (no. DOT19L38RN-5) and a grant under the Progetti di Rilevante Interesse Nazionale (PRIN) 2022, no. 2022TEPX4R.

I extend my heartfelt thanks to my supervisors, Professor Barbara Russo and Professor Antonio Liotta, who provided invaluable insights and struck a perfect balance between granting me independence and guiding me in the right direction. A special thanks goes to Professor Barbara Russo, whose unwavering support made this journey possible. I am also sincerely grateful to Professor Alberto Sangiovanni-Vincentelli for his valuable comments during the first two years of this work.  
My gratitude extends to Dr. Marco Todescato, my tutor during my internship at Fraunhofer Italia Research (FhI), for his guidance and support, and to the entire FhI family for granting me access to their resources.

I deeply appreciate the contributions of the review committee, particularly Professor Anna Perini and Professor Henry Muccini, for their insightful comments and input on the final version of this thesis. Additionally, I would like to thank the Examination Committee members, Professor Henry Muccini, Professor Bruno Rossi, and Professor Barbara Russo, as well as the substitute member, Dr. Jorge Augusto Melegati Goncalves, for evaluating my work and providing their valuable time and expertise.

\newpage
\noindent
\begin{minipage}{\linewidth}
    \raggedleft
    {\fontfamily{pzc}\selectfont \LARGE To My Family}  
\end{minipage}

\vspace{1cm} 

To my wife, Eman, and our two beautiful children, Mohammad and Misk, thank you for your love, patience, and unwavering belief in me.

To my parents, Mohammad and Shatha, for making everything possible through their boundless sacrifices and support, and to my sister Ola, the joy of our family. A special dedication goes to our family’s beloved hero, my brother AbdelRahman, who has endured separation due to the unjust situation we live in under occupation in Palestine, and to his family for their strength and resilience.

Last but not least, to my friends who became a family to me—especially Dr. Refat Othman and Dr. Usman Rafiq—thank you for your companionship, encouragement, and the shared moments that made this journey brighter and more meaningful.

\end{acknowledgements}

\begin{abstract}

Cyber-physical systems (CPS) integrate computational and physical components to interact with humans across various modalities. Online Collaborative Artificial Intelligence Systems (OL-CAIS) are a subset of CPS that collaborate with humans to achieve shared goals through online learning. 
These systems are exposed to environmental changes that may lead to performance degradation (i.e., disruptive events). Decision-makers must develop policies that restore performance (i.e., ensure resilience) while minimizing adverse energy impacts (i.e., ensure greenness). These policies must balance greenness and resilience in collaborative actions.

This research investigates the challenge of balancing greenness and resilience in OL-CAIS under disruptive events. Our objectives are to i) model OL-CAIS resilience for automatic state detection, ii) develop agent-based policies to optimize the trade-off between greenness and resilience, and iii) understand catastrophic forgetting to maintain consistent performance. By addressing these objectives, we equip decision-makers with tools and strategies to manage OL-CAIS effectively.

The research employs a systematic, iterative methodology combining theoretical modeling and empirical evaluation. 
We model OL-CAIS resilience through three operational states: steady, an initial state of autonomous collaboration with humans; disruptive, in which performance degrades and recovering policies are required; and final, in which the system experiences memory degradation after disruption resolution. 
To fill the need for recovering policies in the disruptive state, we introduce the GResilience framework, which recommends recovery actions to balance greenness and resilience. These include: i) one-agent policies using multi-objective optimization, ii) two-agent policies resolving trade-offs via game theory, and iii) reinforcement learning (RL) agent policies maximizing rewards based on greenness-resilience states. 
Then, to evaluate the effectiveness of these agent-based policies in achieving green recovery, we create our measurement framework. The measurement framework defines measurable concepts to evaluate the effectiveness of resilience (i.e., recovery speed and performance steadiness) and greenness (i.e., green efficiency and service autonomy).
The research also explores catastrophic forgetting in online learning, analyzing its effects across multiple disruptions and proposing strategies to maintain steady performance. 
Additionally, we introduce a containerization methodology to optimize resource allocation and reduce energy consumption.

We assess our models and theories through empirical evaluation, which includes real-world and simulated experiments with an industrial collaborative robot learning object classification from human demonstrations. 
Our experiments demonstrate the resilience model's ability to trace system performance evolution across states in runtime during disruptive events.
The effectiveness of the GResilience framework to balance greenness and resilience was evaluated in four real-world experiments and over 800 simulated experiments for each agent-based decision-making policy. 
Results showed that our agents' policies improved green recovery over internal policies, reducing the time needed for recovery, lowering performance fluctuations between acceptable and unacceptable levels, and decreasing human dependency. However, this improvement came at the cost of higher CO$_2$ emissions, resulting from the additional computational power required by the agents. RL-agent policies outperformed two-agent and one-agent policies, respectively.
Furthermore, our experiments revealed catastrophic forgetting caused by repeated disruptions, but our policies ensured consistent performance steadiness. 
Finally, comparing bare-metal and containerized setups, we demonstrate that containerization significantly reduced CO$_2$ emissions, halving those of the bare-metal setup.

In conclusion, this research advances the understanding of green resilient OL-CAIS by providing decision-makers with novel metrics, theoretical models, and actionable strategies. The frameworks and policies developed here empower stakeholders to navigate the trade-offs between resilience and greenness, enabling the design of environmentally sustainable and resilient systems.
\end{abstract}

\tableofcontents
\listoffigures
\listoftables

\printnomenclature

\mainmatter

\myPart{Foundation and Background}

\huge{T}\normalsize{his thesis is divided into four parts to address the different components of our research, \textit{Foundation and Background}, \textit{Research Methodology and Approach}, \textit{Experimental Evaluation}, and \textit{Reflections and Conclusions}.

This first part sets the stage for this PhD research. We first introduce the research context, challenges, and goals, establishing the motivation behind this study. 
Then we delve into the core concepts underpinning this work, such as online collaborative AI systems, resilience, and greenness, while providing insights into decision-making policies and models like optimization, game theory, and reinforcement learning. 
At the end of this part, we present a systematic literature review, offering a comprehensive analysis of previous studies alongside a discussion of related work that connects this research to the broader landscape.
}

\chapter{Introduction}
\label{ch:intro}
Industry 4.0 represents a transformative industrial paradigm, encompassing fully automated facilities equipped with advanced \textit{smart} features. 
These features enable the conversion of traditional systems into intelligent components capable of self-description, communication, environmental sensing, and control. 
These smart components cooperate with other intelligent entities to achieve greater autonomy and efficiency,~\cite{lasi2014industry}.
At the heart of this shift lies the integration of advanced computing and innovative technologies, creating autonomous systems capable of real-time decision-making and continuous adaptation.
A system is a collection of components that cooperate in an organized manner to achieve desired goals,~\cite{dick_requirements_2017}.
A Cyber-Physical System (CPS) is a system that integrates computational and physical capabilities to interact with humans through various modalities,~\cite{baheti2011cyber}.
CPS complexity varies from one domain to another. In recent years, it has evolved into more specialized domains, such as human-centric CPS, which emphasizes human interaction with technology.
In human-centric CPS, the human is not simply a passive user of the system but operates interactively in activities such as knowledge management, learning, and training, to mention a few,~\cite{pinzone_framework_2020}. 
A Collaborative Artificial Intelligence System (CAIS) is a human-centric CPS that performs its tasks in collaboration with humans in a shared environment to achieve a common goal,~\cite{bonfanti_gresilience_2023, kadgien_cais-dma_2024,rimawi_2024_gresilience}.
CAIS is equipped with an Artificial Intelligence (AI) model, which enables the system to learn its tasks from human interactions using continual learning,~\cite{lesort2020continual}.

Continual learning updates the model in continuous iterations of new arriving data instances.
Systems designers turn to continual learning when data has a limited lifetime before being discarded. The main goal is to retain what is significant for the future while forgetting what is not,~\cite{lesort2020continual}.
Continual learning is called either few-shot or online learning, based on the number of data instances arriving at each iteration. In few-shot learning, the training data consists of a few instances at each iteration, while in online learning, only one data point arrives at each iteration for training,~\cite{lesort2020continual}.
Similar to other learning algorithms, online learning operates as one of the following paradigms: 
i) Supervised, data is formed in input-output pairs (labeled data), 
ii) Unsupervised, the data is not labeled, leaving the learner to discover the data structure by its own, and 
iii) Semi-supervised, a mix between supervised and unsupervised techniques is used, where labeled data are used for training to predict the unlabeled data,~\cite{mahesh2020machine}.
In a semi-supervised learning process, the AI model interactively queries the information source (e.g., human) to obtain the desired output labels, calling this process active learning,~\cite{lesort2020continual}.

In our work, we consider CAIS actively learning from humans to classify new data instances online.
We refer to this system as online CAIS, or \textit{OL-CAIS}.
The online learning process forms a cycle of interactions between the human and the autonomous components of the system. 
During this cycle, OL-CAIS develops an internal decision-making policy to perform one of two actions: i) perform its tasks autonomously, or ii) prompt the human to perform the tasks and update the AI model,~\cite{lesort2020continual, hawkins2013probabilistic, ghadirzadeh2020human}.
The internal policy is based on the accuracy of the AI model. 
The higher the accuracy, the more this policy trusts its predictions.
Trusted predictions increase system autonomy, which means the system's ability to perform autonomous actions. While the opposite increases the system's dependency, which means the system's reliance on human actions (i.e., retraining),~\cite{rimawi_modeling_2024}.
To ensure the continuous delivery of OL-CAIS real-time services, it is essential to maximize autonomy.
Thus, in our research, we measure the system's performance by its ability to perform its task autonomously,~\cite{rimawi_modeling_2024, bonfanti_gresilience_2023}.
Losing this ability will cause performance degradation, which means OL-CAIS performance decreases from an acceptable level to a non-acceptable level.

The events that lead to performance degradation are called \textit{disruptive events}. These events can be endogenous, such as system defects, or exogenous, such as environmental changes,~\cite{henry_generic_2012,januario_distributed_2019}.
In this PhD research, we only consider exogenous environmental changes as disruptive events.
Disruptive events, such as power shortages or adversarial attacks, can alter OL-CAIS learning environment settings, affecting the data instances processed by the AI model and potentially hampering its ability to make accurate predictions,~\cite{rimawi_modeling_2024}.
Furthermore, the online learning process exposes the system to \textit{catastrophic forgetting},~\cite{lesort2020continual}.
While catastrophic forgetting typically refers to the phenomenon where a neural network faces a performance degradation on previously learned concepts when trained sequentially on new ones, its implications extend to online learning scenarios. 
In this process, the continuous acquisition of new features, classes, or tasks can also contribute to memory performance degradation,~\cite{lesort2020continual}.
This degradation can be further influenced when OL-CAIS learns the new disrupted environment settings, which may lead to forgetting the original settings. 
Forgetting the original settings when learning the disrupted ones highlights new challenges to making accurate predictions after fixing the disruptive event.
Inaccurate predictions may impose new consequences, such as human intervention to rework the tasks and retrain the AI model. 
Although reworking the task may help the AI model improve its accuracy, this process consumes both energy and time. 
Thus, it is of paramount importance for OL-CAIS decision-makers to optimize this process to achieve the best balance between minimizing the energy adverse effects (i.e., \textit{greenness}) and the time required to restore the performance from degradation (i.e., \textit{resilience}).

In particular, prioritizing resilience without considering energy can lead the system to autonomously execute inaccurate decisions, resulting in wasted operations, increased reliance on human correction, and repeated retraining cycles, all of which compound the energy cost. Therefore, greenness-aware trade-off is essential not to compromise the quality of recovery itself and to reduce the long-term energy footprint associated with inaccurate autonomous actions.
This PhD research acknowledges that greenness might trade-off with other non-functional properties such as reliability, safety, and security, to mention a few. 
Some of these properties, including system failures, and security threats, are considered in this work as sources of disruptive events. However, they may also create trade-offs after the disruption occurs. For example, restoring safety might require disabling certain optimizations, and enhancing security might introduce additional computational overhead. While this work focuses on resilience as the main property that directly affects system performance and energy use, exploring how other non-functional properties interact with greenness remains an important direction for future research.

This research aims to support decision-making with agent-based policies. 
The foundation of these agents is based on the principle of balancing OL-CAIS greenness and resilience.
Following a systematic and iterative methodology, our research explores various agent-based policies within our ``GResilience'' framework,~\cite{bonfanti_gresilience_2023}. Including \textit{one-agent} policies using the weighted sum model, \textit{two-agent} policies using a theoretical game approach, and reinforcement learning agent (i.e., \textit{RL-agent}) policies to optimize this trade-off and help OL-CAIS navigate these complex challenges.
Furthermore, we leverage innovative technologies like component containerization to build a green-resilient OL-CAIS.

\section{Research Challenges}
\label{sec:intro:reseachchallenges}

\begin{figure}[ht!]
    \centering
    \includegraphics[width=\textwidth]{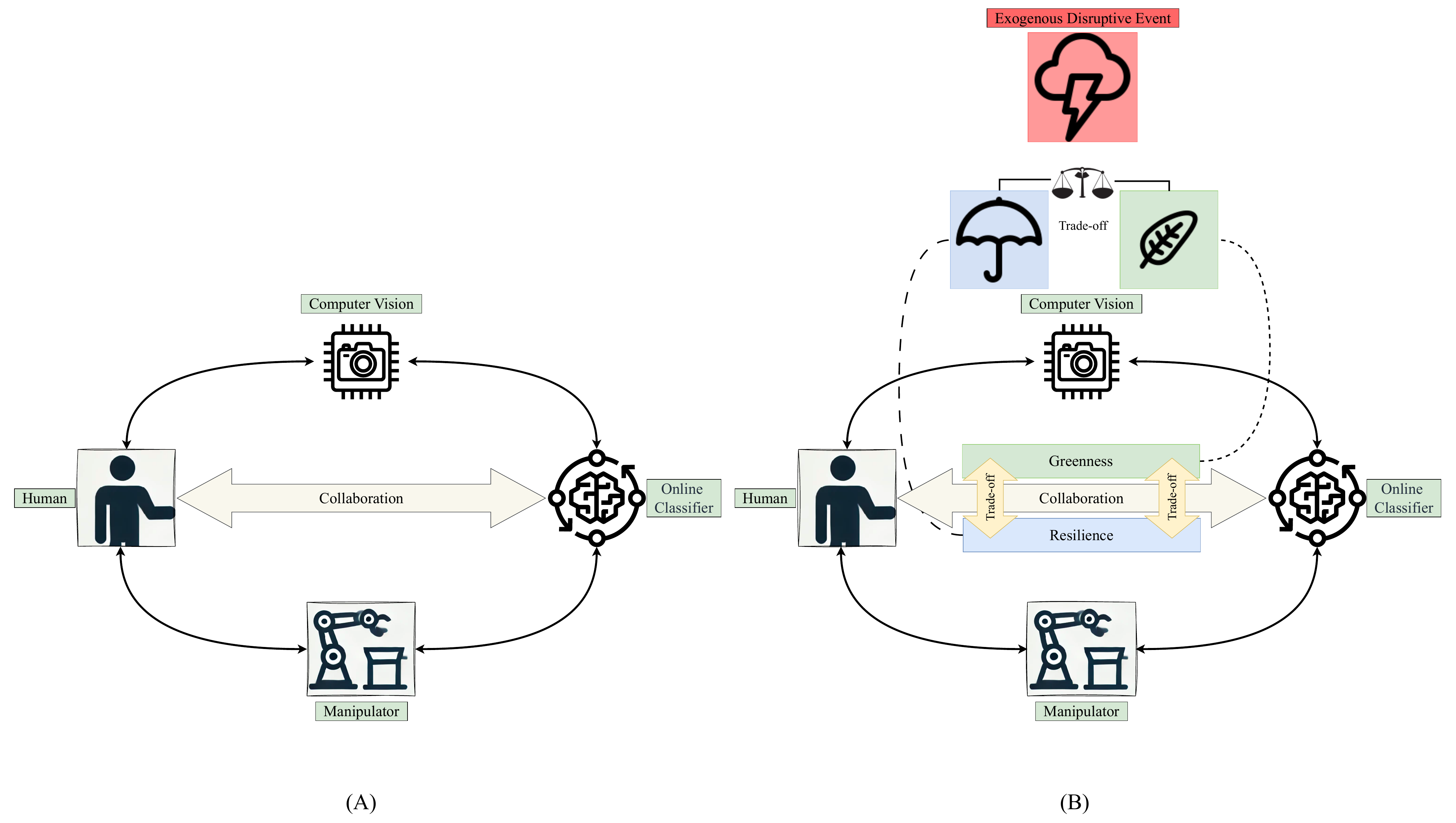}
    \caption{Research challenges that face industrial OL-CAIS: (A) OL-CAIS learns from human collaboration to achieve steady performance. (B) Upon an exogenous disruptive event, its performance may degrade. Thus, decision-makers must balance resilience and greenness while addressing performance degradation.}
    \label{fig:intro:researchchallengestoresilience}
\end{figure}

This research identifies two primary challenges: resilience and greenness. 
Fig.~\ref{fig:intro:researchchallengestoresilience} visually represents the challenges related to OL-CAIS. The system operates within its normal settings, with a steady performance behavior, part (A). 
Being exposed to disruptive events that may lead to performance degradation poses new challenges for decision-makers. They need to develop appropriate policies to restore performance from degradation (i.e., ensure resilience) while managing the energy's adverse effects that may result from these policies (i.e., ensure greenness), part (B).
Thus, these policies must aim to select collaborative actions within OL-CAIS that balance rapid recovery with minimizing energy consumption.

This PhD research addresses the following research challenges.

\begin{itemize}
    \item \textbf{RC1}. \textit{Modeling performance evolution}: 
    The first challenge to face decision-makers when they try to build resilience policies is to understand whether their system performance is currently operating in an acceptable performance or not.
    Thus, it is important to have an automatic instrument that defines the current operating state of the system to provide insights about potentially disruptive events. 
    Disruptive events are either known by decision-makers, such as in-vitro settings, or unforeseen.
    By detecting performance degradation and modeling operational state, we can identify the period in which a disruptive event has occurred, whether it is known or unforeseen.
    \item \textbf{RC2}. \textit{Balancing greenness and resilience upon disruptive events}: The difficulty stems from the natural trade-off between energy efficiency and resilience. 
    The need for more processing capacity for recovery policies frequently results in higher energy consumption, which produces higher energy adverse effects, such as carbon emissions. This leads to the need for different policies that consider a trade-off between rapid recovery and greenness.
    Managing energy-intensive operations, like task reworking or AI retraining, is a significant challenge for real-time systems like OL-CAIS. Therefore, it's also crucial to consider different implementation technologies (e.g., cloud computing) carefully, consider balanced energy utilization, and avoid introducing new energy overheads.
    This brings us to the following sub-challenges:
    \begin{itemize}
        \item \textbf{RC2.1}. Understanding the relation between greenness and resilience: Investigating how resilience policies affect energy consumption and determining the optimal balance between system recovery and sustainability.
        \item \textbf{RC2.2}. Building an automatic decision-making assistant: Implementing a decision-making framework that automatically responds to performance degradation by selecting the best actions to balance rapid recovery with minimizing energy consumption.
        \item \textbf{RC2.3}. Green resource allocation: Utilizing innovative technologies, such as containerization, to optimize both energy use and system performance during recovery actions.
    \end{itemize}
    \item \textbf{RC3}. \textit{Understanding catastrophic forgetting}: Addressing the tendency of the AI model to forget previously learned tasks, specifically when going back from disruptive settings to the original settings.
    This challenge can be broken into the following sub-challenges:
    \begin{itemize}
        \item \textbf{RC3.1}. Understanding if OL-CAIS experiences any catastrophic forgetting: After restoring the original environment settings or fixing the disruptive event, decision-makers need to understand whether their system still has memory from the disruptive environment and forgetting the original system.
        \item \textbf{RC3.2}. Ensuring consistent system performance over time: Decision-makers need to understand what mitigation strategies they can apply to assist their system while forgetting to ensure continuous steady performance with minimal degradation.
    \end{itemize}
\end{itemize}

\section{Research Goals and Questions}
\label{sec:intro:researchgoals}

This PhD research aims to promote green recovery in OL-CAIS while experiencing performance degradation due to unforeseen disruptive events.
Finding a trade-off between greenness and resilience in such systems imposes challenges in detecting performance degradation, balancing greenness and resilience, and understanding catastrophic forgetting.
To address these research challenges, we have derived the following research goals and questions by leveraging the Goal Question Matrix (GQM) approach proposed by \textcite{caldiera_goal_1994}.
These research goals and questions form the foundation of this PhD research.

\begin{itemize}
    \item \textbf{RG1}. \textit{Model the resilience of OL-CAIS behavior to support the decision-makers}. This research goal aims to address the research challenge \textbf{RC1} by answering the following research questions:
    \begin{itemize}
        \item \textbf{RQ1.1}. \textit{What metric can measure OL-CAIS performance to render its behavior during runtime?}
        To answer this question, we follow \textcite{henry_generic_2012} methodology to define a time-dependent quantiﬁable metric, where we define our performance as the system's ability to perform autonomous actions over iterations.
        By populating our function over a sliding window of iterations, we are able to render performance behavior during runtime.
        In particular, we introduce the Autonomous Classification Ratio (ACR), which quantifies system autonomy at any point during runtime. 
        The detailed answer to this question is discussed in our papers, \cite{rimawi_2024_gresilience,rimawi_modeling_2024}.
        \item \textbf{RQ1.2}. \textit{How can we define and distinguish acceptable and unacceptable performance levels in OL-CAIS from the decision-makers' perspective?}
        To answer this question, we monitor OL-CAIS's performance by observing the performance behavior while the system operates under its original settings. Based on this, we establish a threshold for acceptable performance, which is defined as the minimum ACR value observed during this period. The detailed answer to this question is discussed in our papers, \cite{rimawi_2024_gresilience,rimawi_modeling_2024}.
        \item \textbf{RQ1.3}. \textit{What key states and transitions in OL-CAIS performance evolution reflect resilience during and after disruptive events?}
        To answer this question, we analyze the ACR curve during run-time. By examining the curve, we identify patterns that correspond to different system states: a steady state, followed by a disruptive state triggered by a disruptive event, and ends with a final state after resolving the disruption. The detailed answer to this question is discussed in our papers, \cite{rimawi_2024_gresilience,rimawi_modeling_2024}.
    \end{itemize}
    \item \textbf{RG2}. \textit{Develop agent-based policies to balance resilience and greenness in OL-CAIS upon unforeseen disruptive events}. This research goal aims to address the research challenge \textbf{RC2} by answering the following research questions:
    \begin{itemize}
        \item \textbf{RQ2.1}. \textit{What are the different metrics required to measure resilience and greenness in OL-CAIS?}
        To answer this question, we measure resilience by analyzing the system's recovery actions' run-time and assess greenness by tracking CO$_2$ emissions and the number of human interactions. The detailed answer to this question is discussed in our papers, \cite{rimawi_2024_gresilience,bonfanti_gresilience_2023}.
        \item \textbf{RQ2.2}. \textit{What are the necessary components to automatically monitor OL-CAIS behavior and support decision-making?}
        To answer this question, we develop a novel framework to integrate an automatic decision-making assistant into OL-CAIS. The assistant, CAIS-DMA, monitors performance and suggests real-time actions that balance greenness and resilience. The development process simulates disruptive events, automates recovery decisions, and validates performance improvements. The detailed answer to this question is discussed in our papers, \cite{kadgien_cais-dma_2024, rimawi_2024_gresilience}.
        \item \textbf{RQ2.3}. \textit{Are the agent-based policies valuable in supporting OL-CAIS to restore its services while balancing greenness and resilience upon disruptions?}
        To answer this question, we propose one-agent policies based on an optimization model incorporating a weighted sum (WSM) to balance resilience and greenness.
        In two-agent policies, we explore the use of game theory to find a trade-off between greenness and resilience. The game forms two-agent policies, modeling resilience and greenness as two players in a non-cooperative game.
        Finally, in the reinforcement-learning-agent policy, we implement a reinforcement learning agent that continuously adjusts actions to optimize both greenness and resilience during disruptive events. The agent learns from feedback to improve decision-making dynamically.
        The detailed answer to this question is discussed in our papers,  \cite{rimawi_green_2022,bonfanti_gresilience_2023, rimawi_2024_gresilience}.
        \item \textbf{RQ2.4}. \textit{What are the major differences between the agent policies in terms of resilience and greenness?}
        To answer this question, we introduce our measurement framework to compare different decision-making policies during disruptions. We introduce measurable concepts to evaluate resilience and others to evaluate greenness. The policies are evaluated for resilience by their speed in recovering the system performance from degradation and their ability to hold performance from fluctuation outside an acceptable level (i.e., performance steadiness).
        Then, we evaluate the policies for greenness by their efficient energy consumption (i.e., green efficiency) and the policies' ability to improve autonomy of service delivery.
        The detailed answer to this question is discussed in our papers, \cite{rimawi_2024_gresilience,rimawi_modeling_2024}.
        \item \textbf{RQ2.5}. \textit{How can we optimize resource allocation using containerization to ensure efficient energy consumption?}
        To answer this question, we explore how containerization techniques can be applied to optimize resource allocation. Transitioning the system from a bare-metal system to a containerized one ensures more efficient energy usage and reduces its adverse effects.
    \end{itemize}
    \item \textbf{RG3}. \textit{Understand catastrophic forgetting in OL-CAIS in the aftermath of a disruptive event and its resolution}. This research goal aims to address the research challenge \textbf{RC3} by answering the following research questions:
    \begin{itemize}
        \item \textbf{RQ3.1}. \textit{What are the patterns in the resilience model that indicate catastrophic forgetting in OL-CAIS?}
        To answer this question, we trace OL-CAIS performance using our resilience model after resolving a disruptive event. By comparing pre- and post-disruption performance, the system detects whether OL-CAIS retains the original environmental settings or shows signs of catastrophic forgetting. TThe detailed answer to this question is discussed in our papers, \cite{rimawi_modeling_2024,rimawi_2024_gresilience}.
        \item \textbf{RQ3.2}. \textit{How can we ensure that OL-CAIS maintains steady performance over time, even when facing frequent disruptions?}
        To answer this question, we monitor OL-CAIS performance across multiple disruptions and their resolutions. We support the system’s decision-making at each instance of performance degradation to ensure long-term steadiness.
    \end{itemize}
\end{itemize}

\section{Our Contribution}
We initiated this work to understand how different non-functional properties of a cyber-physical system can be traded off to recover from a disruptive event. In particular, we started focusing on two properties, resilience, and greenness and explored game theory techniques,~\cite{rimawi_green_2022}.
Then, we further investigated optimization techniques and sketched the first version of the GResilience framework with the one-agent and two-agent policies,~\cite{bonfanti_gresilience_2023}. 
Later, we implemented an extendable decision-making assistant, CAIS-DMA, that can operate on an OL-CAIS to monitor and actuate the trade-off policies,~\cite{kadgien_cais-dma_2024}. This framework empowers users to create custom policies and select actions to recover an OL-CAIS. 
In~\cite{rimawi_modeling_2024}, we drafted a behavioral model to represent resilience in OL-CAIS and provide the system's managers with appropriate decision-making instruments.
We then leveraged this knowledge and tools to build a holistic methodology that helps decision-makers automatically select recovery actions for an OL-CAIS that ensures a certain level of greenness,~\cite{rimawi_2024_gresilience}.
These papers are the cornerstones of this PhD research, which aggregates all our findings and orchestrated the following key contributions:

\begin{enumerate}
    \item \textit{Resilience Modeling in OL-CAIS}:
    We developed a novel resilience model that characterizes OL-CAIS performance through three operational states: steady, disruptive, and final. This model enables automatic detection of performance states and tracing system performance evolution in runtime during disruptive events.
    \item \textit{The GResilience Framework for Decision-Making}:
    We introduced the GResilience framework to recommend recovery actions that balance greenness and resilience. This framework leverages three agent-based decision-making policies:
    \begin{itemize}
        \item One-agent policies, which apply multi-objective optimization to rank feasible actions.
        \item Two-agent policies, which use game theory to resolve trade-offs between greenness and resilience.
        \item Reinforcement learning policies, which optimize actions based on rewards derived from greenness-resilience states.
    \end{itemize}
    \item \textit{Decision-Making Assistant}:
    We designed an extendable conceptual framework that equips OL-CAIS with the tools and APIs required for performance evolution tracking, invoking decision-making policies, actuating recommended actions, and simulating its environment.
    \item \textit{Measurements Framework}:
    We designed a measurement framework to assess the effectiveness of decision-making policies in achieving green recovery. It introduces measurable concepts, including recovery speed and performance steadiness for resilience and green efficiency and service autonomy for greenness.
    \item \textit{Exploration of Catastrophic Forgetting}:
    We analyzed the impact of catastrophic forgetting in OL-CAIS, particularly during repeated disruptions. We mitigate the forgetting effects to maintain consistent performance by supporting decision-making upon every performance degradation.
    \item \textit{Containerization for Energy Optimization}:
    We introduced a containerization methodology for OL-CAIS, optimizing resource allocation and significantly reducing CO$_2$ emissions compared to traditional bare-metal setups. This approach demonstrated the potential for halving emissions while maintaining system performance.
    \item \textit{Empirical Validation and Findings}:
    Through extensive real-world and simulated experiments, we validated the effectiveness of our models and frameworks. Key findings include:
    \begin{itemize}
        \item The resilience model has effectively reflected the system performance evolution. 
        \item The GResilience framework outperformed internal policies, achieving faster recovery, reducing performance fluctuations, and enhancing autonomy.
        \item RL-agent policies consistently outperformed two-agent and one-agent policies in balancing resilience and greenness.
        \item Containerization demonstrated a marked reduction in CO$_2$ emissions, highlighting its potential for green innovation in OL-CAIS.
    \end{itemize}
\end{enumerate}

\section{Outline}
\label{sec:intro:outline}
This thesis is organized into four main parts, each addressing an aspect of the research. Below, we provide an overview of the structure and goals of each part and its respective chapters.

The first part, \textit{Foundation and Background}, lays the groundwork for this research. Chapter~\ref{ch:intro} introduces the research context, highlighting the key challenges, goals and contributions while establishing the motivation for this work.
Chapter~\ref{ch:background} elaborates on the fundamental concepts relevant to this study, such as Online Collaborative AI Systems, resilience, greenness, and decision-making policies, covering optimization models, game theory, and reinforcement learning. 
Chapter~\ref{ch:lr} presents a comprehensive systematic literature review (SLR), analyzing previous studies, their methodologies, and findings. Additionally, it includes a discussion of related work, situating this research within the scientific landscape.

The second part, \textit{Research Methodology and Approach}, focuses on the methodologies employed in this research.
Chapter~\ref{ch:methods} introduces the overarching methodology used to address the research questions. 
Chapter~\ref{ch:dmframeworks} details the decision-making frameworks, starting with the resilience model and the GResilience framework, followed by the decision-making assistant and measurement frameworks. These frameworks include policies designed to handle various scenarios effectively. 
Chapter~\ref{ch:greenresiliencecps} explains the process of building green resilience in OL-CAIS, by integrating the use containers to create components replicas and reduce energy consumption.

The third part, \textit{Experimental Evaluation}, validates the proposed methodologies through experimental investigations. 
Chapter~\ref{ch:coral} introduces the CORAL case study, showcasing its significance as a testbed. It discusses the dynamics of online learning and disruptive events in CORAL. 
Chapter~\ref{ch:expdesign} outlines the design of experiments, including the setup and execution, focusing on evaluating trade-offs between greenness and resilience and the role of containerization. 
Chapter~\ref{ch:results} presents the experimental results to answer the research questions, drawing comparisons between real-world and simulated scenarios while examining the impact of containerization.

The fourth and final part, \textit{Reflections and Conclusions}, synthesizes the insights gained from the research while reflecting on its limitations and proposing directions for future work. Chapter~\ref{ch:discussion} explores the threats to the validity and outlines avenues for future work, emphasizing the broader implications of this research. Chapter~\ref{ch:conclusion} concludes the thesis with a comprehensive discussion of the findings, summarizing the contributions and highlighting the key takeaways from this work.

\chapter{Background}
\label{ch:background}
In this chapter, we provide an overview of the essential elements of this research context: i) the online collaborative AI system, ii) its qualities, particularly resilience, and greenness, and iii) the decision-making policies, including optimization, game theory, and reinforcement learning.

\section{Online Collaborative AI Systems}
\label{sec:bg:cais}

A system generally refers to an organized set of interconnected components working together to achieve a common goal. These components can be physical, such as machinery, or abstract, like software, and are bound by relationships and interactions that drive the system's overall functionality,~\cite{dick_requirements_2017}.
A cyber-physical system (CPS) is a more advanced configuration of a system, in which, computational and physical processes are deeply integrated. 
CPS consists of embedded computers and networks that monitor and control the physical entities with which they interact in real-time,~\cite{baheti2011cyber}. CPS enables the seamless fusion of the digital and physical worlds, with applications ranging from autonomous vehicles to healthcare and smart manufacturing,~\cite{baheti2011cyber}.
A further evolution of CPS is the human-centric CPS, where the system is designed to interact closely with humans, often in collaborative environments. In such systems, humans and machines work together, with the CPS adapting to human inputs and actions, enhancing decision-making and productivity through continuous learning and interaction,~\cite{pinzone_framework_2020}.
One type of human-centric CPS is the Online Collaborative AI System (OL-CAIS).

OL-CAIS is a human-centric CPS that collaborates with humans within a shared environment to achieve a common goal,~\cite{camilli_risk-driven_2021}.
Humans influence the system by interacting with its AI model. 
The AI model learns from these interactions and replicates them autonomously, enabling collaboration toward a common goal.
In this PhD, we focus on collaborative AI systems learning online from human interactions to solve a classification problem. 
\textit{Online Learning} is a type of continual learning where the learning model updates are done on a single data point basis, or in other words, has a batch size of one data instance, \cite{lesort2020continual}.
Online learning aims to keep necessary information (i.e., learning features) before discarding the data instance.
After the learner receives these features, it estimates the probability for each class in its attempt to predict the features' label, ~\cite{zadrozny_transforming_2002}.

\begin{figure}[ht!]
    \centering
    \centerline{\includegraphics[width = \textwidth]{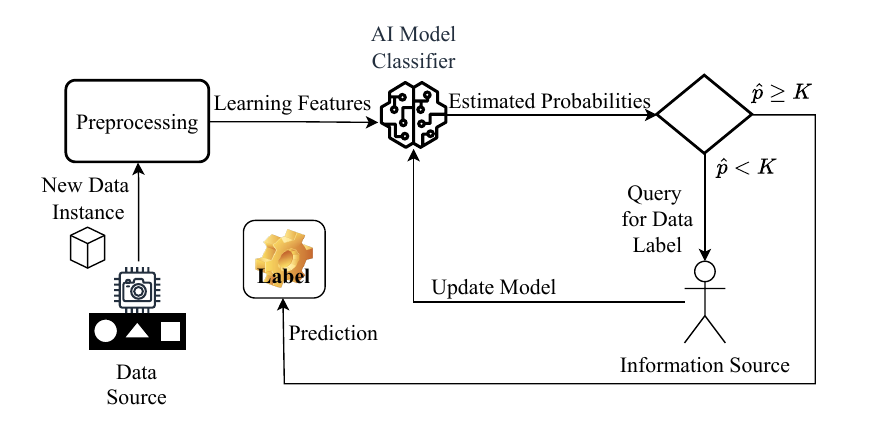}}
    \caption{Decision-making flow to learn new arriving data instances in an online learning fashion.}
    \label{fig:bg:onlinelearnign}
\end{figure}

Fig.~\ref{fig:bg:onlinelearnign} shows the online learning cycle for a new data instance passed for preprocessing to extract the learning features and then to the classifier.
Classifiers often output scores that rank given features based on their likelihood of belonging to a particular class. However, these scores alone are insufficient for decision-making processes requiring precise probability estimates. 
In OL-CAIS, inaccurate probabilities could lead to sub-optimal or unreliable decisions.
Thus, estimating accurate probabilities of class membership in applications like OL-CAIS is particularly important.
In their early stages, methods for calibrating classifiers' scores to obtain probability estimates have been limited to two-class problems. To address this limitation, researchers have developed techniques to extend these methods to multiclass problems.
The key idea is to decompose a multiclass problem into multiple binary classification problems. 
Then, we can derive accurate multiclass probability estimates by calibrating the scores from each binary classifier and combining them.
One innovative approach involves using naive Bayes and support vector machine classifiers,~\cite{zadrozny_transforming_2002}. These classifiers, while effective at ranking, often produce scores that are not well-calibrated. We can transform these scores into reliable probability estimates by applying calibration methods such as isotonic regression or sigmoid fitting. This process ensures that the scores are interpretable and can be effectively used in various high-level decision-making systems.

Once the AI model within OL-CAIS has estimated the probabilities for each class, the system evaluates these probabilities against a pre-defined confidence threshold, $K$. 
This threshold helps decision-makers to operate with some level of confidence in the model's decision,~\cite{huang2016anticipatory}.
If the maximum estimated probability, $\hat{p}$, for a given class exceeds this threshold, $\hat{p} \geq K$, the system proceeds autonomously, taking actions based on the classification.
However, when the maximum estimated probability falls below the confidence threshold, $\hat{p} < K$, the system recognizes insufficient certainty to trust the AI model's prediction. 
In such cases, OL-CAIS leverages active learning principles, which query the information source (i.e., the human collaborator) to provide additional labeled data. 

The human feedback serves two purposes: i) it allows the system to correct any potential misclassifications, and ii) it helps the AI model adapt and update its internal parameters based on new, human-labeled data points.
The human labels the data instances by performing one or more interactions.
These interactions vary from one domain to the other, including hand gestures, voice commands, eye gaze, brain signals, and demonstration, to mention a few, \cite{hu2024human, mukherjee_survey_2022}.
OL-CAIS utilizes its sensors to capture human interactions and convert them into labels to update its AI model.
By selecting one of the two actions (i.e., proceeding autonomously or querying the human) OL-CAIS completes a collaborative \textit{iteration},~\cite{lesort2020continual}. 

\section{Non-Functional Properties}
\label{sec:bg:qualities}

The properties of a system refer to the characteristics that emerge from the interactions among its components. These properties can be classified into functional and non-functional,~\cite{dick_requirements_2017}. Functional properties are directly related to specific tasks or functions that the system must perform, such as processing data or controlling machinery. In contrast, non-functional properties, also known as quality attributes, define how a system performs these functions rather than what it does. Examples of non-functional properties include performance, reliability, usability, and sustainability, among others,~\cite{dick_requirements_2017}.
Among these non-functional properties are resilience and greenness. Resilience refers to the system's ability to recover its performance after an event has degraded it from an unacceptable level to an acceptable level,~\cite{henry_generic_2012}. Greenness, on the other hand, is defined as the system's efficient energy consumption while minimizing its adverse effects,~\cite{kharchenko_concepts_2017}. Balancing these two qualities, especially during disruptions to OL-CAIS, is the primary goal of this PhD research. This section introduces these two qualities in more detail.

\subsection{Resilience}
\label{subsec:bg:qualities:resilience}

Resilience is a non-functional property that enables the system to restore its performance from degradation to an acceptable level~\cite{henry_generic_2012}. For an event to qualify as disruptive, it must cause a measurable decline in the system's ability to perform its intended functions. The resilience research landscape recognizes the dynamic nature of disruptive events and explores various sources and consequences. These events can manifest in diverse forms, either endogenous or exogenous~\cite{mouelhi_predictive_2019}. Endogenous events stem from internal sources such as system defects~\cite{januario_distributed_2019}, while exogenous events arise from external factors like environmental changes or attacks~\cite{henry_generic_2012, zarandi_detection_2020, liu_distributionally_2022}.

In this PhD research, we focus on exogenous disruptive events, which impact the system by reducing the value of its performance metric to an unacceptable level compared to its original state~\cite{henry_generic_2012, hosseini_review_2016}. A performance metric is a measurable criterion used to assess how well the system is functioning, such as response time or throughput. A significant reduction in these metrics is essential for the system to enter a disrupted state, necessitating recovery actions to restore system performance. 
A \textit{disrupted state} is the condition of a system following a disruptive event that significantly impacts its performance. As it absorbs the disruption, the system experiences a loss in performance levels. This state continues until the system initiates recovery actions,~\cite{hosseini_review_2016}.
\textit{Recovery actions} are interventions aimed at restoring the system's performance after a disruption. These actions should reverse performance degradation, bring the system performance to an acceptable level, and ideally reduce both the impact and duration of the disruption state, \cite{henry_generic_2012, hosseini_review_2016}. 
They encompass efforts that absorb the disruption, adapt to new circumstances, and restore performance to an acceptable level, \cite{hosseini_review_2016}.

\begin{figure}[ht!]
    \centering
    \centerline{\includegraphics[width = \textwidth]{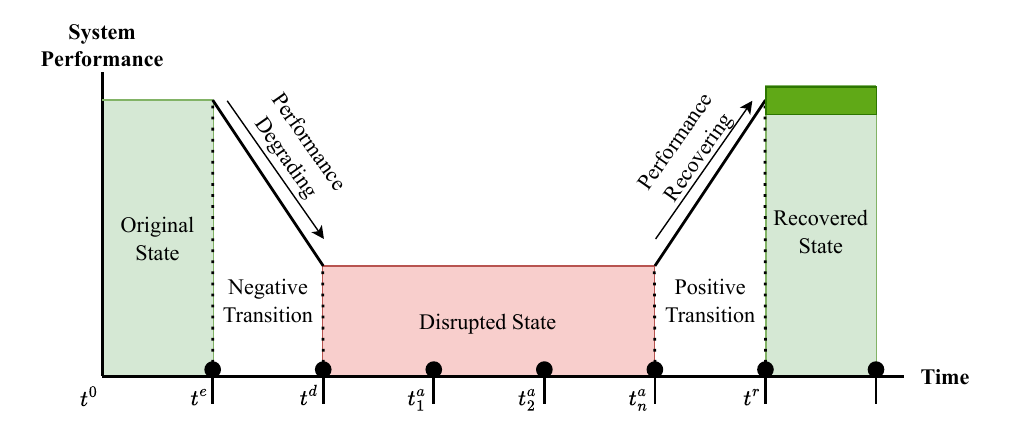}}
    \caption{Resilience model upon disruptive event.}
    \label{fig:bg:resiliencestates}
\end{figure}

Fig.~\ref{fig:bg:resiliencestates} shows the system resilience model by measuring the system performance as a function of time (i.e., performance metric), describing three states: i) Original State, ii) Disrupted State, and iii) Recovered State. 
The system operates in its original state from the initial time, t$^0$, until it faces a disruptive event at event time, t$^e$.
The event leads to a negative transition in the performance metric value entering the disrupted state at a disrupted time, t$^d$.
During the disrupted state, the system absorbs the disruptive settings and then applies one or more recovery actions, $a$,  at action time (t$^a_1$, t$^a_2$, and/or, t$^a_n$) to bring the system's performance value into a positive transition until it recovers to an acceptable value at recovered time, t$^r$. 
In OL-CAIS context, the time is defined by iterations. 

\subsection{Greenness}
\label{subsec:bg:qualities:greenness}

Sustainability in Cyber-Physical Systems (CPS) revolves around the ability of these systems to operate efficiently and continuously, minimizing their environmental impact while maintaining operational performance.
In this context, building a sustainable CPS concerns energy management, self-sustainability, and reducing adverse environmental effects,~\cite{estevez2017green}. 
First, energy management refers to the strategic control of energy resources to ensure system efficiency, balancing energy consumption with operational demands.
Secondly, self-sustainability in CPS involves using energy-harvesting technologies to allow devices to operate independently from external power sources, often through renewable energy collection.
Finally, reducing adverse environmental effects focuses on minimizing the environmental impact, such as pollution and CO$_2$ emissions, by optimizing system energy use and deploying monitoring and controlling mechanisms,~\cite{estevez2017green, kharchenko_concepts_2017}.

Energy consumption, which refers to the amount of energy used by CPS devices to perform computational tasks and physical processes, is at the core of sustainability. 
Efficient energy consumption is critical to reducing these systems' environmental impact.
By managing energy consumption intelligently, CPS systems can reduce carbon emissions.
Carbon emissions, in particular, are a key environmental concern associated with energy consumption in CPS,~\cite{estevez2017green, kharchenko_concepts_2017}.
The CO$_2$ dioxide emission is derived from the energy consumption quantified as kilowatt-hours. The emitted CO$_2$ generated by the amount of energy expressed as kilograms of CO$_2$-equivalents or shortly CO$_2$eq,~\cite{ourworldindataGreenhouseEmissions, energyinstitute2023, mila_codecarbonio_nodate}.
The adverse effects of energy consumption, such as CO$_2$ emissions, are critical targets for green CPS, where minimizing such impacts is part of achieving system greenness,~\cite{estevez2017green}.

Greenness is a non-functional property that directly contributes to sustainability by promoting efficient energy use and minimizing environmental impact,~\cite{kharchenko_concepts_2017}. 
In our context, achieving greenness is not just about reducing energy consumption but also about reducing its adverse effects, particularly CO$_2$ emissions. 
By controlling energy effects, OL-CAIS can achieve green efficiency while recovering its performance from disruptions.
This involves managing direct and indirect impacts on energy use, such as service repetitions. 
In OL-CAIS, performing tasks with low-accuracy autonomous actions may lead to service abortion and repetition, which increases energy consumption and CO$_2$ emissions, negatively impacting greenness. 
Furthermore, repeated tasks prolong recovery time, further undermining system resilience.
Thus, it is critical to identify the actions that best balance greenness and resilience at each decision-making point.

\section{Decision-Making Policies}
\label{sec:bg:dmp}

A decision-making policy in the context of an OL-CAIS refers to the set of rules or strategies that guide the system's behavior in choosing between autonomous actions and human interactions. 
These policies enable the system to operate autonomously while utilizing human input only when needed. This approach ensures that the system performs tasks accurately in real-time, allowing humans to focus on more critical tasks or step in for verification when the system encounters uncertainty.
As OL-CAIS operates in dynamic environments, it continuously learns from its interactions, developing internal policies to enhance performance. 
Such policies are critical for maintaining a system's ability to perform tasks autonomously while ensuring that human collaborators are engaged when necessary.

In OL-CAIS, internal decision-making policies may already consider both greenness and resilience, aiming to maintain efficient energy usage while ensuring the system's performance recovers quickly from degradation. 
However, these policies might not always be optimal when the system encounters significant disruptions. 
In such cases, additional support from intelligent agents is required to navigate the disruption and ensure that the system remains both autonomous and energy-efficient, taking actions that best balance resilience and greenness during these challenging situations.

In this section, we introduce the fundamental background for three types of intelligent agents: i) One-Agent considering optimization model, ii) Two-Agent by defining a theoretical game approach, and iii) RL-Agent by observing the environment with a reinforcement learning agent.

\subsection{Optimization Model}
\label{subsec:bg:dmp:opt}

Optimization models are mathematical frameworks used to find the best possible solution by maximizing or minimizing specific objectives. 
These models can handle either single or multiple objectives. Single-objective optimization focuses on optimizing one criterion, while multi-objective optimization considers trade-offs between several conflicting objectives. 
For example, in engineering, multi-objective optimization might involve optimizing both the cost and the performance of a system, requiring a balance between the two,~\cite{gunantara_review_2018}.

One common approach within multi-objective optimization is the Weighted Sum Model (WSM), which transforms multiple objectives into a single scalar objective by assigning weights to each criterion. This method allows for a prioritized balance of objectives, where higher weights indicate greater importance, leading to solutions that reflect the trade-offs between competing goals,~\cite{gunantara_review_2018}.
WSM finds the optimal solution that maximizes a global score by following five steps: i) building the decision matrix of all solutions and objectives, ii) normalizing the values, iii) calculating the weights of all objectives, iv) calculating the global score for each solution, and v) ranking solutions based on the global score and select the solution with the maximum global score.

We need to define the space and objectives measurements to build the decision matrix. The space represents the set of all possible solutions to explore (i.e., Solutions = \{Sol$_1$, Sol$_2$, ..., Sol$_m$\}). Then, we need to define the objectives measurements that are the functions to measure each objective given the solutions, such that Objective$_n$ function $Obj_n: Solutions \rightarrow \mathbb{R}$. 
Table~\ref{tab:bg:decisionmatrix} shows how to initial the decision matrix, where each row represents the objectives measurements for each solution, where N(*) represents the normalize function, which brings all values between minus one and one, $N: \mathbb{R} \rightarrow [-1, 1]$. 
Normalization is a critical step in optimization to ensure that all objective values are on a comparable scale (e.g., [-1, 1]). 
Normalization techniques include min-max normalization, which scales the values based on the minimum and maximum in the dataset, and norm-based methods such as L1-norm and L2-norm.
L1-norm normalization focuses on the absolute sum of the values, while L2-norm, also known as least squares, minimizes the sum of squared differences between predicted and actual values. 
The least squares method (LSM) is widely used in various fields, ensuring an unbiased, minimal-variance solution when measurements have a normal distribution,~\cite{bektacs2010comparison}.
In our work, we apply L2-norm normalization to minimize the effect of outliers, ensuring that the decision matrix’s values are optimized efficiently.

\begin{table}[ht!]
\caption{General decision matrix of weighted sum model.}
\begin{center}
\begin{tabular}{ccccc}
    \hline
    \textbf{Solutions} & \textbf{Obj$_1$} & \textbf{Obj$_2$} & $\cdots$ & \textbf{Obj$_n$} \\\hline
    \textbf{Sol$_1$} & N(Obj$_1$(Sol$_1$)) & N(Obj$_2$(Sol$_1$)) & $\cdots$ & N(Obj$_n$(Sol$_1$))  \\
    \textbf{Sol$_2$} & N(Obj$_1$(Sol$_2$)) & N(Obj$_2$(Sol$_2$)) & $\cdots$ & N(Obj$_n$(Sol$_2$))  \\
    $\vdots$ & $\vdots$ & $\vdots$ & &$\vdots$   \\
    \textbf{Sol$_m$} & N(Obj$_1$(Sol$_m$)) & N(Obj$_2$(Sol$_m$)) & $\cdots$ & N(Obj$_n$(Sol$_m$))  \\\hline
\end{tabular}
\label{tab:bg:decisionmatrix}
\end{center}
\end{table}

\begin{figure}[ht!]
\centerline{\includegraphics[width=\textwidth]{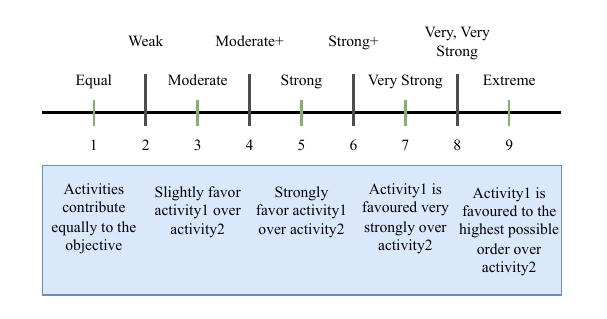}}
\caption{The fundamental scale of absolute numbers.}
\label{fig:bg:absolutenumbers}
\end{figure}

Numerous techniques can be used to calculate the objectives' weights. We will discuss the Analytic Hierarchy Process (AHP), which is the most used with WSM.
AHP is a pairwise comparison to derive ratio scales,~\cite{teknomo_analytic_2006, saaty_decision_2008}. 
The process starts by performing a paired comparison based on the contribution level toward the main goal. The comparison takes values from the fundamental scale of absolute numbers defined by AHP developer \textcite{saaty_decision_2008}. Fig.~\ref{fig:bg:absolutenumbers} shows the scale of absolute numbers starting from one for the equal contribution of the objectives pair to nine for the extreme contribution of the first objective over the second.
The domain expert can use the scale to define the comparison matrix for all objectives. The matrix dimensions are equal to the total number of objectives, n (e.g., $n \times n$ for n objectives). 

Table~\ref{tab:bg:comparisonmatrix} shows an example for n objectives matrix. 
The next step is to write the representative value from the scale of absolute numbers, then normalize these values by dividing each one by the sum of all values.
Then, the weight of each objective is equal to the average value of its corresponding row.
To verify that the weights are consistent and not random, AHP solves the consistency ration by finding the consistency index and dividing it by the
random index. 
Consistency index involves comparing the calculated eigenvalues of the decision matrix to the number of objectives, a higher index indicates greater inconsistency. While the random index is a pre-calculated value based on the size of the decision matrix. It represents the average consistency index for randomly generated matrices.
A consistency ratio less than or equal to 0.1, means a consistent judgments from the decision-maker's, and the weights assigned to the objectives can be used in the analysis.

\begin{table}[ht!]
\caption{Analytic hierarchy process comparison matrix for n-objectives.}
\begin{center}
\begin{tabular}{ccccc}
    \hline
    & \textbf{Obj$_1$} & \textbf{Obj$_1$} & $\cdots$ & \textbf{Obj$_n$} \\\hline
    \textbf{Obj$_1$} & Equal & Strong &$\cdots$ & Moderate  \\
    \textbf{Obj$_2$} & Strong$^{-1}$ & Equal & $\cdots$&Moderate$^{-1}$  \\
    $\vdots$ & $\vdots$ & $\vdots$ & & $\vdots$ \\
    \textbf{Obj$_n$}  & Moderate$^{-1}$ & Moderate &$\cdots$ &Equal  \\\hline
\end{tabular}
\label{tab:bg:comparisonmatrix}
\end{center}
\end{table}

After finding the objectives' weights, the WSM calculates each solution's global score by finding the weighted sum of its normalized objectives values as shown in Eq.~\eqref{eq:bg:wsm}, which finds the global score, G, of the solution, Sol$_m$. Where $\mathbf{w}_{Obj_{i}}$ is the weight of the objective $Obj_i$.

\begin{equation}
    \label{eq:bg:wsm}
     G(Sol_m) = \sum_{i=1}^n \mathbf{w}_{Obj_{i}} \times N(Obj_i(Sol_m))
\end{equation}

The last step is selecting the solution with the highest global score, which represents the optimal solution. This can be expressed using the arg max function as shown in Eq.~\eqref{eq:bg:argmax}, where selects the solution, $Sol$ with the highest global score, $G(Sol)$, where $m$ ranges over all possible solutions.

\begin{equation}
    \label{eq:bg:argmax}
    Sol_{\text{optimal}} = \underset{Sol}{\arg\max} \, (G(Sol_1), G(Sol_2), \cdots, G(Sol_m))
\end{equation}

\subsection{Game Theory}
\label{subsec:bg:dmp:gt}

Game Theory is a branch of mathematics that models the interactions between decision-makers (players) in situations where their choices influence not only their outcomes but also the outcomes of others. It is primarily used to study conflict and cooperation in these decision-making scenarios, often involving multiple competing objectives,~\cite{satapathy_game_2016}.
In non-cooperative games, each player acts independently, aiming to maximize their payoff without collaborating with others.
In cooperative games, players form alliances or coalitions and work together to improve their collective payoffs.
Both cooperative and non-cooperative players may interact with each other. However, in non-cooperative games, they cannot make binding agreements.
In our research, we focus on non-cooperative games with players who share a common goal, but each one has different preferences.

The Battle of Sexes is a non-cooperative game, first introduced by \textcite{luce_games_1989}. 
It is a well-known game involving two players who must choose between two activities for an evening: a football game or a musical show. The players have different preferences but would still rather spend the evening together than apart,~\cite{satapathy_game_2016}.
The boy (player 1) prefers football, while the girl (player 2) prefers the musical.
Table~\ref{tbl:bg:battleofsexes} represents the payoff matrix for this game, where the boy's strategies are listed on the rows, he can choose Football with probability, p, or Music with probability, 1 - p. The girl's strategies are listed in the columns and she can choose either Football with probability, q, or Music with probability, 1 - q.
Each cell in the matrix contains the payoffs for both players in the form (Boy’s payoff, Girl’s payoff). 
The Expected Payoff (EP) for the boy is shown in the last column, based on the girl's strategy (probability q), and the expected payoff for the girl is shown in the last row, based on the boy's strategy (probability p).

\begin{table}[ht!]
\centering
\caption{Payoff matrix for the Battle of Sexes.}
\begin{tabular}{ccccc}
   & & \multicolumn{2}{c}{\textbf{Girl}}& \\
   &  & \textbf{Football} (q) & \textbf{Music} (1 - q)&\textbf{Boy's EP} \\ 
   \cline{2-5}
   \multirow{2}{*}{\rotatebox[origin=c]{90}{\textbf{Boy}}} & \textbf{Football} (p) & (2, 1) & (0, 0) & $q \times 2 + (1-q) \times 0$ \\ 
   & \textbf{Music} (1 - p)& (0, 0) & (1, 2) & $q \times 0 + (1-q) \times 1$ \\
   & \textbf{Girl's EP} & $p \times 1 + (1-p) \times 0$ & $p \times 0 + (1-p) \times 2$ & \\\cline{2-5}
\end{tabular}
\label{tbl:bg:battleofsexes}
\end{table}

The battle of sexes contains more than one solution. First, \textit{pure Nash equilibrium}, where both players act in a way that neither can improve their outcome by unilaterally changing their decision. 
In this game, there are two pure Nash equilibria: i) Both players go to the football game (2,1), which is better for the boy, or ii) Both players attend the musical show (1,2), which is better for the girl.
Neither player has a dominant strategy because their preferences depend on the other player's choices.

Second solution, \textit{mixed strategy Nash equilibrium}, where both players randomize their choices to maximize their expected payoffs. The boy will choose Football with a probability of p, and the girl will choose Football with a probability of q. 
To find the mixed strategy Nash equilibrium, we balance the expected payoffs such that each player is indifferent to their choices. 
This balance reflects situations where coordination is difficult, and both players must account for the other's uncertainty. 
Mathematically it occurs when the expected payoffs from either strategy (Football or Music) are equal.
For the boy, we set the payoff for choosing Football equal to that of choosing Music as Eq.~\eqref{eq:bg:q} shows, where the equation terms represent the boy's EP from Table~\ref{tbl:bg:battleofsexes}.
\begin{equation}
    \label{eq:bg:q}
    \begin{split}
        & 2q = 1 - q \\
        & q = \frac{1}{3}
    \end{split}
\end{equation}

For the girl, we balance the payoffs by setting the payoff for choosing Football equal to that of choosing Music as Eq.~\eqref{eq:bg:p} shows, where the equation terms represent the girl's EP from Table~\ref{tbl:bg:battleofsexes}.
\begin{equation}
    \label{eq:bg:p}
    \begin{split}
        & p = 2 - 2p \\
        & p = \frac{2}{3}
    \end{split}
\end{equation}

Thus, the mixed strategy Nash equilibrium occurs when the boy chooses Football with probability \( p = \frac{2}{3} \) and the girl chooses Football with probability \( q = \frac{1}{3} \).
This mixed strategy equilibrium ensures that neither player has an incentive to deviate from their chosen probability mix, as doing so would not improve their expected payoffs.

\subsection{Reinforcement Learning}
\label{subsec:bg:dmp:rl}

\begin{figure}[ht!]
    \centering
    \centerline{\includegraphics[width = \textwidth]{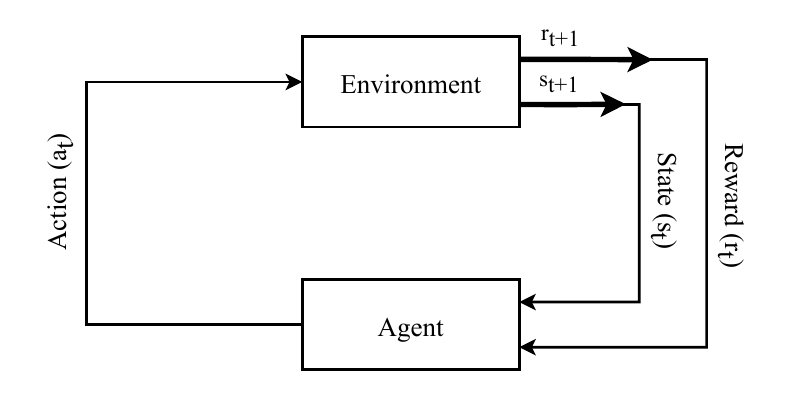}}
    \caption{Reinforcement learning -- agent-environment interaction.}
    \label{fig:bg:reinforcementlearning}
\end{figure}

Reinforcement learning (RL) is a fundamental machine learning paradigm in which an agent learns to map situations to actions with the goal of maximizing a numerical reward signal. 
This approach differs from supervised learning as it focuses on sequential decision-making and long-term reward maximization rather than immediate feedback for each action. 
The learning evolves through an agent's trial-and-error interaction with the system environment, then observing the system's states to maximize its delayed reward,~\cite{sutton2018reinforcement}. 
Specifically, RL models the environment into a Markov Decision Process, which is a tuple of possible states, $S$, possible actions, $A$, 
distribution of current reward given (state, action) pair, $R(r_t|s_t, a_t)$
distribution over next state given (state, action) pair, $P(s_{t+1}|s_t, a_t)$
and the discount factor (determines the value of future rewards), $\gamma$.
Fig.~\ref{fig:bg:reinforcementlearning}, shows the reinforcement agent interaction with the environment, where the agent explore-exploit the action, $a_t$ according to the current state, $s_t$, and reward, $r_t$, then updates the learning model according to the new state, $s_{t+1}$, and the calculated reward, $r_{t+1}$. This process continues until the learning converges (finding the optimal action-value function).

Reinforcement learning algorithms can be classified into two broad categories: \textit{on-policy} and \textit{off-policy} methods. 
In on-policy methods, the agent learns the value of the policy it is currently using to make decisions. In other words, the behavior policy and the target policy are the same,~\cite{sutton2018reinforcement}. 
An example of an on-policy method is the SARSA algorithm~\cite{sutton2018reinforcement}, where the agent updates its value function based on the actions taken by following its current policy.
On the other hand, off-policy methods use separate policies for behavior and learning. The agent learns the value of an optimal target policy while following a different exploratory behavior policy. Q-learning is the most well-known off-policy algorithm, as it updates the value function based on the optimal action-value function (which might not correspond to the actions taken under the exploratory behavior policy),~\cite{sutton2018reinforcement, watkins1992q}.

In this PhD research, we use Q-Learning as an off-policy temporal difference control,~\cite{watkins1992q}. This mechanism learns the action-value function, Q, which approximates the optimal action-value function (q*).  Eq.~\eqref{eq:bg:qfunction} shows the Q function, where $Q(S_t, A_t)$ is the current estimate of the Q-value, $\alpha$ is the learning rate, $\alpha \in (0, 1]$, $R_{t+1}$ is the reward received, $\gamma$ is the discount factor, $\gamma \in [0, 1]$, and the $\max_{a \in A} Q(S_{t+1}, a)$ is the maximum Q-value for the next state given all possible actions,~\cite{sutton2018reinforcement}.

\begin{equation}
Q(S_t, A_t) \gets Q(S_t, A_t) + \alpha \left[ R_{t+1} + \gamma \max_{a \in A} Q(S_{t+1}, a) - Q(S_t, A_t) \right]
\label{eq:bg:qfunction}
\end{equation}

In a tabular setting, Q-learning is commonly referred to as the Q-table. Each entry in this table corresponds to a state-action pair $(S,A)$, and the value of that entry is the expected future reward for taking action, A, in state, S. 
Over time, as the agent explores and updates the Q-table, it learns which actions yield the highest rewards for each state. 
Once the table converges, the agent can act optimally by choosing the action that maximizes the Q-value for the current state,~\cite{sutton2018reinforcement,watkins1992q}.

\chapter{Literature Review}
\label{ch:lr}
In this chapter, we present a comprehensive overview of the existing research relevant to this PhD project, structured into two main sections. 
The first section details a systematic literature review (SLR) conducted at the outset of this PhD to identify critical research gaps in the field of Green Resilient Cyber-Physical Systems. 
This systematic approach rigorously examined the body of knowledge on resilience, greenness, and learning-based methods in CPS, allowing for an in-depth analysis of trends, methodologies, and ongoing challenges. 
The findings from this SLR provided the basis for this dissertation, pinpointing open questions and shaping the research goals, methodology, and contributions of this work.

The second section explores related work, focusing on studies closely aligned with the objectives of this PhD. 
Here, we review recent advancements addressing resilience, greenness, online learning, decision-making mechanisms, and decision-making assistants in CPS, with particular attention to Online Collaborative Artificial Intelligence Systems. 
Together, these sections lay a strong foundation for the research conducted in this PhD, framing the need for innovative approaches to balance greenness and resilience in dynamic, real-world CPS environments.

\section{Systematic Literature Review}
This section presents a systematic literature review (SLR) conducted to identify and analyze research gaps in Green Resilient Cyber-Physical Systems. 
As this PhD research aims to enhance resilience and greenness in CPS through advanced decision-making and learning-based methods, the SLR serves as a foundational study to map the existing landscape, uncover challenges, and inform the research direction. 
Following the software engineering guidelines for reporting secondary studies (SEGRESS) by \textcite{kitchenham_segress_2022}, the SLR ensures a rigorous and methodical approach in identifying, selecting, and synthesizing relevant studies in the field. 

The primary goal of this SLR is to understand the state of research concerning resilience and green aspects in CPS and the use of learning-based techniques to address these challenges. The SLR specifically addresses the following research questions:

\begin{itemize}
    \item \textbf{SLR--RQ1:} What studies exist that investigate resilience in smart CPS?
    \begin{itemize}
        \item \textbf{SLR--RQ1.1:} How do these studies leverage learning-based methods to monitor and mitigate performance degradation in CPS?
        \item \textbf{SLR--RQ1.2:} What methods are used for detecting and recovering from resilience-related issues?
    \end{itemize}
    \item \textbf{SLR--RQ2:} What studies focus on green aspects in CPS, including energy efficiency and CO$_2$ emission reduction?    
    \begin{itemize}
        \item \textbf{SLR--RQ2.1:} What measurements and metrics are used to assess greenness in CPS?
        \item \textbf{SLR--RQ2.2:} Do any studies integrate green strategies when addressing resilience in CPS?
    \end{itemize}
\end{itemize}

\subsection{SLR Methodology}
\label{subsec:lr:slr:method}
The methodology for this SLR follows SEGRESS guidelines \cite{kitchenham_segress_2022} to ensure a systematic approach. 
We outline the eligibility criteria, data sources, search strategy, selection process, data collection, risk of bias assessment, and synthesis methods below.

\paragraph{Eligibility Criteria}
In this SLR, we excluded studies unrelated to computational or technological fields, technical reports, and those that do not address resilience or green issues.  
We included relevant review studies for validation purposes.  
The selection of studies was based on the following inclusion criteria:  
\begin{enumerate}
    \item Studies that directly address resilience issues, such as performance degradation or security attacks, or those that solve green-related challenges.
    \item Studies discussing machine learning techniques for monitoring or controlling resilience, greenness, or both.
    \item Studies conducted within the context of Cyber-Physical Systems (CPS).
    \item Given that this PhD research began in early 2022, we included all studies published in English within five years prior to its start, specifically those published between 2017 and April 2022. We based our decision on the increasing number of documents illustrated by \textcite{colabianchi_discussing_2021}, which shows the year 2016 as the starting year of this interest, and thus we started from the year that follows.
\end{enumerate}
    
\paragraph{Data Sources}
To ensure comprehensive coverage, we gathered studies from four databases: i) Scopus\footnote{Scopus database: \url{https://www.scopus.com/}}, ii) DBLP\footnote{DBLP database: \url{https://dblp.org/}}, iii) IEEE Xplore\footnote{IEEE Xplore database: \url{https://ieeexplore.ieee.org/}}, and iv) ACM Digital Library\footnote{ACM Digital Library database: \url{https://dl.acm.org/}}.

\paragraph{Search Strategy}
We structured the search into three main categories to capture studies focused on resilience, greenness, and machine learning in CPS. 
The following unified search queries were designed to retrieve relevant studies across the four data sources, ensuring coverage while maintaining consistency.
\begin{enumerate}
    \item \textit{Green Studies}: The query focuses on studies addressing green aspects in CPS or exploring the intersection of green and resilience and reads as follows: \textit{(green AND (cyber physical system)) OR (green AND (resilience OR resilient)) AND PUBYEAR > 2016}.
    \item \textit{Machine Learning Studies}: The query includes studies discussing deep or machine learning techniques in CPS and reads as follows:  \textit{((deep OR machine) AND learning AND (cyber physical system)) AND PUBYEAR > 2016}.
\end{enumerate}

\paragraph{Selection Process} 
The selection process involved multiple phases, conducted by the primary researcher (i.e., PhD student) under the supervision of the co-supervisor. In each phase, the primary researcher performed all tasks, and the co-supervisor subsequently reviewed the outcomes. 
To ensure objectivity, the eligibility criteria and selection methodology were predefined and rigorously applied throughout the process. The process began with the removal of duplicate studies. Next, the titles, keywords, and abstracts of the remaining studies were screened to exclude irrelevant studies. Finally, the full texts of the shortlisted studies were reviewed to verify their alignment with the eligibility criteria.

\paragraph{Data Collection} 
We used Zotero to organize and categorize the studies. Tags were assigned to each study, summarizing the problem, aim, and methods used. We grouped the studies into collections based on the following themes: 
\textit{Green}, which includes subcategories such as Carbon CPS, Green and Resilience, Green CPS, and Sustainability; 
\textit{Resilience}, which includes Recover (subcategories: recovery, security) and Resilience and Performance (subcategories: performance, resilience, security); 
and \textit{Machine Learning}, which includes Machine Learning in CPS and Deep Learning in CPS.

\paragraph{Data Items} 
We extracted the following data from each study: year of publication, publication type (journal article, conference paper, or book section), CPS domain, study contributions, results, and whether the study was primary or secondary. 
For each study, we also evaluated and recorded a true or false response to determine whether the study addresses topics such as greenness, energy consumption, resilience, performance, security breaches, or the use of learning techniques, including classification, unsupervised learning, deep learning, reinforcement learning, and regression.

\paragraph{Analysis and Synthesis Methods} 
We performed the analysis and synthesis of the results using graphical representations and tabulation methods to identify key trends, research gaps, and potential future directions in learning-based methods for addressing green resilience in CPS.

\subsection{SLR Results}
This section presents the results produced by our SLR methodology, as outlined earlier in this section. We begin by introducing the selected studies and outlining their primary characteristics.

\begin{figure}[ht!]
\centering
\includegraphics[width=\textwidth]{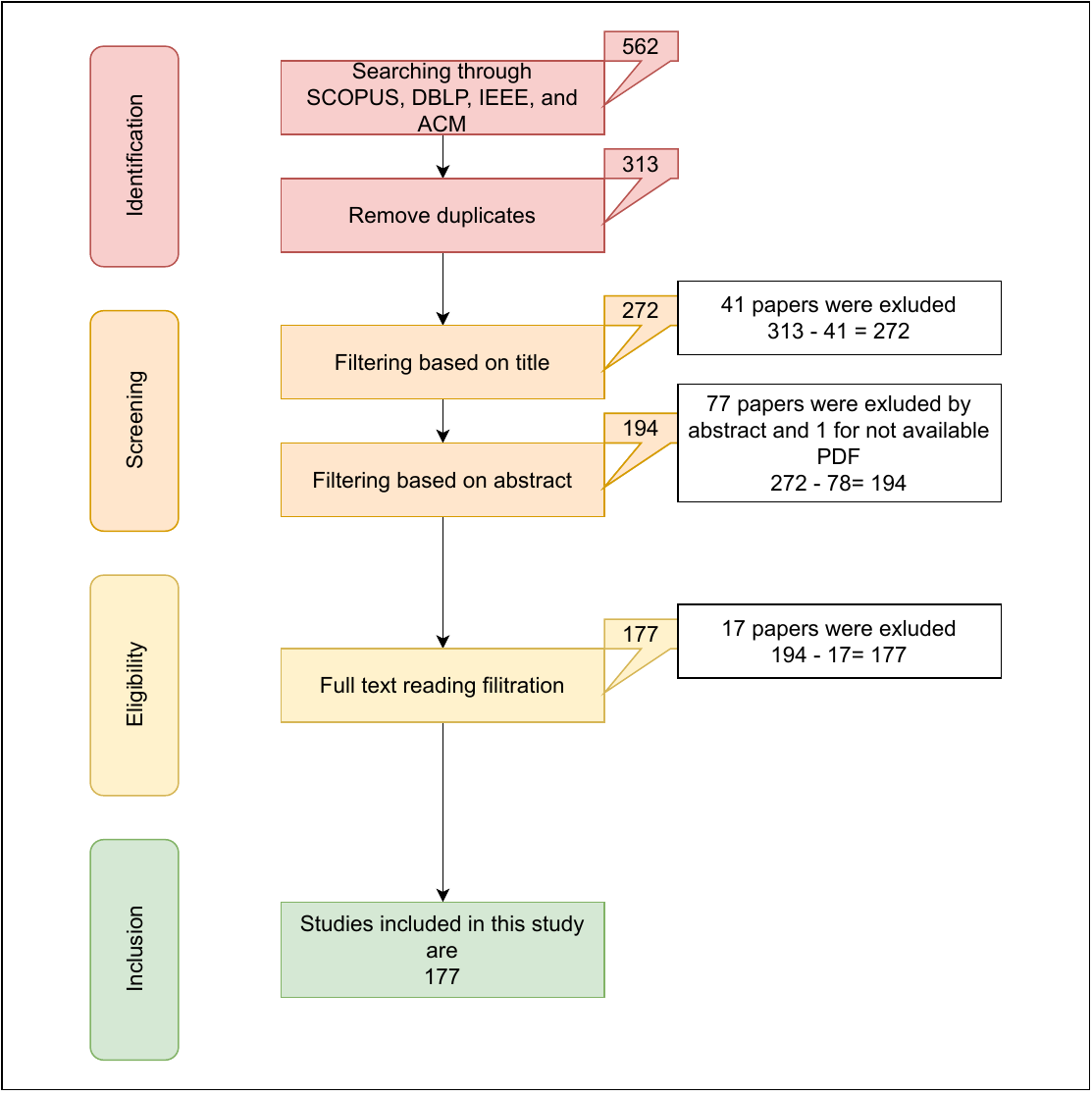}
\caption{The selection process phases and the study counts from each phase.}
\label{fig:lr:slr:SLRSectionProcess}
\index{figures}
\end{figure}

\paragraph{Study Selection}
We conducted the study selection in several phases. 
In the identification phase, we ran the search queries across four databases. This initial search yielded \textit{562} studies. Due to running identical population and intervention queries across different databases, we identified and removed \textit{249} duplicates from this set. 
We then screened the remaining \textit{313} studies by reviewing their titles and keywords, resulting in the exclusion of \textit{41} papers.
Subsequently, we excluded an additional \textit{78} papers after reading their abstracts. 
During the eligibility phase, we reviewed the full text of each remaining study and excluded \textit{17} more, leading to a final set of \textit{177} studies included in this systematic review. 
To provide an overview of the systematic literature review process, Fig.~\ref{fig:lr:slr:SLRSectionProcess} illustrates the four main phases--identification, screening, eligibility, and inclusion--along with the number of studies retained or excluded at each stage.

\paragraph{Study Characteristics}
Studies included in this systematic review are categorized into five groups: studies addressing the green problem, studies focusing on resilience, studies covering performance without explicitly targeting resilience, studies addressing security breaches independently of resilience, and, finally, studies focusing on other properties of cyber-physical systems. 
Each study was evaluated based on its solution to the addressed problem. 
The learning methods commonly found in these solutions, including \textit{deep learning/neural networks}, \textit{reinforcement learning}, \textit{linear regression}, \textit{classification algorithms}, \textit{semi-supervised learning}, and \textit{unsupervised learning}.

To facilitate analysis, the studies were organized into thirty-one sections (S1 - S31), each section resulting from a combination of the study group and the technology applied in the solution the study discussed, such that each study group contains five to six sections. 
For readability, all tables summarizing these characteristics are presented in Appendix~\ref{app:Study_Characteristics_Tables}. The appendix tables detail the studies focused on green, resilience, performance, security, and other CPS properties, along with whether they incorporate machine-learning methods.
Tables~\ref{tab:green} and~\ref{tab:resilience} specifically highlight studies linked to the green and resilience problems, showing that most resilience-focused solutions do not rely on machine-learning approaches. While green and resilience studies are central to the research questions addressed in this review, studies related to performance and security are equally relevant. Tables~\ref{tab:performance} and~\ref{tab:Security} summarize the studies addressing performance and security, respectively.
The final category includes studies exploring additional CPS properties, such as reliability and signal transmission, which provide broader insights and potential avenues for future exploration. Table~\ref{tab:none} presents these studies and specifies whether learning methods were applied in their proposed solutions.

\begin{figure}[ht!]
\centering
\includegraphics[width=\textwidth]{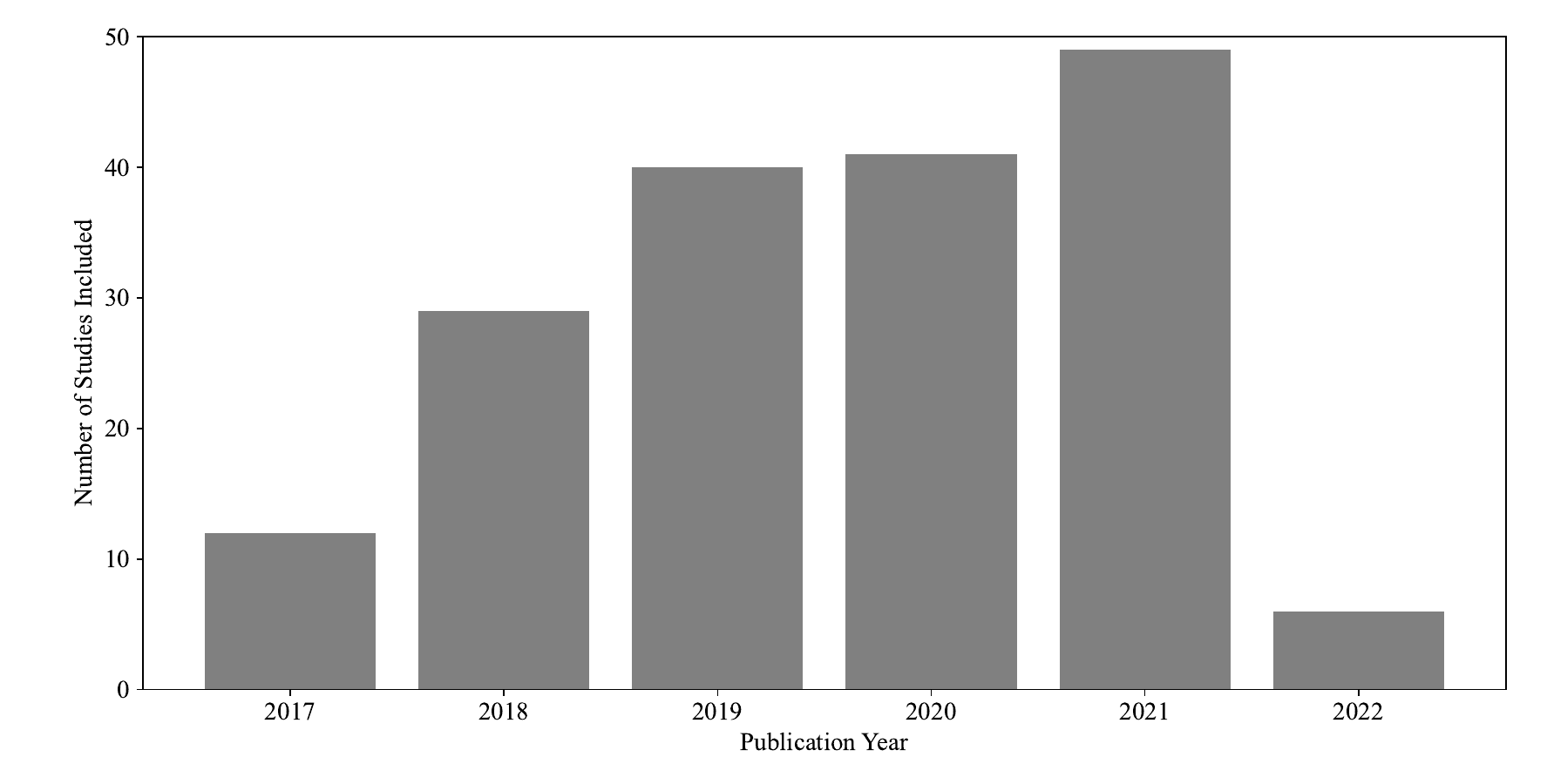}
\caption{Distribution of studies based on year of publication.}
\label{fig:pubyear}
\end{figure}

\paragraph{Results Presentation}
We included studies published between 2017 and 2022 from peer-reviewed conferences, journals, and book sections. 
The studies are mainly from conferences and journals, specifically, \textit{90} conference papers, \textit{82} journal articles, and \textit{5} book sections.
These studies are published between January 1, 2017, and May 6, 2022 (at the beginning of this PhD). 
Fig.~\ref{fig:pubyear} shows the distribution of papers across these years, highlighting a clear increase in the number of studies related to our research. Notably, we only included only the first five months of 2022.
The increasing interest supports the motivation for our research. 

To delve deeper into our analysis and synthesis results, we must clearly understand the state-of-the-art position of cyber-physical systems, resilience, greenness, and green resilience and the solutions used to tackle their problems. 

\paragraph{Cyber-Physical System}
\textcite{pagliari_case_2017} define CPS as engineered systems that integrate computational algorithms with their physical components, a view further confirmed by \textcite{bennaceur_modelling_2019}. 
Additionally, other studies emphasize that CPS involves not only the integration of software and hardware but also the communication networks that connect these components, \cite{bi_novel_2019, murino_resilience_2019, barbeau_metrics_2020}. 
In this work, we build on these definitions by defining CPSs as systems that provide complex real-time services by controlling physical actuators through computational algorithms based on monitored environmental input.

\begin{figure}[ht!]
\centering
\includegraphics[width=\textwidth]{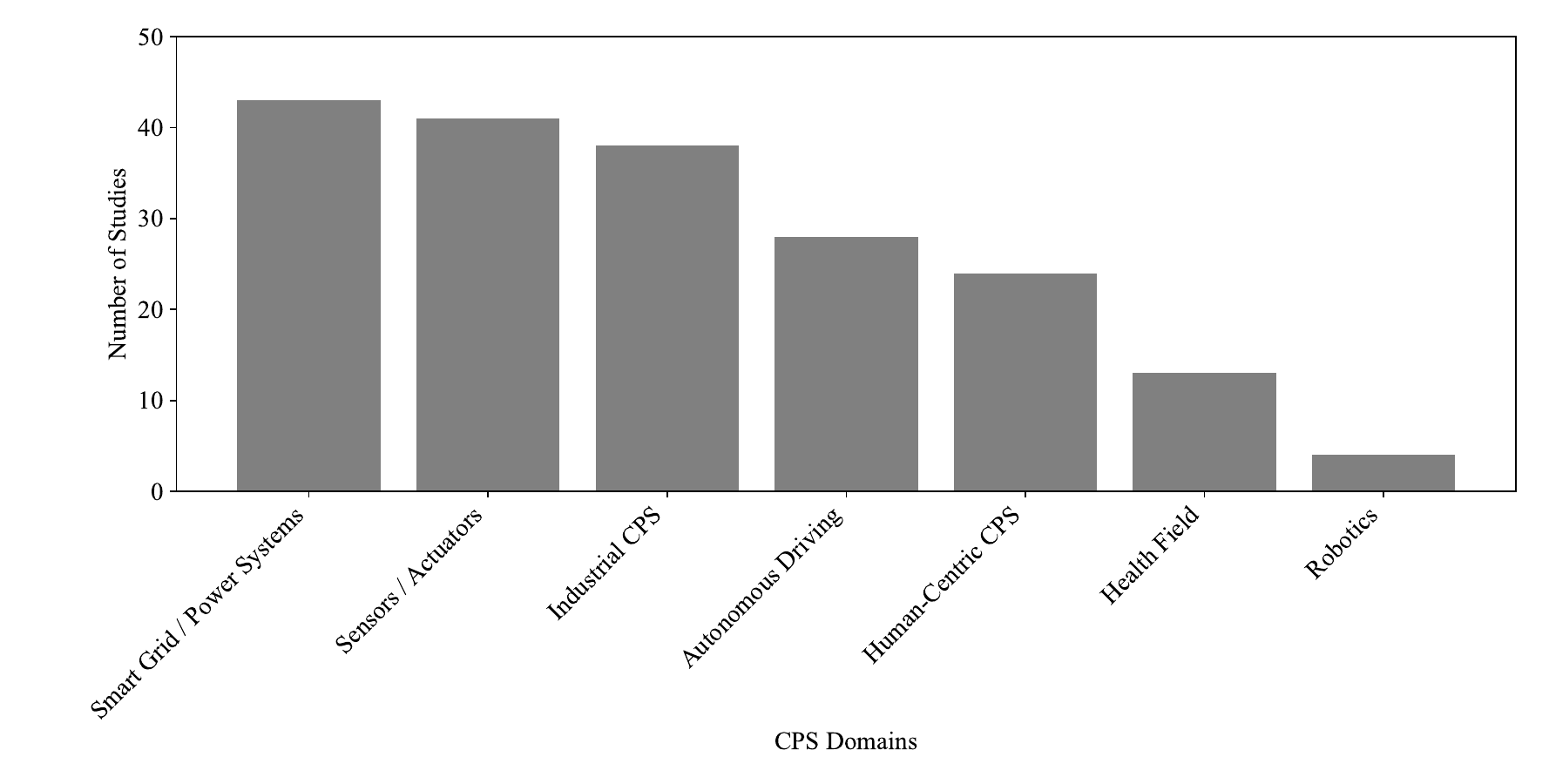}
\caption{Number of studies published based on CPS domains.}
\label{fig:cpsdomains}
\index{figures}
\end{figure}

CPS complexity can vary widely, from decoders \cite{liu_dynamic_2017} and power systems \cite{wu_resilience-based_2021} to smart vehicles \cite{akowuah_recovery-by-learning_2021}, spaceflight systems \cite{iv_feasibility_2018}, and interactive robots \cite{pagliari_case_2017, adigun_collaborative_2022}. Fig.~\ref{fig:cpsdomains} illustrates the CPS domains covered in this study, showing that research predominantly applies the CPS term to grid and power systems, with comparatively few studies focusing on robotic systems.

\paragraph{Resilience}
The term resilience has varying definitions across different fields like psychology, psychiatry, computer science, and others \cite{kouicem_artificial_2019}. In this work, we focus on resilience in the CPS context. 
Some studies describe resilience as the system’s ability to detect and recover from performance anomalies, \cite{haque_contract-based_2018}. Others define it as the capacity to withstand extreme disruptions and return to regular operation, \cite{ti_resilience_2022}.
Further, \textcite{mouelhi_predictive_2019} categorizes resilience into two types according to the source of resilience: \textit{internal autogenous resilience} and \textit{exogenous resilience}. 
Internal resilience refers to the system’s ability to detect and manage faults and attacks, while external resilience refers to the system’s capability to maintain acceptable performance in completing operations within its environmental context.
In our work, we leverage these definitions and define resilience as the system's ability to detect performance degradation upon exogenous disruptions and undertake actions to mitigate it, eventually restoring the system to its original state.

\begin{figure}[hbt!]
\centering
\includegraphics[width=\textwidth]{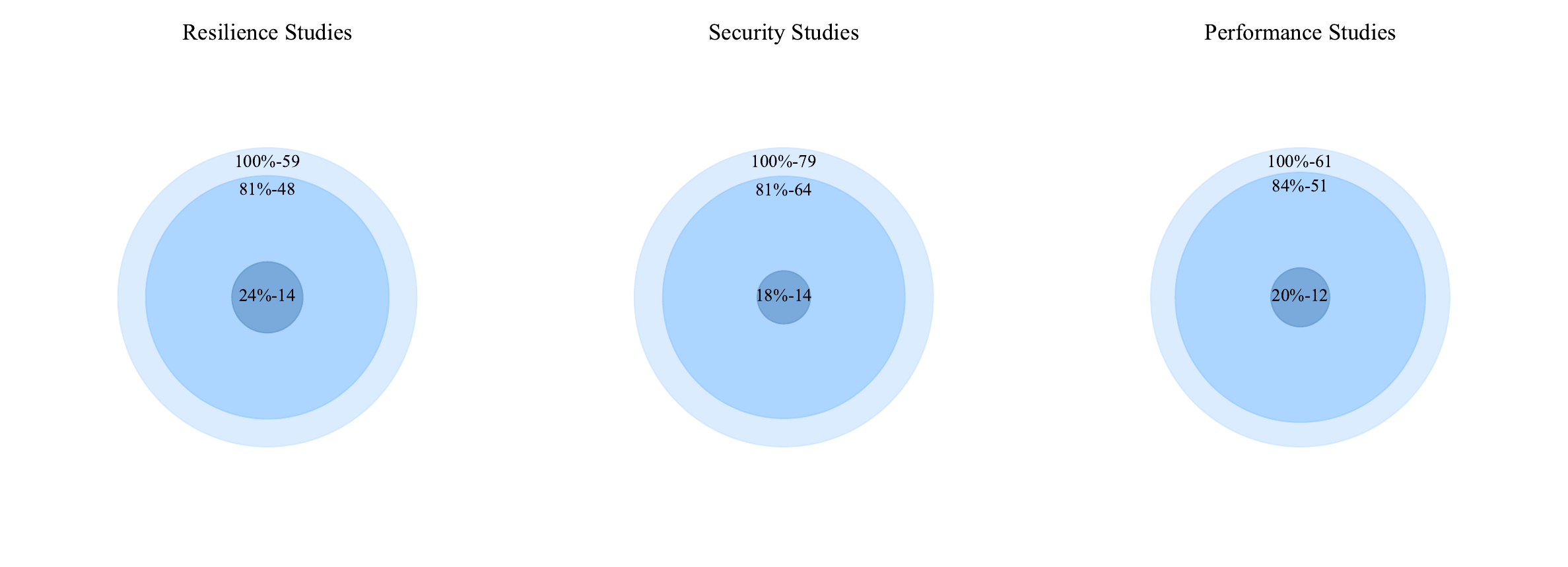}
\caption{Resilience, performance, and security solutions addressing detection, mitigation, or recovery levels.}
\label{fig:respersec}
\index{figures}
\end{figure}

Studies address resilience challenges by introducing detection, mitigation, and/or recovery solutions.
Fig.~\ref{fig:respersec} shows that 59 of our inclusion studies focus on detecting resilience-related issues, such as those by \textcite{alcaraz_cloud-assisted_2018, eke_detection_2020}. Meanwhile, only 81\% of these studies advance further by developing mitigation strategies, \cite{lakshminarayana_performance_2019, yong_switching_2018}, and only 24\% of the total studies take a step further to address mitigation and fully recover back the system to the original state, \cite{kong_cyber-physical_2018, wu_resilience-based_2021}. 
This pattern is also observed in studies on security and performance, where relatively few studies emphasize recovery with 18\% of security solutions and 20\% of performance solutions. 

The resilience landscape in the CPS context introduces various frameworks for designing a resilient CPS. 
Some of these studies turn to optimization techniques to address resilience.
For instance, a hierarchical multi-agent framework is employed to maintain performance during physical disturbances and cyber-attacks by distributing tasks across agents, \cite{januario_distributed_2019}.
This framework, tested on an IPv6-based platform, ensures continued system functionality through task distribution.
Another approach applies a distributionally robust optimization model to identify and mitigate threats, integrated with a constraint-generation algorithm, culminating in an index of critical components for assessing system resilience, \cite{liu_distributionally_2022}. 
This optimization model provides resilience evaluations, particularly for power grid CPS, against malicious threats.

Other studies develop intelligent monitoring components to trace the system's behavior and actuate resilient mechanisms.
For instance, \textcite{bagozi_context-based_2021} addresses resilience in an industrial CPS for the food sector by modeling each component as an intelligent machine with recovery services, including a sensor data API for behavior monitoring and an operator for abnormal condition detection and response initiation. 
Additionally, data summarization helps manage the vast quantities and high speeds of data generated by the sensor data API.
Furthermore, \textcite{xu_automatic_2019} propose an automated, contract-based framework to ensure CPS resilience. 
This framework monitors system components' resilience properties through assume-guarantee contracts and refines these contracts into lower-level components based on input-output dependencies.
Additionally, \textcite{haque_contract-based_2018} proposes a hierarchical solution based on distributed resilience managers. This solution aims to enhance resilience by reducing decision-making bottlenecks through local monitoring of CPS component performance.
These solutions primarily address resilience in the physical components of CPS, relying on sensor input to assess system status.
Conversely, some studies explore resilience in CPS when disruptions originate from these sensors.
For instance, certain studies target resilience against malicious attacks, such as data injection \cite{hopkins_foundations_2020}. 
This approach combines a state-estimation baseline with a threshold to detect and respond to cyber-attack data injections and evaluate vulnerabilities. Similar strategies mitigate sensor attacks and other faults are presented in \cite{zhang_real-time_2020, akowuah_recovery-by-learning_2021, fei_learn--recover_2020, vatanparvar_self-secured_2019, bures_performance_2018, zhang_real-time_2021}.
Furthermore, numerous studies employ learning-based approaches to address CPS resilience, which are discussed in detail later in this section.

\paragraph{Greenness} We adopt the definition of greenness from information technology systems (IT-systems) as “the efficient usage of energy with minimized adverse effects.”
IT-systems, including CPS, are responsible for storing, displaying, transforming, and transferring information, \cite{kharchenko_concepts_2017}.
Some studies on greenness in CPS explore the impact of green strategies on financial growth and organizational resilience, \cite{yang_how_2021}.
Other studies examine greenhouse gas emissions in smart cities and their influence on urban resilience, \cite{eleftheriadou_cities_2021}.
Further studies address the sustainable selection of manufacturing materials for CPS component design, \cite{nandy_carbon_2018, liu_leveraging_2016}.

While these studies do not directly examine the impact of computational processes on energy consumption, \textcite{mohammed_eco-gresilient_2018} propose a three-objective optimization approach to support the meat supply chain with economic, green, and resilient outcomes—termed eco-gresilient. 
This approach optimizes the number of facilities required by balancing economic efficiency, greenness, and resilience through a fuzzy analytical hierarchy process, assigning weights to resilience pillars such as robustness, agility, leanness, and flexibility.

\begin{figure}[ht!] 
\centering \includegraphics[width=\textwidth]{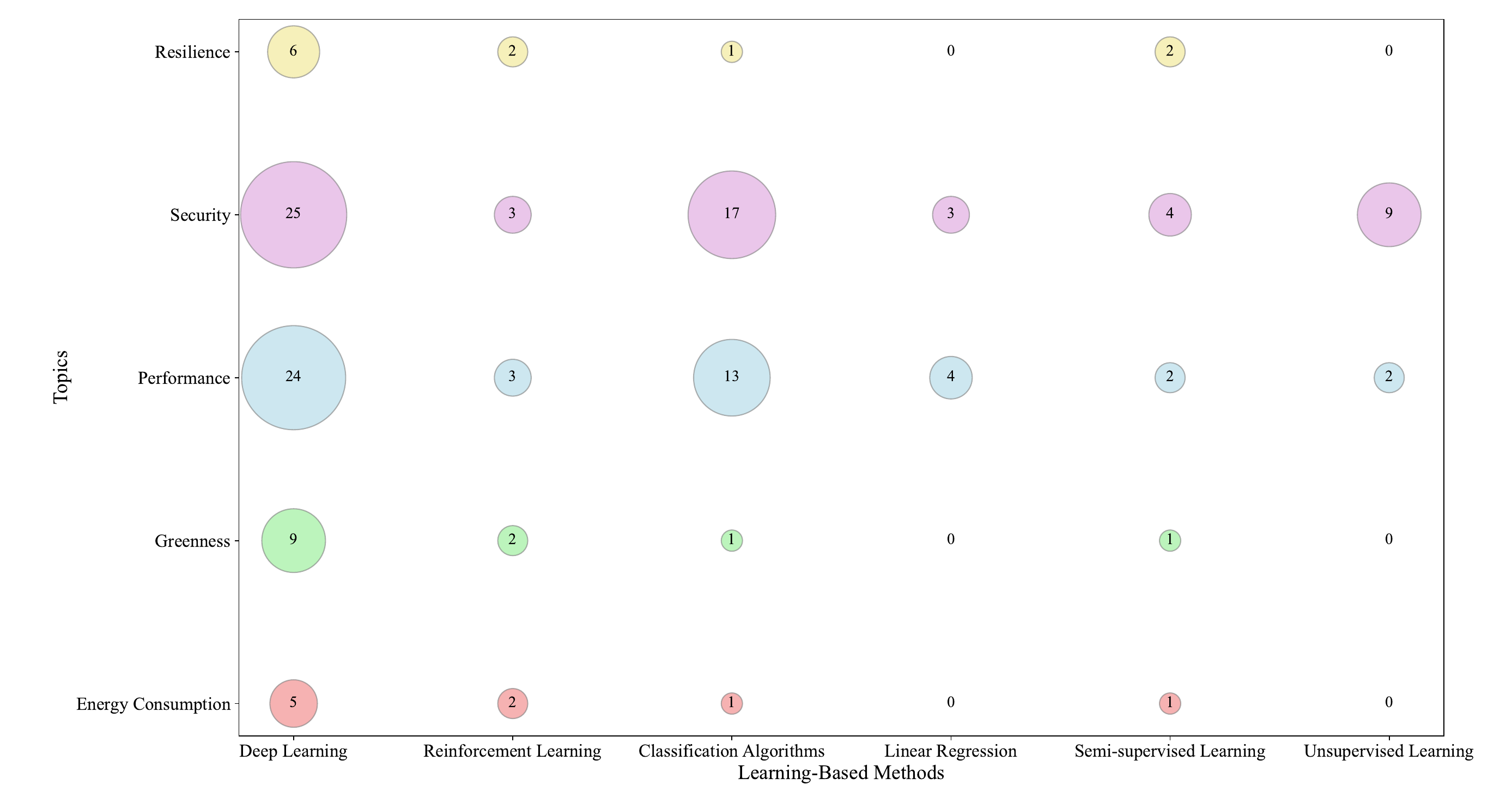} 
\caption{Bubble plot illustrating the distribution of studies across topics (e.g., resilience, greenness, performance, etc.) and the learning methods employed (e.g., deep learning, reinforcement learning, etc.). The size of each bubble represents the relative prevalence of a method in a given topic.} 
\label{fig:bubblechart}
\end{figure}

\paragraph{Learning-Based Methods}
One of the pillars of this PhD research is to utilize innovative technologies to address the greenness resilience balancing problem.
Thus, it is important to understand the machine learning-based solutions introduced by the literature.
Fig.~\ref{fig:bubblechart} illustrates a bubble plot that shows the distribution of studies specifically addressing the terms \textit{Greenness}, \textit{Energy Consumption}, \textit{Performance}, \textit{Security}, and \textit{Resilience}, along with the learning-based methods used to address these topics, namely \textit{Deep Learning}, \textit{Reinforcement Learning}, \textit{Classification Algorithms}, \textit{Linear Regression}, \textit{Semi-supervised Learning}, and \textit{Unsupervised Learning}.

From the bubble plot in Fig.~\ref{fig:bubblechart} we can note several points. First, the largest bubbles come from using deep learning/neural networks across all topics. 
In the resilience dimension, \textcite{zarandi_detection_2020} focuses on cyber-attacks that involve injecting false data through the system's sensors or controllers, impacting system performance. 
Their approach applies deep learning to detect these attacks, followed by resilience control algorithms to isolate the abnormal agent within the system. 
Similarly, \textcite{eke_detection_2020} addresses this issue using a multi-layered deep learning approach, incorporating robustness, resilience, performance, and security.
The significant bubble size for deep learning in security indicates strong research interest in this area. 
Studies explored various CPS attack detections, including enhanced detection in autonomous vehicles \cite{sargolzaei_machine_2017} and wireless data transmission \cite{chen_deep_2018}. 
Another study by \textcite{akowuah_recovery-by-learning_2021} investigates recovery from sensor attacks, addressing both mitigation and system restoration. 
Their Recovery-by-Learning framework includes two components: (i) a state predictor that estimates the CPS state after attack detection using a deep learning prediction model leveraging temporal correlation between heterogeneous sensors, and (ii) a data checkpoint to monitor for attacks.

The second-largest bubble represents studies focused on performance as a core aspect of CPS resilience. 
For instance, \textcite{xu_adaptive_2021} discusses the application of deep reinforcement learning to optimize performance and green objectives in CPS by considering device energy and heating effects on hardware lifespan. 
Furthermore, \textcite{chiu_integrative_2020} explores machine and deep learning methods based on system short- and long-term memory to achieve real-time monitoring and fault detection, expanding on their earlier work in \cite{tsai_apply_2019}.
Additionally, the bubble plot shows the lack of research in reinforcement learning to address these topics; reinforcement learning enables systems to monitor themselves and adjust behavior dynamically, \cite{moghadam_autonomous_2022}.  

The last observation is the prevalence of energy consumption considerations within green studies \cite{hou_cyber-physical_2021, xu_adaptive_2021, pandey_greentpu_2020, pandey_greentpu_2019, deniz_reconfigurable_2020}, represented by a smaller bubble as these studies are a subset of broader green-focused research. These studies also often consider carbon emission, a well-recognized direct and indirect impact of CPS during both development and operation, \cite{kharchenko_concepts_2017}.

\begin{figure}[ht!] 
\centering \includegraphics[width=\textwidth]{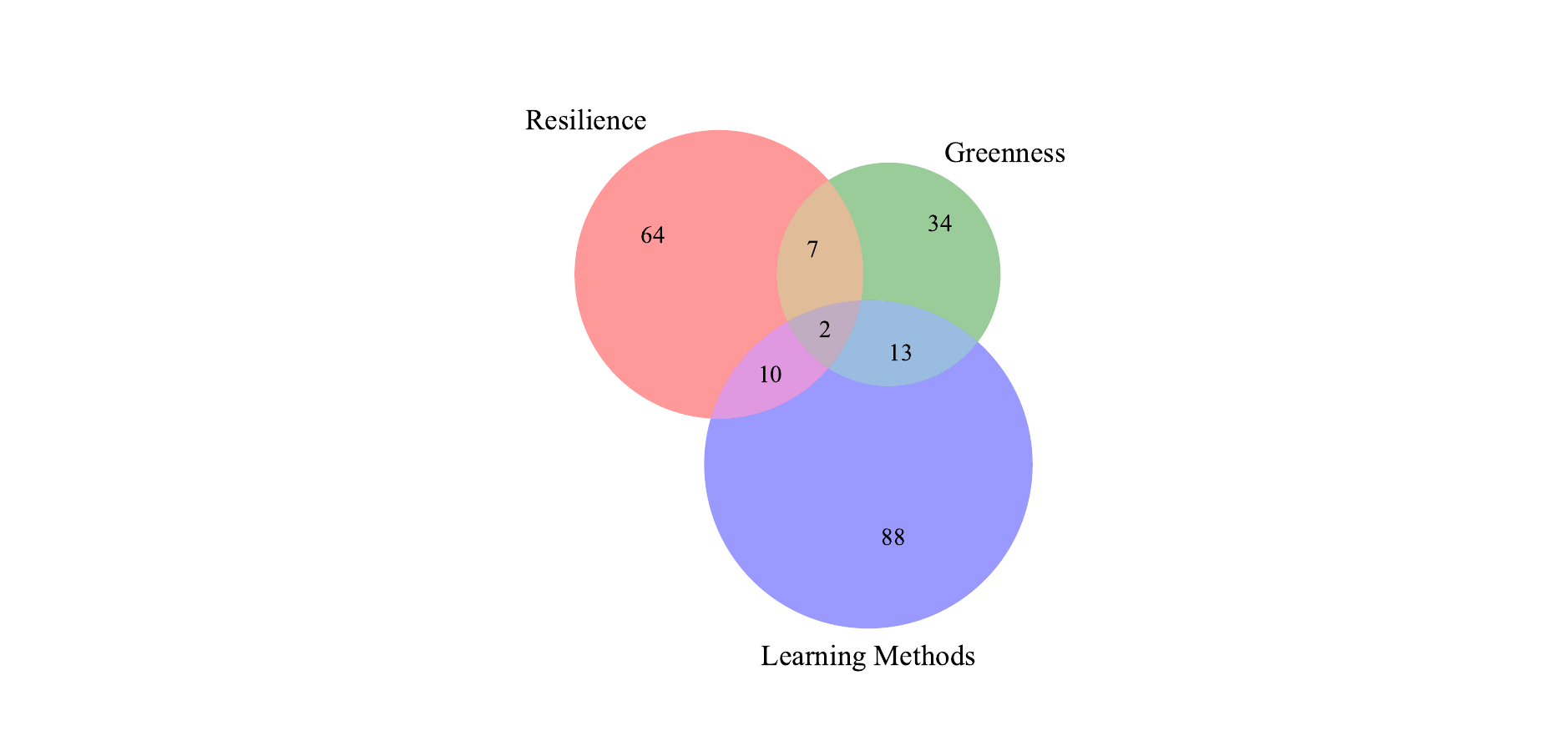} 
\caption{Venn diagram shows the intersection across resilience, greenness, and learning-based methods studies.} 
\label{fig:gressmart}
\end{figure}

\paragraph{Green, Resilience, and Learning Methods}
To synthesize the three core pillars of this SLR (i.e., greenness, resilience, and learning-based methods), we present Fig.~\ref{fig:gressmart}, which illustrates a Venn diagram of the number of studies addressing each pillar and their intersections. 
The SLR includes \textit{64} studies focused on resilience in CPS, \textit{34} on greenness, and \textit{88} utilizing learning-based methods to address one or more non-functional CPS properties.
Of these, only \textit{10} studies integrate resilience and learning-based approaches (listed in Table~\ref{tab:resilience}), while \textit{13} employ learning-based techniques to support greenness in CPS (Table~\ref{tab:green}). 
Additionally, \textit{7} studies cover both greenness and resilience in CPS \cite{konstantinou_chaos_2021, zhu_cyber-physical_2020, mohammed_eco-gresilient_2018, rodriguez_green_2020, pandey_greentpu_2019, pandey_greentpu_2020, zhang_guest_2020}.
Finally, only two studies, by \textcite{pandey_greentpu_2019, pandey_greentpu_2020}, address all three areas simultaneously. These studies investigate Google’s tensor processing unit error patterns, specifically the activation sequences in the systolic array, aiming to maintain minimal accuracy loss while reducing adverse energy impacts, such as carbon emissions, respectively, through low-voltage operations.
 

While the SLR identifies a limited number of studies that jointly address greenness and resilience in CPS using learning-based approaches, the fundamental importance of this integration extends beyond mere research gaps. 
CPS often operate in dynamic and uncertain environments, where maintaining acceptable performance under disruptive conditions requires resilience.
At the same time, many CPS are deployed in energy-constrained settings, making greenness a critical non-functional requirement. Addressing these dual concerns simultaneously is inherently complex, particularly when using traditional control or rule-based systems.
Learning-based solutions offer a promising avenue, as they can autonomously adapt to evolving conditions and optimize responses over time.

\paragraph{Comparison with Existing Secondary Studies}

Our SLR incorporated four relevant secondary studies, after evaluating and excluding four others. 
\textcite{haggi_review_2019} assess extreme-event impacts on power grids, presenting a narrow focus on disaster resilience without broader applicability to CPS. Both \textcite{kim_survey_2021} and \textcite{colabianchi_discussing_2021} concentrate on security and resilience. While \textcite{kim_survey_2021} explores machine-learning applications for CPS attack detection, \textcite{colabianchi_discussing_2021} highlights the need for research in human-CPS resilience. 
Both support our findings around security and resilience, though neither addresses greenness or general performance.
Finally, \textcite{andronie_artificial_2021} emphasizes CPS lifecycle elements, such as IoT integration, big data analysis, and real-time AI-driven decision-making. However, this study’s primary focus on IoT applications for logistics and system monitoring does not explore resilience or green CPS aspects through machine learning or artificial intelligence.

\subsection{SLR Takeaways}

This section synthesizes the primary findings of our SLR, which directly address the SLR research questions (SLR--RQ1, SLR--RQ2) focused on resilience and greenness in CPS. Our findings reveal the landscape of resilience research (SLR--RQ1), exploring learning-based methods for monitoring and mitigating performance issues (SLR--RQ1.1) and examining techniques that go beyond detection to enable recovery (SLR--RQ1.2).
Additionally, these findings address green-focused studies in CPS (SLR--RQ2), including approaches to energy efficiency and CO$_2$ emissions (SLR--RQ2.1) and studies that integrate resilience with green strategies (SLR--RQ2.2).
These findings guide each takeaway below, establishing key directions for further research on balancing resilience and greenness in CPS.

\begin{enumerate}
    \item \textit{From Detection to Recovery} \\
    Many resilience approaches identified in this SLR focus on detection and mitigation, with limited attention to full recovery. While these studies emphasize reaching an acceptable system state, restoring CPS to its original condition is often overlooked. Our findings thus highlight the need for resilience research that advances from detection to complete recovery, enabling fast, comprehensive responses to performance degradation.
    \item \textit{Green, Resilience, and Learning Methods} \\
    The analysis identifies reinforcement learning as a promising tool for self-monitoring CPS behavior, yet additional research is essential to explore a wider variety of learning-based methods across CPS domains. Questions remain regarding which CPS applications align best with specific algorithms and whether unsupervised or semi-supervised learning methods may be suited to unpredictable CPS environments. Most studies on CPS greenness focus on renewable materials, energy efficiency \cite{hou_cyber-physical_2021, xu_adaptive_2021, pandey_greentpu_2020, pandey_greentpu_2019, deniz_reconfigurable_2020}, and carbon emissions \cite{mila_codecarbonio_nodate}. Although this foundation is promising, broadening the scope of green applications in CPS is essential. Notably, only two studies address both resilience and greenness, revealing an opportunity to apply learning-based methods to bridge these domains.
    \item \textit{Expanding CPS Domains} \\
    The SLR results reveal a dominant focus on smart grid, hardware, and industrial CPS applications. This emphasis, while insightful, limits solutions primarily to physical components, with minimal incorporation of cyber functionalities. The findings suggest significant potential in underexplored CPS domains such as robotics, healthcare, and human-CPS interaction, which are vital but less researched. Given the rapid advancement of human-interactive robots and healthcare applications, fostering green and resilient CPS in these areas is critical.
    \item \textit{Innovative Approaches} \\
    Balancing resilience and greenness may extend beyond intelligent techniques to include optimization frameworks and innovative methodologies. For instance, game-theoretic strategies can help decision-makers manage resilience-greenness trade-offs, while virtualization and containerization techniques may improve system greenness and resilience in dynamic environments.
\end{enumerate}

The insights from this SLR reveal a growing but fragmented landscape in CPS research, with substantial efforts addressing either resilience or greenness independently and often focusing on detection and mitigation rather than full recovery. While machine learning-based approaches offer promising directions for self-monitoring and adaptive CPS behavior, especially in the context of resilience, few studies explore their integration with green objectives. Further, the dominance of smart grid, hardware, and industrial CPS solutions highlights gaps in underexplored but critical domains, such as robotics, healthcare, and human-centered CPS.
These findings underscore the need for more cohesive and comprehensive approaches to CPS resilience and greenness, integrating learning-based methods. The next section will delve into related work that further contextualizes the research direction of this PhD.

\section{Related Work}
Considering the insights we have from our SLR, we delve into the related work, building and identifying our contribution to the current state-of-the-art. 
We have reviewed existing literature across five lines of research: i) green and resilience, ii) online learning, iii) decision-making mechanisms, iv) decision-making assistance, and v) reinforcement learning. The following provides an overview of each, followed by our contribution to the community.

\subsection{Green and Resilience}
The literature presents different models and approaches to address resilience in complex systems. 
\citet{vistbakka_modeling_2020} recognize the dynamic nature of multi-agent systems and model resilience through a formal collaborative solution using Event-B. This approach allows agents to dynamically reconfigure their relationships and capabilities to achieve the system’s goals despite unforeseen events. 
\citet{januario_distributed_2019} emphasize the importance of resilience in the cyber-physical system context; they define a hierarchical multi-agent model that aims to maintain system performance during physical disturbance and cyber attacks.
Others model resilience using a tri-optimization model \cite{liu_distributionally_2022} or a deep learning model \cite{zarandi_detection_2020}. 
These models have a common goal to detect performance degradation or disruptive events and recover the system performance to an acceptable level.
According to 
\citet{mouelhi_predictive_2019} disruptive events can take on diverse forms: i) endogenous and ii) exogenous. An endogenous event can be some security vulnerabilities \cite{zarandi_detection_2020, liu_distributionally_2022} or defects \cite{januario_distributed_2019}. While an exogenous event can be caused by environmental factors \cite{henry_generic_2012}.
In summary, the literature recognizes the dynamic nature of disruptive events in a resilience context and explores various sources and consequences.
Our investigation aligns with the literature in considering the dynamic nature of disruptive events that may lead to performance degradation.

The concept of ``Green" includes all activities aimed at monitoring and mitigating energy's adverse effects, \cite{kharchenko_concepts_2017}.
Some studies in the literature examine the greenness of a system as an outcome of improving the system's resilience, describing the relationship between greenness and resilience as a positive correlation, \cite{pandey_greentpu_2020}. 
Others study the resilience or the effect of resilience in green systems developed using green materials or components, \cite{kharchenko_concepts_2017, rodriguez_green_2020}. 
Further, \citet{yang_how_2021} emphasize that green strategies not only enhance financial performance but also improve stability and flexibility within organizations, motivating decisions to integrate green components into a system to enhance resilience against exogenous disruptions.
Our study explores the trade-offs between green and resilience in decision-making to autonomously support the system in balancing the two properties.

\subsection{Online Learning}
Machine learning algorithms vary according to data and purpose, \cite{mahesh2020machine}. \citet{van2022three} define three types of continual learning: i) task-incremental learning, which learns to solve a number of distinct tasks within the same context in a sequential fashion, ii) domain-incremental learning, same as task-incremental learning but within a different context, and iii) class-incremental learning, which learns both the tasks and their context. The three types share the same sequential learning fashion, which distinguishes continual learning from other traditional learning models.
This sequential fashion poses special cases of continual learning, \citet{lesort2020continual} distinct between continual learning special cases based on the data batch needed for learning fusion. They define \textit{Online Learning} as a type of continual learning where the updates are done on a single data point basis, in other words, the batch size is one.
For example, \citet{luong2021incremental} use online learning to continuously update the model with new data as it becomes available, allowing the system to adapt in real-time to improve their mobile robot’s ability to navigate and avoid obstacles over time.
Systems with an environment composed of collaboration between the AI component and humans involve a continuous transfer of knowledge from humans to the system's components \cite{lesort2020continual}, which requires data to be processed on a single data point basis, that is, online collaborative AI systems, \cite{rimawi_green_2022, kadgien_cais-dma_2024, rimawi_modeling_2024}.
\citet{wang_facilitating_2018} states that human-system learning should start with the human teaching the robot through specific interactions, then the robot observes these interactions and updates its model. In their work, they follow a manual method to switch between the learning and testing phases. However, some other applications that use online learning assign a confidence value to the decision made by the model, which ensures sufficient minimum data points to land on the optimal decision to automatically switch between the two stages, \cite{mahesh2020machine, kadgien_cais-dma_2024}.
The type of human-system interactions varies from one system to the other. A recent survey by \citet{mukherjee_survey_2022} lists a number of input modes, such as gaze, gesture, voice commands, facial emotions, and demonstration. 
Notably, our system utilizes human demonstration for object classification, requiring advanced scene understanding. 
The survey emphasizes that most research in this area focuses on the safety and physical well-being of human-system collaboration. This leaves the door open to contributing to other disciplines. The dynamic and continuous nature of the online learning process exposes OL-CAIS to unforeseen disruptive events, and our goal is to support the system in maintaining acceptable performance during the learning process.

\subsection{Decision-Making Mechanisms}
The literature contains a multiple number of studies discussing different mechanisms to support decision-making. An optimization mechanism using a heuristic algorithm has been introduced by \citet{yuan2023new}; they develop a resilience assessment model for human-robot collaborative disassembly of spent lithium-ion batteries, integrating fuzzy Bayesian fusion with an analytical network process.
In the green and resilience context, \citet{mohammed_eco-gresilient_2018} propose the eco-gresilient model. The model tries to balance economic, green, and resilient within a meat supply chain network by finding the optimal number of facilities in the supply network section. The eco-gresilient model uses a multi-objective optimization model along with a fuzzy analytic hierarchy process to find the resilience weight. Further, they measured green by computing the carbon emissions from each facility of the supply chain.

Another mechanism to optimize decision-making is the usage of game theory, which is traditionally applied in decision-making with human actors. \citet{xu_user_2018} develop a novel approach to support decision processes for a recommendation system, emphasizing user satisfaction.
\citet{colabianchi_discussing_2021} discuss the usage of game theory as a decision-making mechanism in the context of cyber-physical systems; they show that game theory has been used to model the interaction between attackers and defenders in this context, considering cybersecurity attacks as one of the major disruptions for cyber-physical system resilience.
Moreover, a review by \citet{satapathy_game_2016} discusses some of the popular game theories, such as the prisoner's dilemma and the battle of the sexes. Then, it defines the most important characteristics of the applications that these theories can solve. 
The prisoner's dilemma helps in making rational choices to reach a sub-optimal outcome, specifically when cooperation is difficult. On the other hand, the battle of the sexes shows a trade-off between coordination and compromise in reaching a mutual decision, even with different preferences. Further, the review emphasizes the importance of using game theory in the robotics context, specifically in coordinating and cooperating in multi-robot systems.

Both optimization and game theoretical mechanisms do not inherently account for history or past decisions unless in some scenarios of repetitive execution. 
On the other hand, as one of the three machine learning paradigms, reinforcement learning is one of the decision-making mechanisms that incorporate continuous environmental feedback to maximize some notion of cumulative reward, \cite{mahesh2020machine}.
\citet{huang2022reinforcement} integrate reinforcement learning with feedback architectures to enhance cyber resilience, specifically targeting posture-related, information-related, and human-related vulnerabilities.
Another study by \citet{zhao2022deep} enhance resilience in cyber-physical systems by using deep reinforcement learning to represent the dynamic scheme of the system as Markov decision process and deep Q-learning to adjust microgrid topologies in real-time.
Further work done by \citet{oliff2020reinforcement}  addresses the disturbance introduced by human operators in collaborating with manufacturing robots. The study uses a multi-layer neural network to understand the environment and human state, integrated with Q-learning to learn the optimal policy.

Building on these ideas, we extend our framework GResilience, which compares optimization and game theory with reinforcement learning for trading off two non-functional properties of OL-CAIS. Specifcally, in this work we com

\subsection{Decision-Making Assistance}
Within CAIS domain, there is a growing body of research exploring automated decision-making support frameworks. These frameworks often use human actions as a reference point for system behavior and develop specialized knowledge to inform decision-making processes. For example, \citet{chen_trust-aware_2020} devised a computational model to assess human trust in CAIS autonomous actions, using this assessment to automatically enhance decision-making by selecting actions that maximize trust values. Similarly, \citet{ghadirzadeh_human-centered_2020} employed deep reinforcement learning to predict human actions, while \citet{quintas_toward_2018} developed an AI agent capable of monitoring human actions and generating descriptive scenarios to support decision-making in CAIS autonomously.
Despite the existence of various decision-making frameworks, none, to our knowledge, offer extendable components backed by the necessary toolbox and utilities.


\myPart{Research Methodology and Approach}

\huge{T}\normalsize{he second part of this thesis consolidates the diverse methods, frameworks, and strategies employed throughout this research. By gathering and structuring each component, we reveal the underlying logic and processes that drive our approach to enhancing greenness and resilience within OL-CAIS.

From the methodologies to the decision-making frameworks and the comprehensive construction of Green Resilience OL-CAIS, this part lays the groundwork for understanding the research flow and the cohesive framework that enables each facet of this study to contribute towards a sustainable and resilient cyber-physical future.}


\chapter{Methodology Overview}
\label{ch:methods}
In this chapter, we present the empirical methodology that guided our research into achieving a balance between greenness and resilience within Cyber-Physical Systems (CPS). 
This approach was not linear but rather evolved through a dynamic, iterative process. 
Each research phase informed and shaped the next, creating a cyclic loop among key methodological components. 
Insights, experimental results, and analyses continually refined our methodological framework, enabling us to develop a holistic understanding of the greenness resilience trade-off problem in CPS.

\begin{figure}[ht!]
    \centering
    \includegraphics[width=\textwidth]{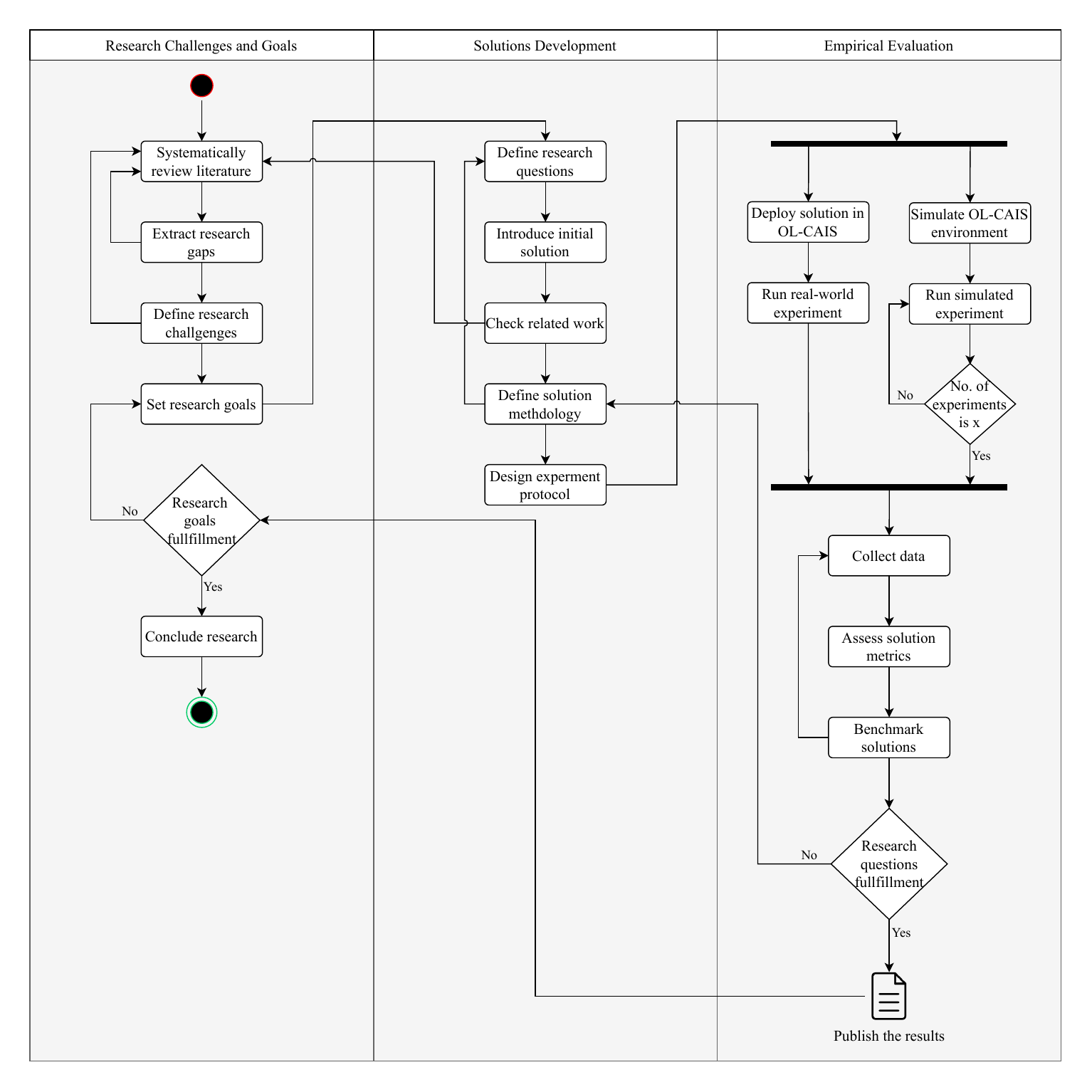}
    \caption{Research methodology.}
    \label{fig:method:method}
\end{figure}

Fig.~\ref{fig:method:method} illustrates the phases and activities undertaken in this PhD research. 
Our research began with a systematic literature review (SLR), a foundational step to uncover current knowledge, gaps, and challenges surrounding green resilience within the CPS context. 
This comprehensive exploration identified specific research gaps, particularly within human-centric CPS and robotic systems, which guided us in defining our research goals and formulating targeted research questions. 
Throughout the research timeline, we were able to iteratively align our solutions with emerging insights from the SLR, revisiting these insights to ensure relevance as our research evolved.

The access to a collaborative robotic system developed at \textit{Fraunhofer Italia Research\footnote{Fraunhofer Italia Research homepage: \url{https://www.fraunhofer.it/}}} enabled us to empirically evaluate resilience and greenness strategies in a real-world setting. 
Observing the collaborative robotic system in these settings enriched our understanding of online learning, human-robot collaboration, and decision-making dynamics. 
These observations informed our theoretical foundation and raised further questions, prompting us to revisit the literature and refine our objectives based on practical constraints and industrial applicability. 
Building on these insights, we developed a theoretical approach leveraging game theory to balance greenness and resilience as two independent agents,~\cite{rimawi_green_2022}. 
Through iterative testing, this model evolved into the GResilience Framework, which contains our policies to achieve optimal trade-offs between greenness and resilience. 
Our framework addresses the trade-off problem from two perspectives: i) two-agent policies using game theory, and ii) one-agent policies through optimization techniques,~\cite{bonfanti_gresilience_2023}.

To equip OL-CAIS with a decision-making framework, we introduced the OL-CAIS Decision-Making Assistant (CAIS-DMA\footnote{CAIS-DMA Github repository: \url{https://github.com/dmrimawi/CAIS-DMA}})~\cite{kadgien_cais-dma_2024}, a tool for real-time monitoring and adaptive action selection within OL-CAIS. 
This decision-making assistant became central to our work, bridging theoretical and applied aspects of resilience. 
We repeatedly tested and adjusted CAIS-DMA under varied conditions, refining our policies to enhance performance during both steady and disruptive states,~\cite{kadgien_cais-dma_2024}. 
Through this iterative process, an empirical resilience model emerged, designed to track the system’s transitions across states and dynamically assess the impact of our policies. 
Implementing this model in real-world trials necessitated the creation of a structured measurement framework to benchmark and systematically compare the effectiveness of each policy,~\cite{rimawi_modeling_2024}. 
By revisiting each model and metric in response to empirical findings, we established a feedback loop that strengthened our analytical foundation and provided a comprehensive assessment of our green resilience strategies,~\cite{rimawi_2024_gresilience}.

To address the limitations inherent in real-world testing, we constructed a simulator for our OL-CAIS environment. 
This simulator facilitated extensive experimentation by replicating disruptive events and enabling decision-making analysis under diverse scenarios,~\cite{kadgien_cais-dma_2024}. 
The simulation phase completed the empirical cycle, allowing us to test hypotheses and perform parameter testing on solutions developed in previous stages, validating our findings through repeated trials. 
Insights from this phase led us back to the literature, where we explored advanced decision-making approaches, including reinforcement learning (RL), as an optimization tool to identify actions that best balance greenness and resilience trade-offs.

Additional solutions emerged during an internship at Fraunhofer Italia Research, where we were able to experiment with containerizing components of our collaborative robotic system. 
This architectural shift provided additional resilience by enabling flexible component replication and significantly reducing energy consumption and carbon emissions, further improving our green resilience research.

Finally, empirical observations revealed a post-disruption catastrophic forgetting, in which the system forgets its steady state and reverts to a disruptive state. 
This critical insight led to adjustments in the CAIS-DMA framework to provide continuous support and ensure consistent performance. 

By employing an iterative experimentation, analysis, and refinement process, we developed a comprehensive methodology that systematically addresses green resilience across diverse scenarios within CPS.
Our methodology encompasses literature review, real-world experimentation, simulation, and continuous refinement to guide decision-makers in developing a green resilient OL-CAIS.

In the following two chapters, we delve into the core components of our methodology. 
The next chapter provides an in-depth exploration of the decision-making frameworks that underpin adaptive resilience and greenness within OL-CAIS. 
We examine the theoretical foundations, development process, and evaluation criteria that inform our approach to real-time decision-making under disruptive conditions. 
Subsequently, we introduce the strategies and architectural elements that contribute to building a green resilient CPS. 

\section{Takeaways}
This chapter outlined the dynamic and iterative methodology developed throughout this PhD research to address the greenness–resilience trade-off in Cyber-Physical Systems. This chapter has highlighted the following key insights:

\begin{itemize}
    \item We employed a dynamic and iterative methodology that continuously evolved through feedback loops between theoretical modeling, empirical experimentation, and simulation. Our approach ensures that hypotheses, models, and experiments were adapted in response to observed outcomes and grounded in both literature and real-world insights.

    \item We introduced and validated a set of decision-making components: the \textit{Resilience Model}, the \textit{GResilience Framework}, and the \textit{CAIS-DMA}. These components serve as a cohesive toolkit that supports green and resilient behavior in OL-CAIS under disruptive conditions.

    \item We leveraged simulation not only to replicate disruptive events, but also to stress-test policies and validate them across diverse operational scenarios, enabling generalization beyond isolated experiments.

    \item Ultimately, we laid a methodological foundation for implementing and evaluating green resilience strategies in Cyber-Physical Systems. Our methodology is practical, extensible, and grounded in empirical validation.
\end{itemize}


\chapter{Decision-Making Components}
\label{ch:dmframeworks}
\begin{figure}[ht!]
    \centering
    \includegraphics[width=\textwidth]{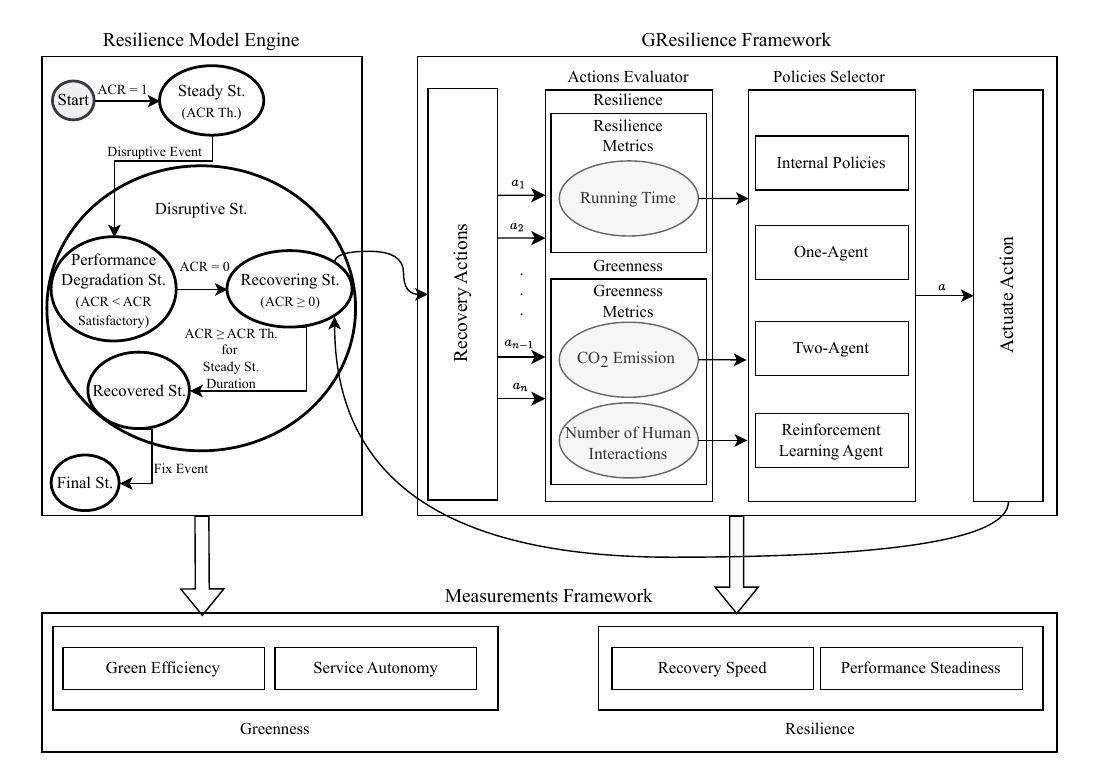}
    \caption{The detailed representation of our decision-making components.}
    \label{fig:method:methodscomponents}
\end{figure}

In this chapter, we present the decision-making components developed to support green recovery in OL-CAIS.
Fig.~\ref{fig:method:methodscomponents} shows the detailed representation of our decision-making cornerstone components.
The first component describes the \textit{resilience model engine} to detect disruptive events dynamically.
While in a recovering state, the resilience model creates an overtime evaluation by associating to the second component, i.e., \textit{the GResilience framework}. The GResilience framework identifies feasible recovery actions for evaluation. The actions evaluator measures the actions' resilience using their running time and greenness by combining the CO$_2$ emission and the number of human interactions.
The policies selector then uses these metrics with the pre-selected policies to recommend one of these actions to be actuated.
After actuating the recommended action, we complete the cycle with the resilience model to verify if the system performance has recovered.
The data from the resilience model and the GResilience evaluation are passed to the \textit{Measurements framework}. 
This third component evaluates the overall effectiveness of the decision-making policies in achieving greenness and resilience by comparing the different policies for recovery speed, performance steadiness, green efficiency, and service autonomy.
Furthermore, in this chapter, we introduce our \textit{Decision-Making Assistant (CAIS-DMA)}, an extendable tool designed for real-time action selection and monitoring equipped with our decision-making components. 

\section{Resilience Model}
\label{sec:dm:resiliencemodel}

The resilience model aims to support decision-makers by rendering OL-CAIS performance evolution over time.
This model addresses the need to automatically identify the timing of disruptive events that affect system performance.
Our approach leverages the iterative nature of online learning to track and quantify system behavior.
To this end, we introduce the \textit{Autonomous Classification Ratio (ACR)}. 
ACR measures the ratio of autonomous actions over a sliding window of m interactions, where $ACR \in [0, 1]$. 
ACR serves as an indicator of system autonomy, where higher ACR values indicate that the system operates autonomously. In contrast, lower values mean increased reliance on human intervention, often signaling performance degradation and human-dependency. 
A value of 0.5 means that both the human and the system have the same likelihood of performing an action.

We model the performance evolution of OL-CAIS as a time series of ACR values, each computed over the previous m-iterations. 
Alg.~\ref{algo:method:acrcalculation} shows the algorithmic steps to calculate ACR in OL-CAIS, where each iteration starts with the arrival of a new data instance.
The estimated probabilities by the OL-CAIS classifier are collected to determine whether the maximum estimated probability (i.e., $\hat{p}$) can be trusted by comparing it to the confidence threshold (i.e., $K$) to proceed with an autonomous action or human intervention is needed.
An autonomous action inserts one into a first-in-first-out queue and zero in case of human intervention.
At the end, the ACR values are calculated by dividing the sum of the queue over m.
At the start of a new iteration, we remove the top of the queue to ensure consecutive sliding.
Alg.~\ref{algo:method:acrcalculation} also shows that all ACR values are stored in the vector, $\mathbf{v}$, to form the ACR time series for further evaluation. 

\begin{algorithm}[htb!]
\caption{Calculating the autonomous classification ratio and defining the ACR values time series}
\label{algo:method:acrcalculation}
\footnotesize
\begin{algorithmic}[1]
\State \textbf{Parameters:}
\State $m \gets 5$  \Comment{5: is an example of sliding window size}
\State $\mathbf{v} \gets$ Vector() \Comment{v: is an empty vector to store series of ACR values}
\State $\mathbf{q} \gets$ Queue() \Comment{q: is an empty queue}
\BlankLine
\State \textbf{Initialization:} 
\For{$i = 0 \to m$}  
    \State $\mathbf{q}$.enqueue(0) \Comment{Insert 0 to the top of the queue}
\EndFor
\BlankLine
\State \textbf{Main Loop:}
\While{True} \Comment{Loop for each iteration of OL-CAIS}
    \State waitUntilReceiveNewDataInstance() \Comment{Holds the loop until new data instance arrives}
    \State $\hat{p} \gets$ $max($getClassifierEstimatedProbabilities()$)$ \Comment{$\hat{p}$: the classifier's maximum estimated probability}
    \State $\mathbf{q}$.dequeue() \Comment{Remove the top of the queue}
    \If{$\hat{p} \geq K$}
        \State $\mathbf{q}$.enqueue(1)
    \Else
        \State $\mathbf{q}$.enqueue(0)
    \EndIf
    \State $acr \gets \mathbf{q}.sum() / m$ \Comment{Calculate new ACR value}
    \State $\mathbf{v}$.append(acr) \Comment{Append the last ACR value to ACR vector $\mathbf{v}$}
    \State $state \gets$ evaluateTheCurrentState($\mathbf{v}$) \Comment{This function tracks the ACR values over iteration to understand at which state the OL-CAIS is in}
\EndWhile
\end{algorithmic}
\end{algorithm}

The primary objective of the resilience model is to identify the operational state the system stands in at any iteration during runtime.
To this aim, we define distinct \textit{states} as patterns of performance behavior by analyizing the ACR curve. 
The sequence of these states is the building block of our resilience model.
Similar to the resilience model described in Sec.~\ref{subsec:bg:qualities:resilience}, the ACR curve defines distinct operational states within OL-CAIS’s performance trajectory. 
The performance trajectory drawn by the ACR curve identifies an initial state where the system attempts to autonomously collaborate with humans to learn the OL-CAIS tasks, i.e., \textit{Steady State}.
As a disruptive event occurs, the curve continuously drops, indicating the start of the \textit{Disruptive State}.
The system recovers when ACR exceeds a defined threshold, and thus, it enters the \textit{Recovered State}. The \textit{ACR Threshold} is computed as the minimum ACR point in the initial steady state preceding the disruptive event.
The ACR Threshold also identifies if the system operates with an acceptable performance level, $ACR \ge ACRThrshold$, or not $ACR < ACRThrshold$.
The system transfers to a \textit{Final State} after resolving the disruptive event. We consider fixing the disruptive event as a second disruptive event, as it leads to a second performance degradation due to the memory of the AI model after learning the disruptive settings.

In some cases, disruptive events can be known to OL-CAIS's managers or they might require an automatic estimation criteria. 
In the latter case, we estimate when it has occurred by monitoring the effect on the system's performance. 
To this extent, we define a satisfactory level of ACR (e.g., $0.5$), which indicates the beginning of performance degradation and ends when ACR reaches zero (i.e., Performance Degradation State). 
This state estimates the upper bound of the period in which a disruptive event has occurred.
Alg.~\ref{algo:method:states} defines a state evaluation function that tracks ACR values to determine the OL-CAIS’s current operational state in response to performance changes. This function evaluates the ACR values in the received time series of ACR values and returns the current state of the system with respect to the following definitions of state.

\begin{enumerate}
    \item \textbf{Steady State}: The OL-CAIS has learned to make decisions autonomously. This state starts with the first sliding window of $ACR = 1$, and ends when a disruptive event occurs. In this state, the ACR can vary, and the ACR Threshold is computed as the minimum ACR value within this state.
    \item \textbf{Disruptive State}: This contains three sub-states, each capturing a different phase of the resilience responses.
    \begin{enumerate}
        \item \textit{Performance Degradation State}: OL-CAIS performance exhibits a continuous decline. It starts when ACR goes under the satisfactory level (e.g., $ACR = 0.5$) and lasts until it reaches zero.
        \item \textit{Recovering State}: The ACR values fluctuate above and below the ACR threshold, indicating attempts to recover the system by applying specific policies. When the degradation due to the disruptive event is at its minimum, $ACR=0$, the Recovering State starts and ends with the start of the Recovered State.
        \item \textit{Recovered State}: 
        When the ACR values stop fluctuating below the ACR threshold for at least the same length as the steady state, the Recovered State starts and ends when resolving the disruptive event.
    \end{enumerate}
    \item \textbf{Final State}: Similar to the disruptive state, this state has three sub-states as follows.
    \begin{enumerate}
        \item \textit{Second Performance Degradation State}: OL-CAIS performance exhibits a second continuous decline after fixing the disruptive event. Similar to the Performance Degradation State, it starts when the ACR goes under the satisfactory level and lasts until the ACR reaches its minimum $ACR = 0$.
        \item \textit{Second Disruptive State}: Fixing a disruptive event may have consequences on the performance. If this happens, the system experiences a second performance degradation, and then ACR values fluctuate to recover back to an acceptable level.
        \item \textit{Second Steady State}: This state's characteristics are the same as the recovered state, where OL-CAIS performance recovered from degradation to surpass the ACR Threshold in the original environmental settings and continues to operate normally.
    \end{enumerate}
\end{enumerate}

\begin{algorithm}[htb!]
\caption{Function to evaluate the current state}
\label{algo:method:states}
\footnotesize
\begin{algorithmic}[1]
\State \textbf{Arguments:}
\State $\mathbf{v} \gets$ Vector() \Comment{v: is a vector received from the function call}
\BlankLine
\State \textbf{Parameters:}
\State $curr \gets -1$ \Comment{Set the current state to -1 (no state)}
\State $satisfactory \gets 0.5$ \Comment{The satisfactory level of ACR (e.g., 0.5)}
\State $t_0 \gets 0$ \Comment{$t_0$: is the initial time of the steady state}
\State $t_e \gets 0$ \Comment{$t_e$: is the event time and the end of the steady state}
\State $t_a \gets 0$ \Comment{$t_a$: indicates the beginning of the recovering state (selecting recovery actions)}
\State $t_r \gets 0$ \Comment{$t_r$: indicates the beginning of the recovered state}
\State $acr\_threshold \gets 0$ \Comment{Set the ACR Threshold to 0}
\State $recovered\_flag \gets 0$ \Comment{A flag to mark the start of a potential recovered state}
\BlankLine
\State \textbf{States Loop:}
\For{$i = 0 \to \mathbf{v}.length()$}   
    \If{$curr = -1$ and $\mathbf{v}_i = 1$}
        \State $curr \gets 0$ \Comment{0: Steady State}
        \State $t_0 \gets i$ \Comment{Set the initial time to iteration number i}
    \ElsIf{$curr = 0$ and $\mathbf{v}_i \geq satisfactory$}
        \State $t_e \gets 0$ \Comment{Reset the event time}
    \ElsIf{$curr = 0$ and $\mathbf{v}_i < satisfactory$}
        \State $t_e \gets i$ \Comment{Set the event time to iteration number i}
    \ElsIf{$curr = 0$ and $\mathbf{v}_i = 0$}
        \State $curr \gets 1$ \Comment{1: Performance Degradation State}
        \If{$acr\_threshold = 0$}
            \State $acr\_threshold \gets min(\mathbf{v}_{t_0:t_e})$ \Comment{ACR Threshold is the minimum between $t_0$ and $t_e$}
        \EndIf
    \ElsIf{$curr = 1$}
        \State $curr \gets 2$ \Comment{2: Recovering State}
        \State $t_a \gets i$ \Comment{Set the start of recovery actions time to iteration number i}
    \ElsIf{$curr = 2$ and $\mathbf{v}_i < acr\_threshold$}
        \State $t_r \gets 0$ \Comment{Reset the recovered state start}
        \State $recovered\_flag \gets i + 1$ \Comment{Set the recovered flag to the next iteration}
    \ElsIf{$curr = 2$ and $\mathbf{v}_i \geq acr\_threshold$}
        \State $t_r \gets i$ \Comment{Set the recovered state start at iteration i}
        \If{$t_r - recovered\_flag = t_e - t_0$}
            \State $curr \gets 3$ \Comment{3: Recovered State}
        \EndIf
    \ElsIf{$curr = 3$} \Comment{For more than one disruptive event, reset state}
        \State $curr \gets 0$ \Comment{0: Steady State}
    \EndIf
\EndFor
\BlankLine
\State \textbf{Return:}
\State \textbf{\textit{return}} $curr$
\end{algorithmic}
\end{algorithm}

\begin{figure}[ht!]
    \centering
    \includegraphics[width=\textwidth]{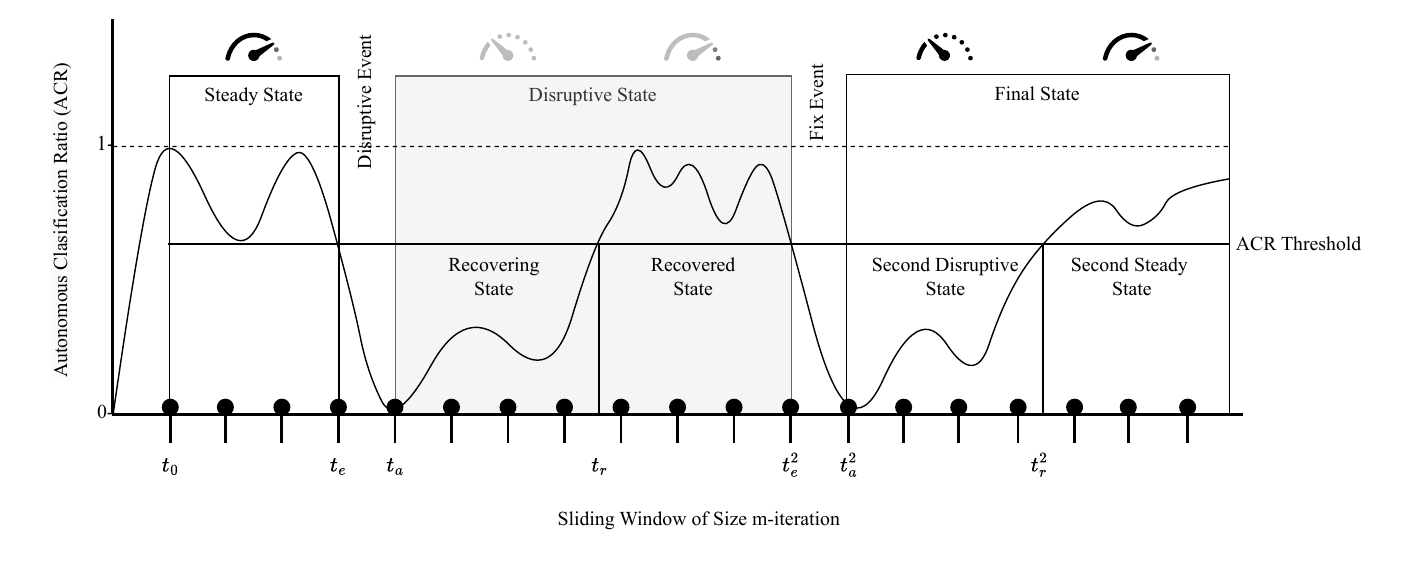}
    \caption{OL-CAIS resilience model.}
    \label{fig:method:resiliencemodel}
\end{figure}

Fig.~\ref{fig:method:resiliencemodel} illustrates OL-CAIS performance states and transitions across a disruptive event cycle. 
The figure describes the evolution of performance at two levels. 
An acceptable performance level is above the ACR Threshold, and a performance degradation level is under the threshold.
By observing the ACR curve, we can see the different states of the system as the curve moves above and under the threshold.
Starting with the \textit{Steady State}, the system learns to be able to perform actions autonomously, considering the threshold as the minimum ACR point during this state.
After a \textit{Disruptive Event}, the system experiences performance degradation and enters the \textit{Disruptive State}. In this state, the AI model may not be able to operate autonomously. 
Thus, it turns to its policies, entering the \textit{Recovering State} to recover the performance into the \textit{Recovered State}. 
Finally, the system enters the \textit{Final State} after fixing the cause of the disruptive event, which is considered a second disruptive state due to the forgetting of the original settings. 
The AI model may reenter a \textit{Second Disruptive State} and may need to interact with the human to restore its performance to the \textit{Second Steady State}. 
This state cycle applies to each disruptive event, which means there is another disruptive state after a disruptive event, another second disruptive state after fixing it, and so on.
Further, in Fig.~\ref{fig:method:resiliencemodel}, we only focus on one cycle of states to describe OL-CAIS behavior for a specific disruptive event. However, OL-CAIS may enter a repeated cycle of degradation, recovery, and stabilization in response to each disruptive event.

The resilience model not only aids in monitoring OL-CAIS stability but also provides a structured basis for implementing resilience-oriented policies. 
With the states defined and tracked via ACR, decision-makers have a systematic approach to managing and mitigating disruptive impacts on OL-CAIS performance.
Thus, during the recovering state, the resilience model invokes the GResilience framework, which operates in cycles to support the OL-CAIS to enter the recovered state.

\section{GResilience Framework}
\label{sec:dm:gresilienceframework}

\begin{figure}[ht!]
    \centering
    \includegraphics[width=\textwidth]{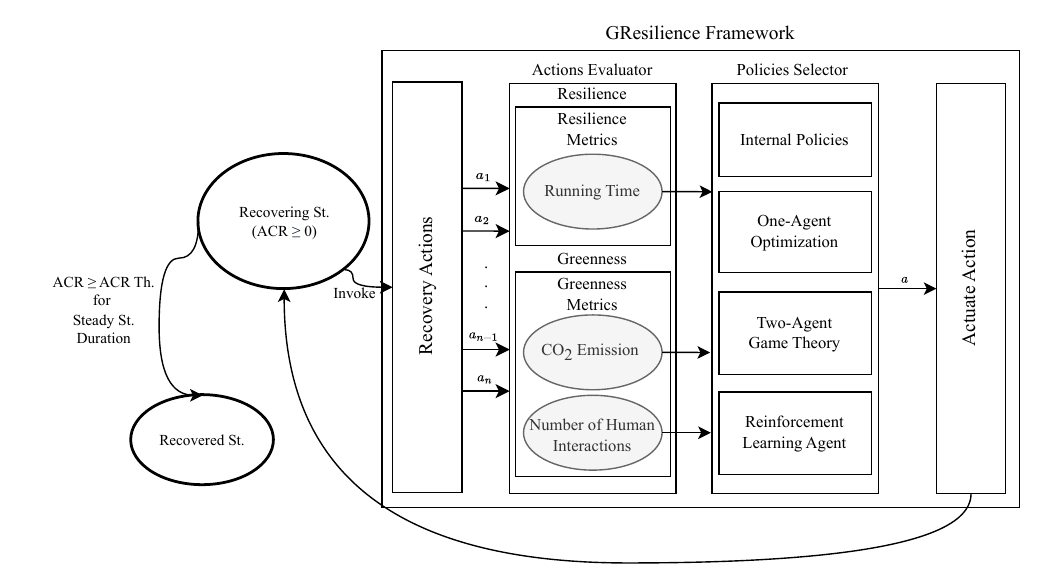}
    \caption{The GResilience framework components, illustrating the iterative execution during the recovery state of OL-CAIS.}
    \label{fig:method:gresiliencecomponents}
\end{figure}

Following the detection of performance degradation, OL-CAIS activates its recovery mechanisms, guided by internal policies, to restore performance to an acceptable level, as discussed in Sec.~\ref{sec:bg:cais}. 
The \textit{GResilience Framework} is designed to support OL-CAIS in selecting recovery actions that balance greenness and resilience during the recovery process. 
When the system enters a \textit{Recovering State}, the framework is invoked to evaluate recovery actions and apply a selected decision-making policy. The framework provides three policy options: \textit{one-agent}, \textit{two-agent}, and \textit{RL-agent} policies, each addressing trade-offs between greenness and resilience. Once the selected policy successfully restores system performance above the \textit{ACR Threshold}, the system transitions into the \textit{Recovered State}.
This section is structured to detail the GResilience framework components shown in Fig.~\ref{fig:method:gresiliencecomponents}: We first introduce the \textit{action evaluator}, then discuss the \textit{policies selector}, including \textit{one-agent policies}, \textit{two-agent policies}, and \textit{RL-agent policies}.

\subsection{Actions Evaluator}
\label{subsec:measure}
This component assesses feasible recovery actions for their contribution to the resilience and greenness attributes.
A recovery action is an action that is called after the system experienced a performance degradation due to a disruptive event,~\cite{henry_generic_2012}.
In our work, the GResilience Framework begins with a set of recovery actions $A = {a_1, a_2, ..., a_n}$, where each action is represented as a tuple of attributes.
The set, A, contains actions that can be executed either by autonomous system components or through human intervention.
The action's attributes play an essential role in determining the overall balance between greenness and resilience. 
By evaluating these attributes, our framework systematically identifies trade-offs that guide decision-making processes. 
These attributes enable the framework to adaptively respond to performance degradation, aiming to restore system stability while minimizing environmental impact.
We quantify each of these attributes by defining actions' metrics such as \textit{run time}, which contributes to the resilience attribute, and \textit{CO$_2$ emission} and \textit{number of human interactions} contribute to the greenness attribute.
The following defines these attributes.
\begin{itemize}
    \item \textit{Running Time}: We define the running time of an action as the duration from its initiation to completion. 
    Using the running time value directly in the decision-making process introduces randomness in estimating run time for future iterations. Thus, it is essential to incorporate historical information from previous iterations to guarantee more accurate estimations.
    We estimate the running time, $\hat{t}$, for each iteration and action by \textit{exponential smoothing}, which incorporates the history from the past iterations. 
    Eq.~\eqref{eq:forecastingexponentialsmoothing}, shows how we estimate the running time for the coming iteration $i+1$, $\hat{t}'$, where $t$ is the actual running time of the action at the current iteration $i$, $\hat{t}$ is the estimated running time at the current iteration, and $\alpha$ is the smoothing constant.
    The smoothing constant controls the influence of the past observations, a small smoothing constant decays the past observations, whereas a high smoothing constant boosts the past observations. Thus, we chose a smoothing constant that averages the past with the present, $\alpha = 0.5$.
    \begin{equation}
    \label{eq:forecastingexponentialsmoothing}
       \hat{t}' = \hat{t} + \alpha \cdot (t - \hat{t})
    \end{equation}

    \item \textit{CO$_2$ Emission}: This metric describes the CO$_2$ dioxide emitted from the CPU energy consumed by the AI model computations to estimate the probability (i.e., prediction) or update the model (i.e., learning from the human).
    To accurately measure the energy consumed by our AI model and isolate other possible routines running in the background, we define two energy points and measure their difference.
    The first reads the energy consumption before calling the AI model. The second reads the energy consumption immediately after estimating the probability or completing the model update if required.
    Energy consumption is quantified in kilowatt-hours and then multiplied by the carbon intensity (i.e., grams of CO$_2$ emitted per kilowatt-hour of electricity). 
    We take the carbon intensity value depending on the country of the electricity source, which is Italy in our case. The carbon intensity in Italy in the year 2023 is \textit{330.718} based on~\cite{ourworldindataGreenhouseEmissions, energyinstitute2023, mila_codecarbonio_nodate}. 
    The multiplication of carbon intensity and the energy consumption gives the CO$_2$ dioxide emitted in kilograms of CO$_2$equivalents (CO$_2$eq).
    We have used the \textit{CodeCarbon} \cite{mila_codecarbonio_nodate} to automatically collect and perform the carbon calculations.
    
    Similar to the running time, we estimate the CO$_2$ Emission, $\hat{c}$, for each iteration and action by exponential smoothing.
    Eq.~\eqref{eq:carbonfootprintestimation}, shows how to estimate the CO$_2$ Emission for the coming iteration $i+1$, where $c$ is the actual CO$_2$ Emission at the current iteration $i$, $\hat{c}$ is the estimated one for the iteration $i$, and $\alpha$ is the smoothing constant, again chosen as $0.5$.
    \begin{equation}
    \label{eq:carbonfootprintestimation}
       \hat{c}' = \hat{c} + \alpha \cdot (c - \hat{c})
    \end{equation}
    
    \item \textit{Number of Human Interactions.} This is the number of human interactions to complete an action.
    In practice, the total number of interactions for a task must be limited; otherwise, the human becomes overly dependent on the system.
    Thus, the number of human interactions represents the number of interactions still available, $h$, for each iteration and action.
    Eq.~\eqref{eq:humanlaborcost}, shows how to measure the remaining number of human interactions for the coming iteration $i+1$, $h'$, where $h_{max}$ is the maximum number of human interactions allowed, $h_{done}$ is the total number of human interaction already done for the action, and $h$ is the number of human interactions needed to perform the action.
    \begin{equation}
    \label{eq:humanlaborcost}
       h' = h_{max} - h_{done} - h
    \end{equation}
\end{itemize}

Finally, we represent each action in the set of actions, $A$, as a tuple of three metrics, making $A = \{(\hat{t}'_{a_1}, \hat{c}'_{a_1}, h'_{a_1}), (\hat{t}'_{a_2}, \hat{c}'_{a_2}, h'_{a_2}), \cdots, (\hat{t}'_{a_n}, \hat{c}'_{a_n}, h'_{a_n})\}$, where $a_n$ is the selected recovery action at iteration $n$.
The set of actions, $A$, is then passed to the policies selector component.

\subsection{Policies Selector}
The policies selector works buffer to pass the set of actions, $A$, toward preselected policies by the decision-makers.
Each of these policies activates different mechanisms to select the action that best trade-off between the greenness resilience metrics.
The rest of this section will discuss each of these policies in more detail.

\begin{figure}[ht!]
    \centering
    \includegraphics[width=\textwidth]{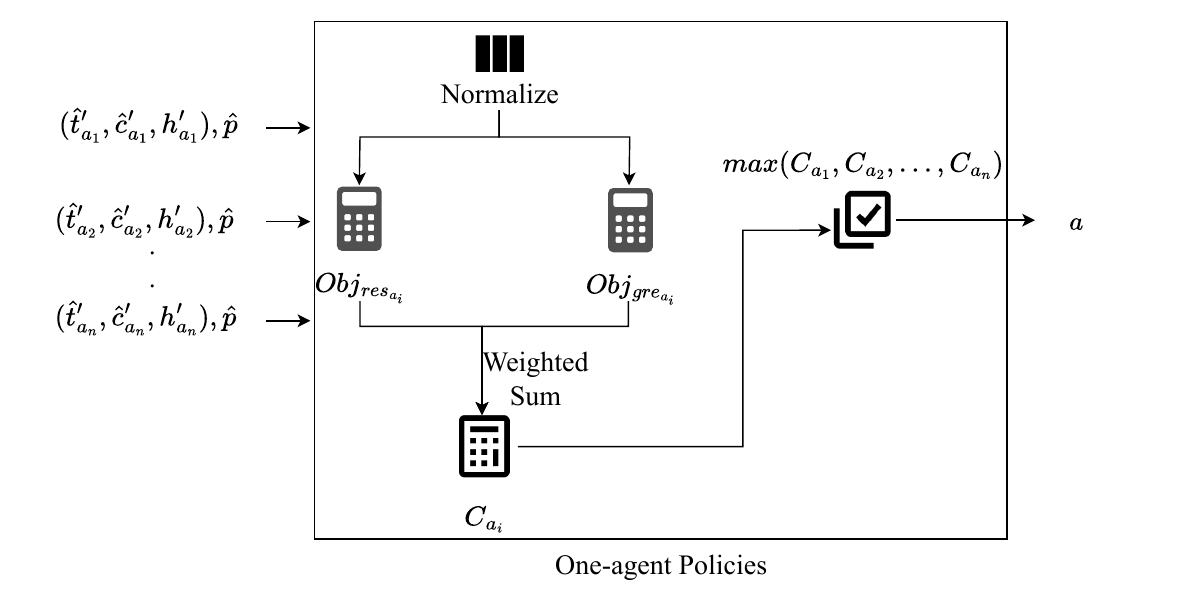}
    \caption{Flowchart of one-agent policies for selecting recovery actions. We normalize the action metrics to compute the resilience and greenness costs for each action. Then, we calculate a weighted sum of these costs and select the action with the highest combined score.}
    \label{fig:method:wsmflowchart}
\end{figure}

\paragraph{One-Agent Policies.} 
\label{subsec:dm:oneagent}
Fig.~\ref{fig:method:wsmflowchart} illustrates the structured decision-making process within the \textit{Weighted Sum Model (WSM)} approach for selecting recovery actions. The flowchart visually represents how the model evaluates potential actions by assessing their impact on both resilience and greenness. Each action is analyzed based on its estimated run time, CO$_2$ emissions, remaining human interactions, and classifier confidence. These attributes are normalized and combined into a weighted cost function, which ranks the actions and selects the one that balances the trade-off between resilience and greenness.

In one-agent policies, we define a cost function leveraging the ``Weighted Sum Model'' (WSM) discussed in Sec.~\ref{subsec:bg:dmp:opt}. 
This cost function forms a one-agent representation of the trade-off problem between greenness and resilience, combining the functions of the two properties to find a single score per action.
Our goal is to find the optimal action that maximizes our cost function. 
In OL-CAIS context, the possible recovery actions at each iteration are \{$a_1:$ \textit{classify the new data instance autonomously}, $a_2:$ \textit{inquire the human to classify the new data instance}\}.
These two actions represent the solutions in the WSM decision matrix. Table~\ref{tab:dm:decisionmatrix} rows show these solutions, and their columns show the objective function for each solution.
The resilience objective function, $Obj_{res}: \mathbb{R} \times \mathbb{R} \rightarrow \mathbb{R}$, takes the maximum estimated probability from the OL-CAIS classifier, $\hat{p}$, and the action estimated run time, $\hat{t}'$, and finds the production between $\hat{p}$ and the normalized inverse of  $\hat{t}'$. We find the inverse of $\hat{t}'$ as we maximize the value of resilience, and the fastest action leads to faster operation.
Noting that $N(*)$ represents the normalized value of $*$. Where we use the L2-norm to normalize these values.
Similarly, the greenness objective function, $Obj_{gre}: \mathbb{R} \times \mathbb{R} \times \mathbb{R} \rightarrow \mathbb{R}$, takes the maximum estimated probability from the OL-CAIS classifier, $\hat{p}$, the remaining number of human interactions, $h'$, and the estimated CO$_2$ emission, $\hat{c}'$, and finds the production between complementary probability of $\hat{p}$ and the sum of the normalized values of $h'$ and the inverse of $\hat{c}'$.

\begin{table}[ht!]
\caption{GResilience one-agent decision matrix of weighted sum model.}
\begin{center}
\begin{tabular}{ccc}
    \hline
    \textbf{Solutions} & \textbf{Obj$_{res}$} & \textbf{Obj$_{gre}$} \\\hline
    \textbf{$a_1$} & $\hat{p} \cdot N(1/\hat{t}'_{a_{1}})$ & $(1-\hat{p}) \cdot (N(h'_{a_{1}}) + N(1/\hat{c}'_{a_{1}}))$ \\
    \textbf{$a_2$} & $\hat{p} \cdot N(1/\hat{t}'_{a_{2}})$ & $(1-\hat{p}) \cdot (N(h'_{a_{2}}) + N(1/\hat{c}'_{a_{2}}))$ \\\hline
\end{tabular}
\label{tab:dm:decisionmatrix}
\end{center}
\end{table}

To find the optimal action to balance the resilience and greenness objectives, we find the weighted sum of the two objectives and then rank them to select the action with the maximum score.
Formally, let $C: \mathbb{R} \times \mathbb{R} \times \mathbb{R} \times \mathbb{R} \rightarrow \mathbb{R}$ be the cost function mapping the maximum estimated probability and the action's attributes to real numbers. 
The cost function $C$ in the next iteration $i+1$, defined as Eq.~\eqref{eq:globalscore}, such that $w_r$ and $w_g$ are the weights for resilience and greenness objectives, respectively.
Since our goal is maximizing both greenness and resilience equally, we set the weights to $0.5$ each. 
The action to select is the action with the maximum cost, $max(C(\hat{p}, \hat{t}'_{a_{1}}, h'_{a_{1}}, \hat{c}'_{a_{1}}), C(\hat{p}, \hat{t}'_{a_{2}}, h'_{a_{2}}, \hat{c}'_{a_{2}}))$, at each iteration $i+1$, which requires decision-making support.

\begin{equation}
    \label{eq:globalscore}
C(\hat{p}, \hat{t}'_{a}, h'_{a}, \hat{c}'_{a}) = w_r \cdot \left[\hat{p} \cdot N\left(1/\hat{t}'_{a}\right)\right] + w_g \cdot \left[(1 - \hat{p}) \cdot \left(N(h'_{a}) + N\left(1/\hat{c}'_{a}\right)\right)\right]
\end{equation}

\begin{figure}[ht!]
    \centering
    \includegraphics[width=\textwidth]{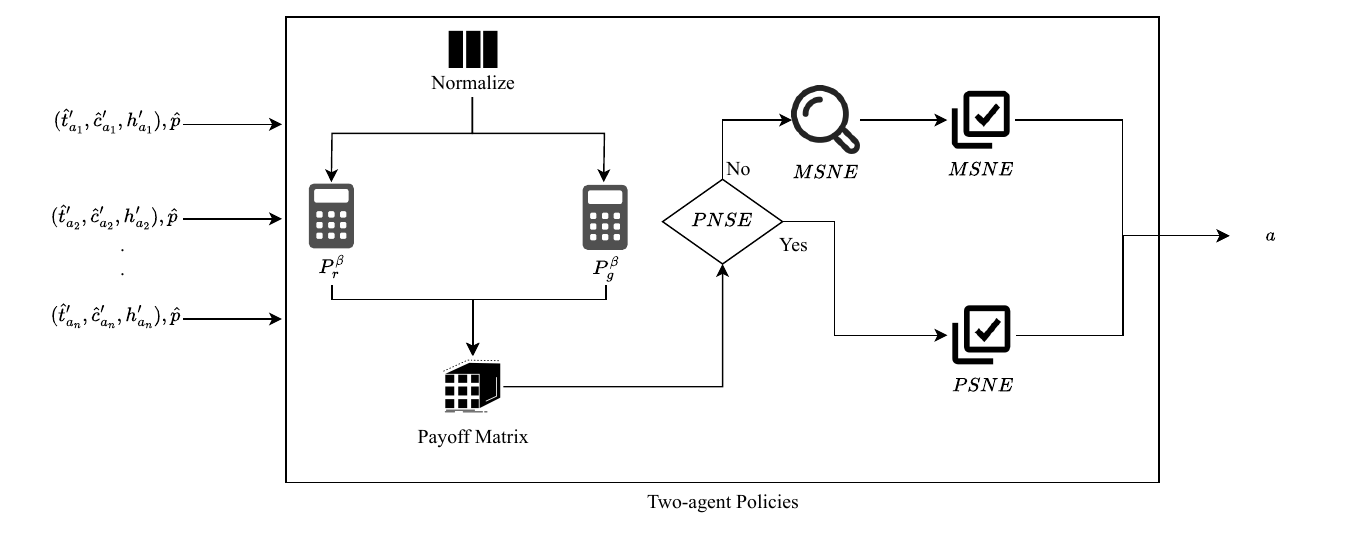}
    \caption{Flowchart of two-agent policies for selecting recovery actions. We normalize the action metrics to compute the payoffs for both the resilience and greenness players, then construct the payoff matrix. Using this matrix, we search for a Pure-Strategy Nash Equilibrium (PSNE) and select the action with the higher payoff. If no PSNE exists, we compute the Mixed-Strategy Nash Equilibrium (MSNE) and select the action with the highest probability.}
    \label{fig:method:twoagentflowchart}
\end{figure}

\paragraph{Two-Agent Policies.} 
\label{subsec:dm:gresiliencegame}
Fig.~\ref{fig:method:twoagentflowchart} provides a structured flowchart of the decision-making process within the \textit{Two-Agent Policies}, illustrating how the GResilience Game formulates the trade-off between greenness and resilience. The flowchart outlines the sequential steps where the resilience player ($P_r$) and the greenness player ($P_g$) evaluate available actions, compute payoffs, and determine an optimal strategy using either Pure Strategy Nash Equilibria (PSNE) or Mixed Strategy Nash Equilibria (MSNE). This approach ensures that the selected recovery action aligns with both resilience and greenness objectives while maintaining the system's ability to recover from disruptive events.

Our game, The GResilience Game, leverages the non-cooperative game ``The Battle of Sexes'', \cite{stowe_cheating_2010}. 
The GResilience Game is a two-agent representation of the trade-off problem: $P_{g}$ and $P_{r}$, the greenness player and the resilience player, respectively. 
The goal of $P_{g}$ is to maximize OL-CAIS greenness by minimizing the CO$_2$ emission by relying more on human collaboration. In contrast, the goal of $P_{r}$ is to maximize the system resilience by minimizing the running time. 
In addition, both players share the same goal of recovering the system.
The players must adopt a strategy to recover the system from the performance degradation caused by the disruptive event while trying to maintain their private goals.
The GResilience Game uses a payoff matrix structure similar to The Battle of the Sexes, discussed in Sec.~\ref{subsec:bg:dmp:gt}, where it has multiple solutions, two Pure Strategies Nash Equilibria (PSNE) when the two players agree on the same action, and it may have another Mixed Strategy Nash Equilibrium (MSNE), by randomizing the choice based on the probability of each player's action,~\cite{satapathy_game_2016, bonfanti_gresilience_2023, rimawi_green_2022, stowe_cheating_2010}.
Each player has multiple actions to choose from. Table~\ref{tbl:dm:gresiliencegame} represents the payoff matrix and expected payoffs (EP) of the GResilience game for greenness and resilience players with two actions each.
The first action, $a_1$, classifies the new data instance autonomously preferred by the resilience player. The second action, $a_2$, inquires the human to classify the new data instance preferred by the greenness player.
We pass the action itself as an argument to the payoff functions for readability. However, the action attributes and the maximum estimated probability, $\hat{p}$, are the actual arguments of the functions.

\begin{table}[ht!]
\centering
\caption{Payoff matrix for the GResilience Game in case of two actions: Representation of player strategies and outcomes in the decision-making process for system recovery.}
\footnotesize
\begin{tabular}{ccccc}
   & & \multicolumn{2}{c}{\textbf{$P_g$}}& \\
   &  & \textbf{$a_1$} (p) & \textbf{$a_2$} (1 - p)&\textbf{$P_r$ EP} \\ 
   \cline{2-5}
   \multirow{2}{*}{\rotatebox[origin=c]{90}{\textbf{$P_r$}}} & \textbf{$a_1$} (q) &  $P^{2}_{r}(a_1), P^{2}_{g}(a_1)$ &  $P^{1}_{r}(a_1), P^{1}_{g}(a_2)$ & $p P^{2}_{r}(a_1)  + (1 - p)  P^{1}_{r}(a_1)$ \\ 
   & \textbf{$a_2$} (1 - q)&  $P^{1}_{r}(a_2), P^{1}_{g}(a_1)$ &  $P^{2}_{r}(a_2), P^{2}_{g}(a_2)$ & $p  P^{1}_{r}(a_2)  + (1 - p)  P^{2}_{r}(a_2)$ \\
   & \textbf{$P_g$ EP} & $q  P^{2}_{g}(a_1)  + (1 - q)  P^{1}_{g}(a_1)$ & $q  P^{1}_{g}(a_2)  + (1 - q)  P^{2}_{g}(a_2)$ & \\\cline{2-5}
\end{tabular}
\label{tbl:dm:gresiliencegame}
\end{table}

Both Eq.~\eqref{eq:resiliencepayoff}~and~\eqref{eq:greennesspayoff} show how to calculate the payoff for each player, where the resilience player payoff function $P^{\beta}_{r}: \mathbb{R} \times \mathbb{R} \rightarrow \mathbb{R}$, finding the product between the maximum estimated probability, $\hat{p}$, and the normalized inverse of the action estimated time, $\hat{t}'$.
While the greenness player payoff function $P^{\beta}_{g}: \mathbb{R} \times \mathbb{R} \times \mathbb{R} \rightarrow \mathbb{R}$, finding the product between the complementary probability of $\hat{p}$, and the normalized inverse of the action estimated time, $\hat{t}'$, and remaining number of human interactions, $h'$.
In both payoff functions, $\beta$ represents the matching index that is $1$ in case the players selected different actions and $2$ in case of pure equilibrium. The matching index serves as a magnifier of the payoff function in the case of pure equilibrium.
In case of mixed equilibrium, player $P_r$ chooses $a_1$ with probability $q$ and $a_2$ with probability $1-q$, while $P_g$ chooses $a_1$ with probability $p$ and $a_2$ with probability $1-p$, which results the \textit{expected payoff}. Thus, to find the probability $q$ (resp. $p$) with MSNE, we equal the expected payoffs of $P_g$ (resp. $P_r$) for $a_1$ and $a_2$ and solve the resulting equation for $q$ (resp. $p$).
The payoff matrix can be generalized for any number of actions, with the matrix diagonal as PSNE, and the probabilities of all actions sum to $1$ for MSNE.

\begin{equation}
\label{eq:resiliencepayoff}
    P^{\beta}_{r}(\hat{p}, \hat{t}'_{a}) = \beta \cdot \hat{p}  \cdot N(1/\hat{t}'_{a})
\end{equation}

\begin{equation}
\label{eq:greennesspayoff}
    P^{\beta}_{g}(\hat{p}, h'_{a}, \hat{c}'_{a}) = \beta \cdot (1 - \hat{p} ) \cdot  N(1/h'_{a}) \cdot N(1/\hat{c}'_{a})
\end{equation}

The GResilience Game selects the action based on the existing equilibrium of each run. 
The priority is to choose the pure equilibrium with the maximum payoffs combined. 
In the case of pure equilibrium absence, we choose the mixed equilibrium with probabilities $q$ and $p \in [0, 1]$. We then multiply the probabilities in the payoff matrix to find $q \cdot p$, $q \cdot (1 - p)$,  $p \cdot (1 - q)$, and $(1 - q) \cdot (1 - p)$, and then we choose the solution with maximum payoff.

\paragraph{Reinforcement Learning Agent Policies.}
\label{subsec:dm:rlagent}
We use Q-Learning as an off-policy temporal difference control,~\cite{watkins1992q}. This mechanism learns the action-value function, Q, which approximates the optimal action-value function ($q_*$).  
Eq.~\eqref{eq:qfunction} shows the Q function, where $Q(S_t, A_t)$ is the current estimate of the Q-value, $\alpha$ is the learning rate, $\alpha \in (0, 1]$, $R_{t+1}$ is the reward received, $\gamma$ is the discount factor, $\gamma \in [0, 1]$, and the $\max_{a \in A} Q(S_{t+1}, a)$ is the maximum Q-value for the next state given all possible actions,~\cite{sutton2018reinforcement}.

\begin{equation}
Q(S_t, A_t) \gets Q(S_t, A_t) + \alpha \left[ R_{t+1} + \gamma \max_{a \in A} Q(S_{t+1}, a) - Q(S_t, A_t) \right]
\label{eq:qfunction}
\end{equation}

\begin{figure}[ht!]
\centering
\includegraphics[width=\textwidth]{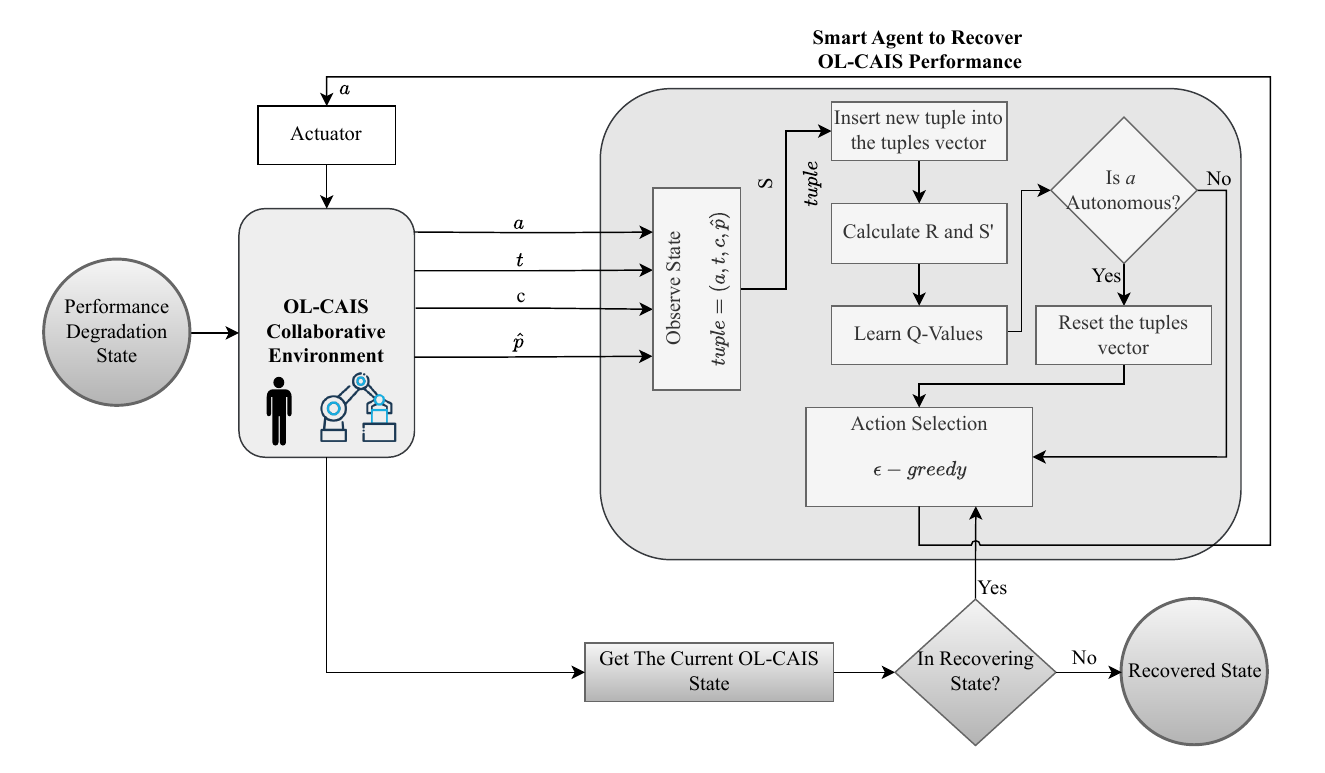}
\caption{Reinforcement learning agent architecture.}
\label{fig:rlarch}
\end{figure}

During a performance degradation state, a vector of selected actions (i.e., autonomous/human actions), the actual run time of each action, $t$, the actual CO$_2$ emission, $c$, and the maximum estimated probability at each iteration, $\hat{p}$, are passed to the RL-agent to calculate the initial state.
The states in our RL-agent measure the greenness resilience balanced state of the system.
The RL-agent calculates the states by computing the weighted sum of accumulated CO$_2$ emissions (i.e., greenness) and run times (i.e., resilience) for all actions in the vector.
Fig.~\ref{fig:rlarch} shows the RL-agent architecture, which operated during the recovering operational state. 
To create its policies, the RL-agent initiates the Q-table to keep track of Q-values at each state.
At each iteration, the agent selects one action using the $\epsilon-greedy$ function, which is a function that selects a random action with $\epsilon$ probability a number of times and uses its policies for the rest of the times. Using $\epsilon-greedy$ ensures a trade-off between exploration-exploitation of the different actions.

The RL-agent then observes the collaborative environment of the OL-CAIS to create a tuple of the action's $t$, $c$, and the classifier's $\hat{p}$. 
The tuple is then appended to the vector of actions to calculate the new state and the reward obtained from this action.
The reward is designed with two goals: i) restore the system's ability to perform its tasks autonomously after a performance degradation, and ii) improve OL-CAIS AI model accuracy to maintain an acceptable level of performance after it recovers. 
Thus, we calculate the reward by finding the product of all $\hat{p}$ in the vector divided by the vector length.
Then, we reset the actions' vector when the last actuated action in the collaborative environment is autonomous.
This ensures that autonomous execution is returned when achieving higher accuracy.
The Q-table is then updated by computing the temporal difference as shown in Eq.~\eqref{eq:qfunction}.
This approach enables the RL-agent to iteratively enhance autonomy and accuracy, supporting the overall goal of balancing greenness and resilience.

The outlined process for action selection, reward calculation, and Q-table learning is detailed in Alg.~\ref{algo:performancemeasure}.
The algorithm illustrates the steps to estimate the optimal policy, $q_*$.
The algorithm shows the expected arguments and their definition. Then, it defines the RL states to combine both greenness and resilience measures and the reward to the product of the classifier maximum estimated probability, $\hat{p}$.
The RL-agent loops in episodes and steps. The episodes loop until the OL-CAIS operational state changes from recovering to recovered. 
The steps loop stops when selecting an autonomous action or when the number of steps is equal to the number of iterations in the steady state.
During each step, the RL-agent observes the collaborative environment to collect the explored-exploited action's attributes and $\hat{p}$ to find the state, reward, and learn Q-values.

\begin{algorithm}[ht!]
\caption{Q-Learning function for estimating $q_{*}$ for OL-CAIS upon disruption}
\label{algo:performancemeasure}
\footnotesize
\begin{algorithmic}[1]
\State \textbf{Arguments:}
\State $\alpha \in (0, 1]$ \Comment{$\alpha$: learning rate}
\State $\epsilon > 0$ \Comment{$\epsilon$: the exploration-exploitation constant, $\epsilon$, should be small}
\State $\mathbf{pd} = [(a_1, c_{a_1}, t_{a_1}, \hat{p}_{a_1}), \cdots, (a_i, c_{a_i}, t_{a_i}, \hat{p}_{a_i})]$ \Comment{$\mathbf{pd}$: is a vector of all actions executed during the Performance Degradation State}
\State $ol\_cais\_state \in $ [0: `Steady State', 1: `Performance Degradation State', 2: `Recovering State', 3: `Recovered State']
\State $steady\_duration > 0$ \Comment{Number of iterations in the Steady State}
\BlankLine
\State \textbf{Initialization:} 
\State $S \gets  w_g \sum_{i=0}^{|\mathbf{pd}|}\mathbf{pd}_{i,c}$  + $w_r \sum_{i=0}^{|\mathbf{pd}|}\mathbf{pd}_{i,t}$
\State $Q(s, a) = 0, \forall s \in S, a \in A$ 
\BlankLine
\State \textbf{Main Loop:}
\While{$ol\_cais\_state = 2$} \Comment{Episodes loop (2: `Recovering State')}
    \State $\mathbf{av} \gets  vector<tuple>$ \Comment{Create an empty vector of tuples}
    \State $steps\_count \gets 0$
    \While{$steps\_count < steady\_duration$} \Comment{Steps loop}
        \State $a \gets  \epsilon$-greedy() \Comment{Choose action $a$ based on exploration-exploitation policy}
        \State $run(a)$ \Comment{Actuate action $a$, and observe the collaborative environment}
        \State $c \gets a_{CO_2}$ \Comment{Observe CO$_2$ emitted by action $a$}
        \State $t \gets a_t$ \Comment{Observe action $a$ run time}
        \State $\hat{p} \gets$ $max($getClassifierEstimatedProbabilities()$)$ \Comment{$\hat{p}$: the maximum estimated probability}
        \State $tuple \gets  (a, c, t, \hat{p})$ \Comment{Create a tuple of observed values}
        \State $\mathbf{av} \gets \mathbf{av} \cup \{tuple\}$ \Comment{Append the tuple to the vector}
        \State $S' \gets  w_g \sum_{i=0}^{|\mathbf{av}|} \mathbf{av}_{i,c} + w_r \sum_{i=0}^{|\mathbf{av}|} \mathbf{av}_{i,t}$ \Comment{Calculate the new state}
        \State $R \gets  (1 / |\mathbf{av}|) \prod_{i=0}^{|\mathbf{av}|} \mathbf{av}_{i,\hat{p}}$ \Comment{Calculate the reward}
        \State $Q(S, a) \gets  Q(S, a) + \alpha [ R + \gamma \max_{a'}Q(S', a') - Q(S, a) ]$ \Comment{Update Q-table using temporal difference}
        \State $S \gets  S'$ \Comment{Update current state}
        \If{$a = Autonomous$} 
            \State break \Comment{Break the steps loop on the first autonomous action}
        \Else
            \State $steps\_count \gets steps\_count + 1$
        \EndIf
    \EndWhile
    \State $ol\_cais\_state \gets $getTheCurrentState() \Comment{Get the current state of the OL-CAIS under test}
\EndWhile
\end{algorithmic}
\end{algorithm}

At the end of the GResilience framework, the action selected by the decision-making policies is actuated. Then, close the cycle with the resilience model engine to verify whether the OL-CAIS has recovered.
The resilience model and the decision-making iterations are passed to the measurement framework to assess the effectiveness of the policies to the overall greenness and resilience.

\subsection{Measurements Framework}
\label{subsec:method:measures}
The third component of our decision-making components is the Measurements Framework, which supports a comprehensive evaluation of decision-making policies for OL-CAIS under specific disruptions. 
This framework measures specific concepts designed from our perspective that we gained during this research period. 
These concepts offer decision-makers insights into selecting the optimal policy based on the system’s resilience and greenness requirements.
To understand decision-making policies' effectiveness toward resilience, it is important to measure both \textit{recovery speed} and \textit{performance steadiness}.
The recovery speed concept defines the policies' abilities to transfer the OL-CAIS performance from a recovering state to a recovered state.
To measure the recovery speed, we compute the duration ratio of the recovering state to the total disruptive state.
The policies minimizing this ratio are the ones with the highest recovery speed.
Although recovery speed is a very important concept for resilience, performance steadiness is equally important. The performance steadiness concept describes the performance behavior during the disruptive state. It indicates the decision-making policies that keep the performance behavior from fluctuating.
We measure this concept by finding the fluctuation ratio between the number of ACR points below the ACR Threshold to the points above it. The policies with a lower fluctuation ratio are the ones achieving better performance steadiness.

To understand decision-making policies' effectiveness toward greenness, we define both \textit{green efficiency} and \textit{autonomy} concepts.
The green efficiency concept defines the carbon sustainability of the decision-making policies by measuring the computational energy consumption's carbon emission. 
In particular, we measure the mean CO$_2$ emission emitted by the classifier of OL-CAIS during the disruptive state. The lower the mean, the better the policies for achieving green efficiency.
Emitting low CO$_2$ on the cost of increasing the OL-CAIS dependency on human interactions contradicts the industrial goals of OL-CAIS. Here comes the importance of the autonomy concept, which defines the system's ability to make autonomous actions. We measure autonomy by finding the system human-dependency, which is the average human actions during the disruptive state.
The following details the measurement framework concepts measures:

\begin{enumerate}
    \item \textit{Recovering State Duration Ratio}. This metric quantifies the duration of the recovering period as a percentage of the total disruptive state length, indicating how quickly the system regains its performance after a disruption. 
    Lower values reflect faster recovery, as described by Eq.~\eqref{eq:disper}, where $|*|$ denotes the length of a state.
    \begin{equation}
    \label{eq:disper}
            \text{DurationRatio}_{RecoveringState} = 
            \frac{|\text{RecoveringState}|}{|\text{DisruptiveState}|}
    \end{equation}
    \item \textit{Performance Fluctuation}. Performance fluctuation measures the performance instability during the disruptive state. A lower ratio indicates less fluctuation, reflecting greater steadiness. It is calculated as the ratio of ACR values below the threshold to those above it, per Eq.~\eqref{eq:put_pat_ratio}. Where ACR$_{\text{DisruptiveState}}$ is the set of ACR values during the Disruptive State.
    \begin{equation}
    \label{eq:put_pat_ratio}
        \text{Fluctuation\_Ratio} =
        \frac{\textit{Count}(\text{acr} < \textit{ACRThreshold})}{\textit{Count}(\text{acr} \geq \textit{ACRThreshold})}, \text{for acr $\in$ ACR$_{\text{DisruptiveState}}$}
    \end{equation}
    \item \textit{CO$_2$ Emission Mean}. This metric represents the mean CO$_2$ emissions during the disruptive state, highlighting policies that minimize environmental impact. It is computed as the total CO$_2$ emissions in the disruptive state divided by its length, as shown in Eq.~\eqref{eq:co2_emission_mean}, where $\mu(\text{CO$_2$})$ is the CO$_2$ emission mean, and $\text{CO}_{\text{DisruptiveState}}$ is the set of CO$_2$ emitted values during the Disruptive State.
    \begin{equation}
    \label{eq:co2_emission_mean}
        \mu(\text{CO$_2$}) = 
        \frac{\sum_{\text{co}_2 \in \text{CO}_{\text{DisruptiveState}}} \text{co}_2}{|\text{DisruptiveState}|}
    \end{equation}
    \item \textit{Human-Dependency}. Human-dependency assesses the degree of reliance on human interactions to restore performance. Lower dependency scores indicate higher autonomy. It is calculated as the total number of human interactions in the disruptive state divided by its length, per Eq.~\eqref{eq:human_interactions_mean}, where $\text{HI}_{\text{DisruptiveState}}$ is the set of human interactions during the Disruptive State.
    \begin{equation}
    \label{eq:human_interactions_mean}
        \text{Human\_Dependency\_Ratio} =
        \frac{\sum_{h \in \text{HI}_{\text{DisruptiveState}}} h}{|\text{DisruptiveState}|}
    \end{equation}
\end{enumerate}

By aiming to minimize each metric, decision-makers can identify the policy that best supports rapid recovery, steadiness, green efficiency, and autonomy. 
However, selecting the optimal policy may also require a balance between metrics, as improvements in one area may affect performance in another.
This measurement framework, therefore, equips decision-makers with the flexibility to prioritize and adapt policies to meet the specific requirements of their OL-CAIS applications.

\section{Decision-Making Assistant}
\label{sec:method:cais-dma}

The Decision-Making Assistant for OL-CAIS (CAIS-DMA) is developed to provide an extendable framework that equips OL-CAIS with essential monitoring and actuation capabilities, enabling automatic policy invocation upon disruptions. 
This conceptual architecture not only supports real-time adaptability but also creates a simulated environment for OL-CAIS, allowing in-depth experimentation and stress-testing of greenness and resilience strategies. 
For decision-makers and managers, CAIS-DMA offers a controlled platform to explore the effects of various disruptions on the AI components of OL-CAIS, refining their strategies without impacting the physical system.

\begin{figure}[ht!]
\includegraphics[width=\textwidth]{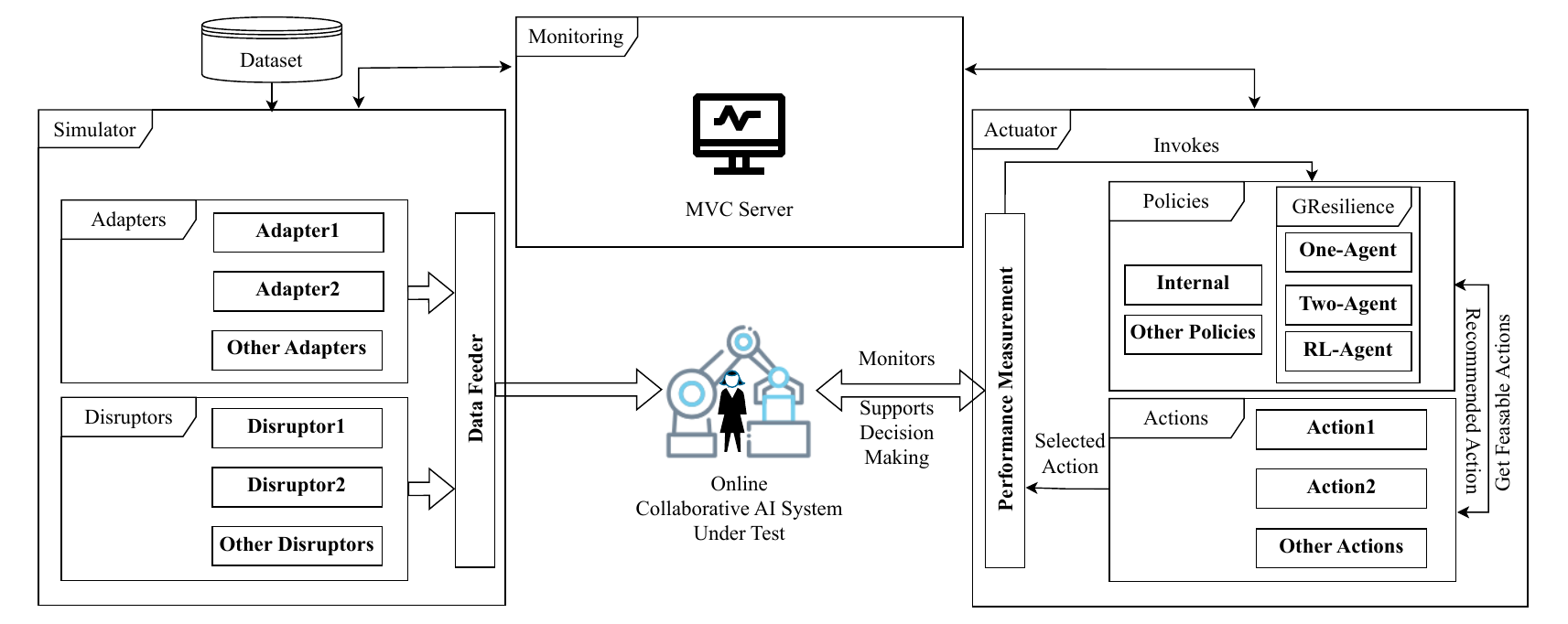}
\caption{CAIS-DMA conceptual architecture.} \label{fig:caisdma}
\end{figure}

The framework consists of three components: i) a data simulator, ii) a decision-making actuator, and iii) a monitoring component. Fig.~\ref{fig:caisdma}, shows the conceptual architecture of CAIS-DMA components interacting with the OL-CAIS under test. 
The rest of this section will discuss each of these components.

\subsection{Simulator}
\label{subsec:simulator}

The simulator component aims to simulate the AI model's learning data in OL-CAIS, edit the data to represent a specific disruptive event effect, re-structure the data to the same format as expected by the AI model, and stream it one data instance at a time to align with online learning concepts. 
To achieve this, we designed the simulator to include a \textit{dataset} collected from the OL-CAIS environment, an \textit{adapters} module, a \textit{disruptors} module, and a \textit{data feeder}, as shown in Fig.~\ref{fig:caisdma}.

The dataset contains the training data collected from the actual environment of the OL-CAIS under test. For example, in an industrial environment with an object sorter based on shape or color, it is important to collect images of the objects that are or will be used when the sorter runs.
Additionally, to guarantee a realistic simulation of the system actions, it is important to define the actions, their run time, CO$_2$ emission, and the number of interactions needed from the human to complete the action.
Furthermore, this dataset must include the ground truth for the classifier, as it is also used to simulate human action, assuming that it is the information source in online learning.

The adapters and disruptors modules then interact with the dataset to perform one or more modifications to transform the raw data collected to suit the AI model's learning data format.
The adapters module defines the data structure in which the AI model receives data. For example, a JSON structure with $x$, $y$, and $z$ as fields, such that $x$ has a title string, $y$ has a \textit{base64} encoded image, and $z$ has an integer with the class number.
To ensure enough abstraction and extendability, we created an abstract class that defines the necessary methods of all adapters and a factory method that returns the requested adapter.
The disruptors module follows a similar implementation strategy. However, the disruptors have a different purpose, as they manipulate the AI learning data to simulate an effect similar to a disruptive event.
For example, in classifiers that depend on images, a disruptor can apply one or more filters to change the image histogram.

The data feeder is another module responsible for passing the data to the AI model one instance at a time, simulating the online learning process of handling one data instance in each iteration.
Furthermore, it passes the data instances in normal and disrupted modes.
The data instances in normal mode are passed without the effect of the disruptors simulating the original environment of the system. 
For example, the raw data can be structured in the AI model format without applying any filters to change the image histogram. The disrupted mode passes the data instances in the AI model format but also with the filters that change the histogram.
The purpose of having these two modes is to simulate resilience model states. The data feeder operates in normal mode at the start, allowing the system to enter a steady state. Then, it switches to the disrupted mode to simulate a disruptive event that may lead to a disruptive state, and finally, it goes back to normal mode to enter the final state.

The simulator component functions as an experiment moderator, enabling control over iteration count and disruptor types for testing OL-CAIS under various scenarios. 
This capability also provides OL-CAIS managers with a practical tool for empirically evaluating system performance across different disruptive events and durations.

\subsection{Actuator}
\label{subsec:actuator}
The actuator component monitors the OL-CAIS AI model estimated probabilities and supports decision-making when needed.
It achieves these two tasks through the \textit{performance measurement} module, which reads the estimated probabilities of the classifier to trace the performance evolution through the autonomous classification ratio (ACR). 
By tracking the ACR curve, CAIS-DMA can automatically assess when the system requires assistance.
Assistance is triggered upon detecting a performance degradation state.
The performance measurement module invokes a \textit{pre-defined policy} to support the OL-CAIS decision-making.
Decision-makers establish pre-defined policies for production or testing purposes. If an internal policy is selected, the actuator models resilience without further assistance. 
For other policies, the actuator creates an instance of the specified decision-making policy, which then associates feasible \textit{actions} and selects the optimal action recommended by the invoked policy.

Each of the policies and action modules shown in Fig.~\ref{fig:caisdma} are invoked through a factory method that receives the name of the policy/action.
This design pattern supports a straightforward extension of the module, enabling the addition of new policies and actions as needed.
Thus, we equip the CAIS-DMA policies module with our GResilience Framework policies (i.e., one-agent, two-agent, and RL-agent).
Further, the actions module holds the system actions, such as the human and autonomous actions. Each action structure stores the action's attributes collected from the system environment.

The actuator operates in two modes: real-world and simulation.
In real-world settings, the actuator is the only CAIS-DMA component necessary to support decision-making.
The performance measurement module just needs to be tuned to the data format of the AI model.
This simply represents the APIs to ensure smooth communication between our framework and the bare-metal system.
In simulation mode, the action attributes are drawn from the simulator's dataset, with the data feeder module providing these attributes to ensure the actuator operates on realistic data collected from the live system, as discussed earlier.

\subsection{Monitoring}
\label{subsec:monitor}
The monitoring component is a web-based application that provides framework users with tools to visually analyze OL-CAIS performance in real-time and adjust framework variables for running experiments on the OL-CAIS under test.
The web application follows the Model, View, Controller (MVC) architecture.

The view provides an in-browser user interface to tune the running parameters, such as the number of iterations in three states, the GResilience Framework policy to use, the size of the sliding window to compute ACR, the confidence level of the classifier estimated probabilities, the disruptors and adapters to be used, and additional configurable parameters.
The selected parameters are passed through application routes to generate command-line instructions, which are executed via the controllers.

The experiment results, including the ACR curve annotated with the resilience model states, selected actions and their attributes at each iteration, and all logs, are dumped in a specific directory within the monitoring component.
These outputs are accessible through various views on the web interface.
No specific models have been created, as all data are stored in comma-separated values files to allow for further data analysis.

\section{Takeaways}

This chapter introduced the core decision-making components essential for achieving green resilience in OL-CAIS. We presented a comprehensive methodology that enables run-time monitoring, response to performance degradation, and recovery policy evaluation. The following summarize our contributions:

\begin{itemize}
    \item We introduced the \textbf{Resilience Model}, which continuously monitors OL-CAIS performance using the ACR. This model enables rapid detection of disruptive events and transitions across system states, enabling the system to trigger recovery actions when needed during runtime.
    
    \item We developed the \textbf{GResilience Framework} to support recovery decision-making by evaluating feasible actions using multiple agent-based policies: \textit{one-agent}, \textit{two-agent}, and \textit{RL-agent}. These policies support balancing greenness and resilience under different disruptions.
    
    \item We designed the \textbf{Measurements Framework} to provide actionable insights for evaluating policy effectiveness. By quantifying recovery speed, performance steadiness, CO$_2$ emission, and autonomy, this component empowers decision-makers to compare and select the most appropriate policies for their application needs.
    
    \item We implemented the \textbf{CAIS-DMA} as a modular toolkit that integrates simulation, runtime monitoring, and policy actuation. CAIS-DMA enables both experimentation in synthetic environments and deployment in real-world OL-CAIS setups, facilitating iterative development and safe testing of decision strategies.
\end{itemize}

\chapter{Building Green Resilience OL-CAIS}
\label{ch:greenresiliencecps}
This chapter introduces the methodology we applied to improve greenness and resilience in OL-CAIS by transitioning from a bare-metal, multi-machine setup to a containerized system.
Specifically, we containerized the system’s components, hosted on separate machines for machine learning, data instance detection, human demonstration tracking, and hardware control, into containers running on a single machine.
This approach facilitated resource efficiency and introduced a scalable framework to enhance both non-functional properties.

\begin{figure}[ht!]
    \centering
    \includegraphics[width=\textwidth]{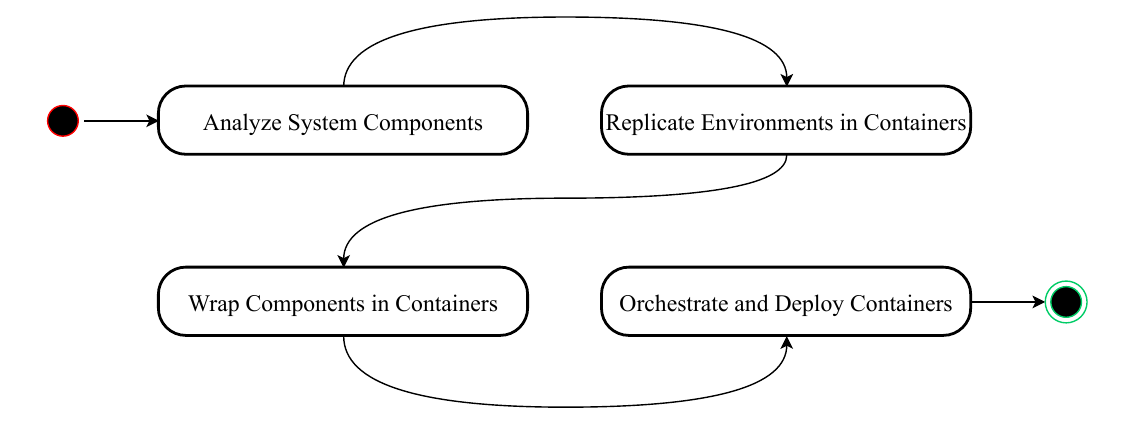}
    \caption{OL-CAIS containerization methodology.}
    \label{fig:gr:method}
\end{figure}

Fig.~\ref{fig:gr:method} illustrates our methodology in containerizing OL-CAIS, which begins by analyzing the bare-metal system to understand the various components, their operating environments, and inter-component communication methods. 
Next, we replicate each component’s environment within containers, including specifications like OS type, dependencies, port configurations, real-time requirements, etc. 
We then encapsulate each system component within its container environment. 
Finally, we utilize orchestration to run all containers concurrently, creating the necessary replicas and ensuring that communication across the network is redirected to localhost addresses for seamless integration.

\paragraph{Analyze System Components.} 
The first step in our methodology involves a detailed analysis of the existing bare-metal system, which may house separate machines for tasks such as machine learning, data detection, human tracking, and hardware control. 
In this analysis, we identify each component's dependencies and runtime environments to determine operating system compatibility, software requirements, and communication protocols. 
Additionally, we identify how the components interact, including the technology they use, like streaming data across the network, and the interaction structure, such as the data fields, their type, and content. 
This foundational step ensures an accurate reflection of the existing system's architecture, preserving functional interconnections.

\paragraph{Replicate Environments in Containers.} 
Following the system analysis, we replicate each component’s environment within containers. 
This involves implementing the containers, including specifying the appropriate operating system, required system and software dependencies, and specific port settings for each component.
We also identify any real-time constraints to determine whether specific optimizations are necessary.
By replicating these environments, we ensure that each component retains its original requirements within an isolated, portable container setting. This provides compatibility with the original system while preparing it for deployment within a containerized infrastructure.
We use Dockerfiles to define and automate the creation of each component's environment, ensuring consistency and repeatability during deployment.

\paragraph{Wrap Components in Containers.}
After replicating the environments, the next step involves encapsulating each system component within its container.
To ensure full automation of the booting process, we wrap each component's code within its designated container. This ensures that each component is self-contained and ready to boot as required.
At this stage, each container represents a standalone component that can replace the bare-metal component by setting its network configuration.
At the end of this step, it is important to test each containerized component with the bare-metal OL-CAIS to ensure that we can safely replace the original component.

\paragraph{Orchestrate and Deploy Containers.}
The final step in the methodology involves orchestrating and deploying all containers to function as a cohesive system. 
This step uses tools such as Docker Compose to manage the containers, ensuring that they start, stop, and interact in an organized manner.
Through orchestration, we can run all containers in the required order, create necessary replicas, and allocate resources to meet the system’s resilience requirements.
In this step, we update the network configurations, replacing external IP-based communication with localhost-based interactions to minimize latency and simplify the network structure. 
This orchestration guarantees the seamless integration of all system components and supports improving resilience.

On the other hand, to empirically evaluate the containerized system's greenness improvement, we perform a comparative study by running the same tasks within the same time frame, once with the bare-metal system and the other with the containerized system.
During this period, we monitor the system's overall energy consumption and calculate the carbon emissions for both versions.
We measure energy consumption using a Wattmeter connected to the power source, and then we plug all the system components directly through the Wattmeter.

Finally, for a Green Resilient OL-CAIS that addresses both properties in design and at runtime, we propose integrating our decision-making toolbox as an intermediate container.
This container would monitor interactions between system components and dynamically redirect decisions to balance greenness and resilience. This will be further discussed in future work.

\section{Takeaways}

This chapter has highlighted the following key insights:

\begin{itemize}
    \item We designed and implemented a containerization methodology that transitioned OL-CAIS from a distributed bare-metal architecture to a unified container-based setup, enabling a more modular, energy-efficient, and resilient system foundation. The methodology consists of four main activities:
    \begin{itemize}
        \item We analyzed the system components, such as the machine learning engine, data instance detector, human demonstration tracker, and hardware controller. This analysis results in a better understanding of the runtime environments and communication protocols, thereby ensuring accurate replication in containers.
        
        \item We used Dockers to replicate the execution environment of each component, including system dependencies, operating system constraints, port mappings, and real-time execution requirements, which allowed us to preserve behavior and timing alignment.

        \item We encapsulated each replicated environment into a self-contained container, enabling modularity, portability, and automated boot-up, while preserving compatibility with the original bare-metal deployment.

        \item We orchestrated the containerized components using Docker Compose to run them concurrently, allocate resources efficiently, and manage replicas, thereby enabling fault recovery and runtime resilience through controlled replica activation.
    \end{itemize}

    \item We developed a comparative experimental setup to monitor and quantify energy consumption and estimate carbon emissions in both the bare-metal and containerized configurations of OL-CAIS, establishing the foundation for empirical greenness evaluation.
\end{itemize}



\myPart{Experimental Evaluation}
\huge{T}\normalsize{his part focuses on validating the proposed methodologies through experimental investigations. 
First, we introduce the CORAL case study, highlighting its relevance and significance as a testbed. It also discusses the dynamics of online learning and disruptive events in CORAL. Then we outline the experiment protocol and design, including the setup, execution, and focus on evaluating trade-offs between greenness and resilience and the role of containerization. 
Finally, we present the results of these experiments, drawing comparisons between real-world and simulated scenarios and examining the impact of containerization.}

\chapter{Case study: CORAL}
\label{ch:coral}
A case study is defined as an empirical investigation of a case, with the investigator or investigators not actively participating in the case under investigation, employing a variety of data collection techniques to examine a current phenomenon in the real-world setting,~\cite{rainer2023case}. 
In alignment with this definition, we present the case study of ``CORAL'' a collaborative robot designed by \textit{Fraunhofer Italia Research} as part of the ARENA Lab initiative\footnote{CORAL homepage: \url{https://www.fraunhofer.it/en/Application-Centre-ARENA/Preliminary-research-and-applications/coral.html}}. 
CORAL serves as an industrial robotic demonstrator, showcasing the automation of real-world tasks such as sorting and assembling components based on their characteristics. 
By integrating online learning, CORAL exemplifies the current phenomenon of OL-CAIS operating in real-world industrial settings. This chapter provides an in-depth exploration of CORAL, detailing its components and functionalities.
This chapter explores CORAL's architecture, online learning mechanisms, and risk assessment, setting the stage for evaluating its greenness and resilience under disruptive events.

\section{CORAL Overview}

\begin{figure}[htbp]
\centering
\centerline{\includegraphics[width = \textwidth]{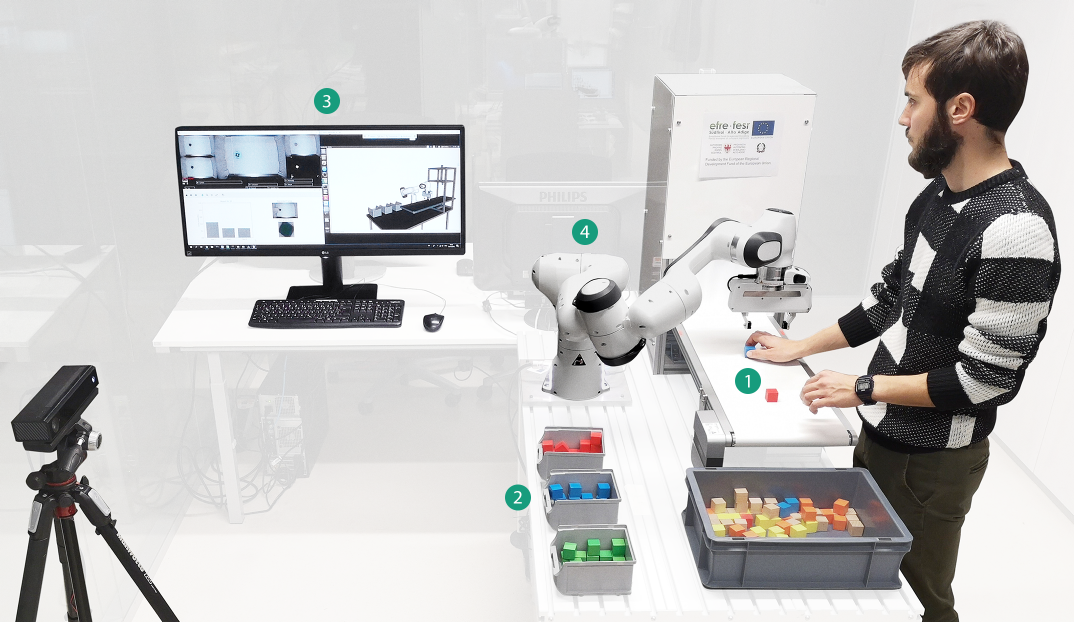}}
\caption{Collaborative robot learning by demonstration.}
\label{fig:coral}
\end{figure}

CORAL, shown in Fig.~\ref{fig:coral}, is a collaborative robot system equipped with online learning capabilities for sorting objects arriving on a conveyor belt (1). Each object is classified by its color and placed into corresponding boxes labeled red, green, or blue (2). 
The system uses an RGB sensor installed above the conveyor belt to detect objects. This sensor is connected to object detection software that isolates the objects from their background (3). 
Extracted objects are then streamed to a machine-learning classifier, which processes their color histograms as feature vectors. These features are mapped to the corresponding box label provided by the human. Both the object detection and classification components are operating on the same machine (3).
Additionally, the system comprises a robotic arm (4) that performs sorting tasks with human collaboration. Two types of actions are employed to complete the task: either the human places the object in the corresponding box, marking the class for learning purposes, or the robotic arm performs the action autonomously.
The robotic arm is controlled in real-time through software running on a second machine equipped with a robot operating system.
To enable human-robot interaction, a depth sensor captures human skeletal movements and estimates arm dimensions. This sensor operates on a third machine, enabling the system to track human actions effectively.

According to the six-level taxonomy of human-robot collaboration~\cite{mukherjee_survey_2022}, which ranges from fully programmed (level 0) to fully autonomous (level 5), CORAL is classified at level 4.  This level represents an intersection between human actions $A_h$ and system/robot actions $A_s$. 
The intersection between the two sets of actions is represented by placing the object in the corresponding box, as either of the two operators can perform this action.

In summary, CORAL is a heterogeneous OL-CAIS comprising four functional components distributed across three separate machines: i) the object detector and machine learning classifier, ii) the robotic arm controller, and iii) the human skeleton tracker. Each component communicates via specific service IPs and ports, enabling seamless integration and data sharing.
This architecture enables CORAL to perform adaptive decision-making through online learning. 
The next section explores the online learning process implemented in CORAL and how it is employed in its internal policies.

\section{Online Learning in CORAL}
\label{sec:coral:onlinelearning}

By leveraging online learning, CORAL adapts during runtime, continuously improving its object classification accuracy through effective human-robot collaboration. This adaptability is achieved using a classifier that implements regularized linear models with stochastic gradient descent (SGD) learning. Specifically, CORAL employs the \textit{SGDClassifier} developed by~\textcite{scikitlearn2011}. 
The classifier iteratively updates its model parameters by estimating the loss gradient for each batch and applying a decreasing strength schedule (i.e., learning rate). 
The SGDClassifier's mini-batch feature enables the system to perform online learning by processing one data point per iteration.
CORAL operates in a collaborative cycle between human and autonomous components. The robot learns from human demonstrations~\cite{rimawi_modeling_2024, lesort2020continual}, and based on this learning, the classifier provides $n$-estimated-probabilities, where $n$ is the number of classes (i.e., number of classes in CORAL --red, green, and blue-- $n=3$).

\begin{figure}[th]
\centering
\centerline{\includegraphics[width = \textwidth]{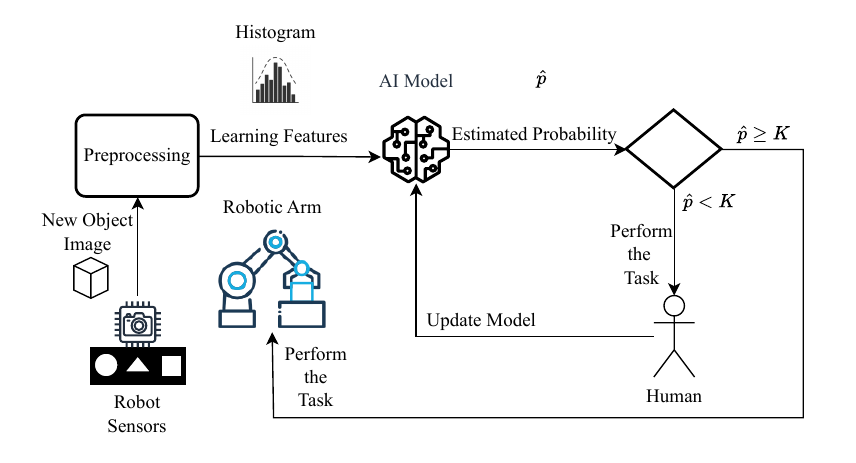}}
\caption{Schematic representation of the online learning process.}
\label{fig:onlinelearningprocess}
\end{figure}

Similar to the decision-making flow to learning new arriving data instances in an online learning fashion detailed in Sec.~\ref{sec:bg:cais}, CORAL follows the same schematic representation flow as Fig.~\ref{fig:onlinelearningprocess} shows.
The online learning process in CORAL starts when the robot’s sensors collect data instances and preprocess them to extract the relevant learning features (e.g., the object's color histogram). 
The AI model (i.e., the classifier) estimates the probability for each classification class (corresponding to the labeled boxes: red, green, or blue), ensuring that the sum of probabilities equals one.
The classifier then calculates the confidence level of prediction, which is the \textit{maximum estimated probability}, denoted as $\hat{p} \in [0, 1]$,~\cite{zadrozny_transforming_2002}. 

The maximum estimated probability, $\hat{p}$, is compared to a predefined confidence threshold, $K$, which serves as the foundation of CORAL's internal decision-making policies.
The confidence threshold $K$ establishes the minimum level of certainty required for CORAL to classify and sort an object autonomously. 
If the confidence level $\hat{p}$ surpasses $K$, the robotic arm autonomously classifies and places the object. Conversely, if $\hat{p}$ is below $K$, the system requests human intervention. The human performs the classification task, and the system updates its model by incorporating the new labeled data into its learning process.
Based on this conditional flow we can categorize CORAL's actions as follows:

\begin{enumerate}
    \item \textit{Autonomous action}: The robotic arm autonomously classifies and places an object when $\hat{p} \geq K$.
    \item \textit{Human action}: The system prompts human intervention to ensure accurate classification when $\hat{p} < K$.
\end{enumerate}

\begin{figure}[th]
\centering
\centerline{\includegraphics[width = \textwidth]{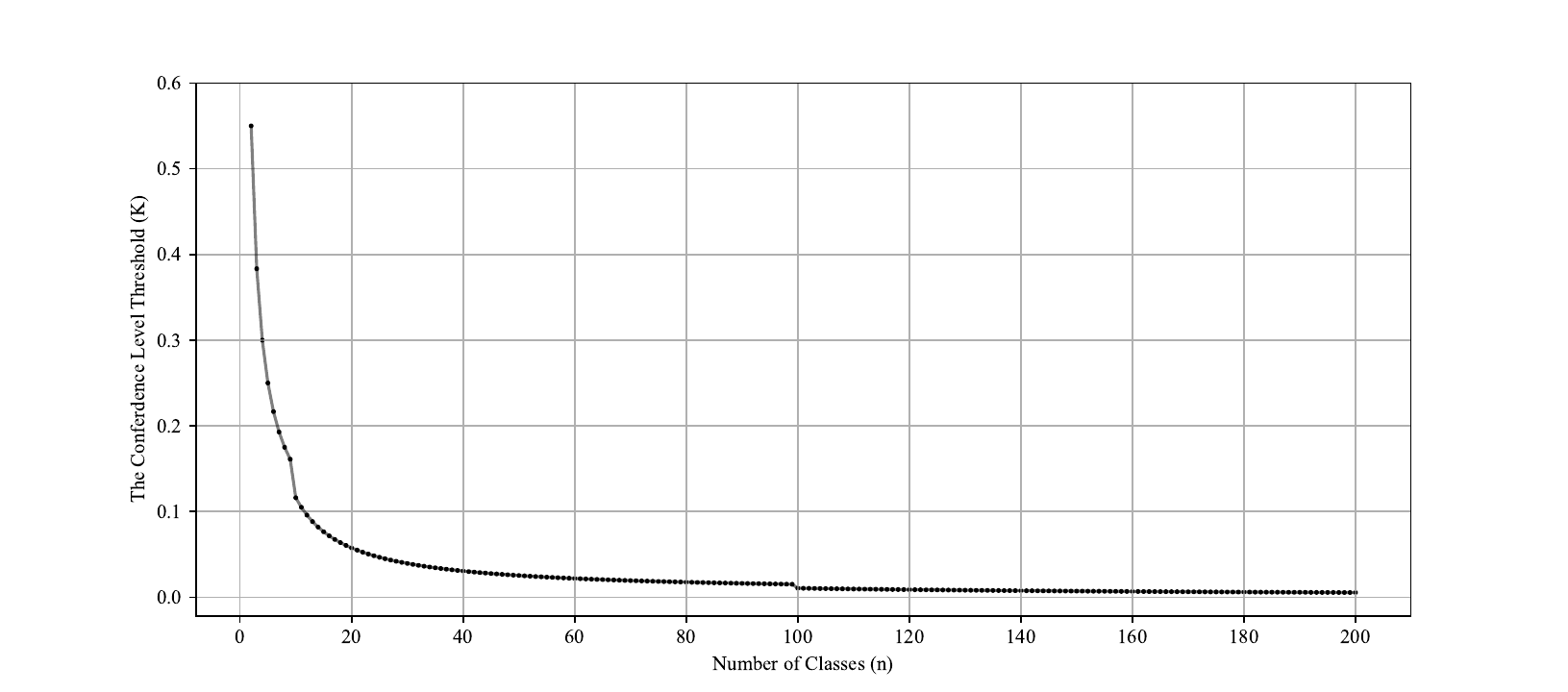}}
\caption{Variation of confidence level threshold $K$ with the number of classes ($n$), for $n > 1$.}
\label{fig:kvalues}
\end{figure}

The threshold $K$ is carefully designed to avoid scenarios where the system stalls due to equal or closely aligned probabilities for multiple classes. 
The minimum value of $K$ is derived as shown in Eq.~\eqref{eq:kvalue}, which represents the reciprocal of the number of classes, $n$, augmented by a discrimination factor to ensure adequate separation between probabilities. 
This ensures that $K$ scales appropriately with the classification task's complexity.
For example in CORAL, with $n = 3$, $K$ is calculated as approximately $0.38$. When $\hat{p} \geq 0.38$, the robotic arm proceeds autonomously.
Otherwise, the system queries the human operator, ensuring accurate classification and updating the AI model with the new learning data.
Fig.~\ref{fig:kvalues} illustrates how $K$ varies with the number of classification classes, demonstrating its adaptability across different scenarios.

\begin{equation}
\label{eq:kvalue}
   K = (1/n) + 0.5 \times 10^{-(\lfloor \log_{10}(n) \rfloor + 1)}, \text{ for } n > 1.
\end{equation}

The confidence threshold $K$ acts as the core of CORAL's internal decision-making policies to guide the human-robot collaboration.
It ensures that the system can operate independently when confident while leveraging human expertise to improve its learning when uncertain. 
This approach enhances classification accuracy by enabling CORAL to adapt dynamically to changes in task complexity.
By employing this adaptive decision-making policy, CORAL exemplifies how OL-CAIS can effectively integrate human and system collaboration to achieve steady performance.
However, it is important to understand the possible disruption events that may lead to CORAL's performance degradation and whether its internal policies are sufficient to ensure a green recovery during these events or require further support.

\section{Disruptive Events in CORAL: A Risk Assessment}

As we discussed in Sec.~\ref{sec:bg:qualities}, a disruptive event is defined as one that transforms system performance from a steady state to a disruptive state, resulting in performance degradation. 
Understanding the potential threats to CORAL is critical for ensuring its steady performance and resilience. 
To this end, we conducted a risk assessment to identify possible disruptive events that could impact CORAL's operation and evaluated their risks based on CORAL's components and dependencies.

The risk assessment follows the Plan-Do-Check-Act (PDCA) principle and adopts a comprehensive framework for security risk management tailored for Cyber-Physical Systems (CPS),~\cite{mokalled2019comprehensive}. 
To extract the potential disruptive events we followed the risk assessment focuses on the plan phase of the PDCA framework and consists of three key steps.
\begin{enumerate}
    \item \textit{System Functional Modeling}: Identify CORAL's critical components and their dependencies.
    \item \textit{Threat Selection and Modeling}: Identify relevant threats based on potential risks to CORAL's operations.
    \item \textit{Risk Analysis and Treatment Plan}: Evaluate the likelihood and impact of identified threats and establish mitigation strategies.
\end{enumerate}

Table~\ref{tab:components-threats} 
lists CORAL's components, whether hardware or software, dependencies, and the associated criticality levels based on the Confidentiality, Integrity, and Availability (CIA) model. This analysis highlights the areas of highest risk, which inform the selection of disruptive events for further experimentation. Each threat is assigned a criticality level: High (H), Medium (M), or Low (L). 
The assigned levels were taken and verified by CORAL's manager.
For example, the \textit{classifier} (D) software requires the extracted object image coming from the \textit{object detector} (C), and in case it inquiries the human to classify the object, it requires the chosen box coming from the \textit{skeleton tracker} (I).
The classifier confidentiality has a medium criticality in terms of confidentiality and integrity and high criticality in terms of availability.

\begin{table}[!ht]
    \centering
    \caption{Threats to CORAL components based on the CIA model.}
    \label{tab:components-threats}
    \begin{tabular}{lllllll}
    \hline
        \textbf{ID} & \textbf{Component} & \textbf{Hardware/Software} & \textbf{Dependencies} & \textbf{C} & \textbf{I} & \textbf{A} \\ \hline
        A & Lights & Hardware & ~ & M & M & H \\ 
        B & RGB Sensor & Hardware & A & M & M & H \\ 
        C & Object Detector & Software & B, H & L & M & M \\ 
        D & Classifier & Software & C, I & M & M & H \\ 
        E & Depth Sensor & Hardware & J & L & M & H \\ 
        F & Robotic Controller & Software & D, E, G, H, I & M & H & H \\
        G & Robot & Hardware & F & L & H & H \\ 
        H & Conveyor Belt & Hardware & F & L & M & H \\ 
        I & Skeleton Tracker & Software &  E, J & L & M & M \\
        J & Human & Human &  D, I & L & M & M \\ \hline
    \end{tabular}
\end{table}

By identifying the components, their dependencies, and the criticality value for the CIA model, we need to select the threats that affect our collaborative robot performance.
The European Union Agency for Cybersecurity, ENISA, publishes an annual assessment on the condition of the cybersecurity threat landscape called the ENISA Threat Landscape (ETL) report,~\cite{ENISA_2022}.
The PDCA refines the ETL report to classify threats related to cyber-physical systems, including:
\begin{enumerate}
    \item \textit{Human errors}: From configuration errors, operator/user errors, and loss of hardware to non-compliance with policies or procedures.
    \item \textit{Third party failures}: Including internet service provider, cloud service provider, utilities (power/gas/water), remote maintenance provider, and security testing companies.
    \item \textit{Malicious actions}: Such as denial of service attack, exploitation of (known or unknown) software vulnerabilities, misuse of authority/authorization, network interception attacks, and social attacks.
    \item \textit{System failures}: Including failures of devices or systems, failures or disruptions of communication links (communication networks), failure of parts of devices, failures or disruptions of main supply, and failures or disruptions of the power supply.
    \item \textit{Natural phenomena}: Such as earthquakes, floods, solar flares, volcano explosions, and nuclear incidents.
\end{enumerate}

By associating the CPS threats, the criticality value of the CIA model, and our goals in this PhD research.
We based our selection on the disruptive events on the threats associated with the decision-making of CORAL.
In particular, we select the classifier component as it estimates the maximum classification probability (i.e., $\hat{p}$), which is the key value to make an autonomous or human action.
The objects to classify by the classifier component come from the object detector component, which uses the RGB Sensor supported by the Lights components to detect this object.
Thus, what can disrupt the classifier and lead to uncertainty in decision-making is manipulating the objects arriving.
We selected two threats as disruptive events for further experimentation: a hardware failure in the lights supporting the RGB sensor darkness of the detected objects and a malicious action that manipulates the histogram of the detected object. We employ these threats in our experiments to evaluate their impact on CORAL's performance and use our theories to treat these disruptive events. The following details the selected disruptive events.

\begin{enumerate}
    \item \textit{Hardware Failure (Lights Failure)}: A malfunction in the lighting system supporting the RGB sensor results in dark images of objects. 
    This disruption affects the classifier's accuracy in identifying objects, increasing reliance on human intervention for reclassification and lowering the performance measures (i.e., ACR).
    \item \textit{Adversarial Image Attack (Histogram Equalization)}: A simulation of a malicious attack spoofs the image data streamed from the object detector to the classifier components. This attack manipulates the object's histogram features, affecting the classifier's decision-making accuracy. 
\end{enumerate}

These events represent real-world challenges that CORAL may face and may degrade its performance, providing a basis for testing our theories in achieving green recovery.

\section{Takeaways}

This chapter has highlighted the following key insights from our case study on CORAL:

\begin{itemize}
    \item We presented CORAL as a real-world OL-CAIS case study demonstrating how collaborative robotics can integrate human interaction and online learning in an industrial setting. CORAL is composed of heterogeneous components including an RGB sensor, object detector, classifier, and robotic controller, all distributed across three machines. It uses the \textit{SGDClassifier} as its learning model, which is continuously updated through a human-in-the-loop online learning process, enabling runtime classification based on color histograms derived from the RGB sensor input.

    \item We designed and deployed an internal policy based on a confidence threshold $K$, which allowed the classifier to autonomously decide whether to act or request human intervention depending on prediction certainty. This threshold formed the core of CORAL’s internal decision-making policy, showing how OL-CAIS can embed minimal yet effective intelligence to balance autonomy and accuracy in uncertain environments.

    \item We conducted a risk assessment following the PDCA framework to identify threats that could affect CORAL’s performance. Each component was evaluated using the Confidentiality, Integrity, and Availability (CIA) model, while threat types were derived from the ENISA Threat Landscape. From this analysis, we selected two representative disruptive events for evaluation: a hardware failure in the lighting system and an adversarial attack using histogram equalization.

    \item We connected these real-world disruptions with our theoretical models to examine whether CORAL’s internal policies are sufficient to maintain resilience and greenness or require additional decision-making support. This process established CORAL and its simulated environment as testbeds that bridge theory, simulation, and real-world experimentation in OL-CAIS, thereby laying the foundation for the evaluations that follow.
\end{itemize}

\chapter{Experiments Design and Execution}
\label{ch:expdesign}
In this chapter, we provide a comprehensive overview of our experiment design. Throughout this PhD research, we conducted experiments to evaluate our decision-making toolbox, designed to balance the greenness and resilience of OL-CAIS under disruptive events. These experiments assess the behavior of OL-CAIS with and without the decision-making support provided by our framework.
We also analyze how transitioning system components from bare-metal, multi-machinery setups to containerized environments affects resilience and greenness.
The first section of this chapter details the experimental protocol and design for evaluating the trade-off between greenness and resilience. The second section explores the experiment conducted to analyze the effects of containerization on OL-CAIS performance.

\section{Evaluating the Trade-Off Between Greenness and Resilience}

We designed and executed a series of experiments following a structured protocol to systematically evaluate the balance between greenness and resilience in OL-CAIS under disruptive events.

The protocol includes defining the experimental environment, executing the experiments iteratively, and analyzing the resulting data to assess system performance and decision-making effectiveness.
The protocol is divided into three stages, each with a specific role in ensuring the reliability and validity of the experiments: setup, iterative execution and data collection, and data analysis. Below, we detail each stage of the experimental protocol.

\begin{enumerate}
    \item \textit{Setup}: We setup the experimental environment by defining the following variables:
    \begin{enumerate}
        \item Disruptive event: Select the disruptive event to enforce during the experiment and its enforcement procedure (for example, but not limited to intentional hardware shutdown or running an automated script in case of real-world experiments and applying filters to the collected data in a simulated environment).
        \item Recovery actions: We identify feasible recovery actions, which include both autonomous and human interventions.
        \item Sliding window size: Define the $m$-iteration sliding window size to compute ACR values continuously.
        \item Steady State length: Define a number of iterations to keep the OL-CAIS in a steady state to ensure a similar training level across all experiments.
        Steady State length is essential to detect the Recovered State during the Disruptive State and the Final State.
        \item Decision-making policies: Each experiment is intended to evaluate specific decision-making policies (i.e., internal, one-agent, two-agent, RL-agent policies). Thus, it is important to set the decision-making policies in the setup stage.        
    \end{enumerate}
    \item \textit{Iterative Execution and Data Collection}: The experiments are executed over iterations, with our framework selecting one action for each iteration.
    For each selected action, we collect i) the action's running time, ii) the number of human interactions utilized, iii) the CO$_2$ emission emitted by the action, and iv) the maximum estimated probability by the classifier (i.e., $\hat{p}$).
    
    We track the number of consecutive iterations with standalone decisions, which describes the number of decisions made by the classifier with $\hat{p}$ above the confidence threshold, $\hat{p} \geq K$.
    
    We stop our framework support after we reach a number of consecutive standalone decisions equal to the sliding window size, \textit{m}.
    We monitored the system for at least the Steady State length to ensure it maintained a Recovered State after support stopped.
    \item \textit{Data Analysis}: Following the iterative execution stage, we analyze the collected data. We assess the ability of the selected decision-making model to recover the OL-CAIS under test performance by analyzing the resilience model. We employ our measurements framework to measure greenness and resilience and compare decision-making policies. 
    This analysis allows us to identify patterns or trends from the experimental results.
\end{enumerate}

To empirically evaluate our theories, we conducted two types of controlled experiments. The following details the design of each of these experiments:

\begin{enumerate}
    \item \textit{Real-world experiments}  were conducted using our collaborative robot (i.e., CORAL).
    All experiments utilize the same environmental setup but employ different decision-making policies, as follows:
    \begin{enumerate}
        \item Experiment with internal policies: This experiment serves as a baseline for further comparison.
        \item Experiment with one-agent policies: In this experiment, we support OL-CAIS's decision-making by enforcing new policies built on one-agent solution.
        \item Experiment with two-agent policies: We support OL-CAIS's decision-making by enforcing new policies built on two-agent solutions.
        \item Experiment with RL-agent policies: We deploy an RL agent that observes the OL-CAIS environment and explore-exploit recovery actions upon disruption.
    \end{enumerate}
    \item \textit{Simulated experiments} simulate CORAL's environment for further generalization of the results.
    By simulating CORAL's environment, we reproduce the results of real-world experiments. Additionally, we can repeat each experiment one hundred times.
    The simulated results reduce random results and outliers, providing a comprehensive overview of the four experiments.
\end{enumerate}

In our OL-CAIS, CORAL is equipped with a vision component that captures objects on a conveyor belt, classifies them based on their color histograms, and determines their placement in designated boxes.
Each experiment follows the same experimental protocol, with variations introduced through the decision-making policies and disruptive events applied. These experiments aim to evaluate the system's ability to balance greenness and resilience under disruption.

\subsection{Experimental Setup}
In each experiment, a disruptive event is introduced to degrade CORAL’s performance and test the effectiveness of the intelligent agent policies:
\begin{itemize}
    \item Real-world experiment: The disruptive event involves physically turning off the supporting light above the conveyor belt, fading the objects' colors, and reducing the learning model's ability to classify them autonomously.
    \item Simulated experiment: Two types of simulated disruptions are applied:
    \begin{enumerate}
        \item Darkness filter: A filter is applied to the images to simulate the effects of turning off the light.
        \item Adversarial attack: A histogram equalization transformation is applied to simulate an adversarial attack.
    \end{enumerate}
\end{itemize}

In both scenarios, feasible \textit{recovery actions} involve object classification, performed either autonomously by the robotic arm or manually by a human, to enhance the classifier's learning.
The ACR value is computed using a \textit{sliding window} of five consecutive iterations ($m=5$). This size ensures that the system's performance is continuously evaluated without overly stressing the framework. 
The experiments maintain the same \textit{steady state length} (i.e., 30 iterations) across all scenarios to ensure consistent learning levels.
Finally, the \textit{decision-making policies} under test in each experiment include: i) Internal policies, where there is no external support; ii) one-agent policies, support is provided by solving an optimization problem; iii) two-agent policies, support is extended using game theory, and iv) RL-agent policies, a reinforcement learning agent dynamically optimizes decisions to balance resilience and greenness.

The experiment progresses iteratively, with each iteration starting with the arrival of a new object on the conveyor belt. 
The object is classified and placed in its designated box, either autonomously or with human support. The system monitors its performance, models the performance behavior, and applies the decision-making policies for the experiment in case of performance degradation.

\subsection{Experiment Execution}
\label{subsec:expsetup}

The experiments focus on supporting our collaborative robot online learning component by re-establishing the relation between its autonomous components and the human operator after encountering a disruptive event. The goal is to select actions that maximize greenness and resilience during transitions between states. This subsection provides an integrated view of the experimental process, participant preparation, and data collection, building on the experimental protocol described earlier.

Initially, the system starts with an untrained classifier, allowing CORAL to learn tasks autonomously in a controlled environment during the \textit{Steady State}. 
After sufficient learning, we induce a disruptive event according to the experiment type which transitions the system into the \textit{Disruptive State}.
Our resilience model monitors CORAL's performance behavior through the ACR value, detecting the transferring of our collaborative robot through a \textit{Performance Degradation State}.
To mitigate the impact of the disruption, the GResilience framework is activated, applying decision-making policies, such as internal, one-agent, two-agent, or RL-agent, to aid CORAL in recovering its performance.
Feasible actions are recommended and actuated based on the selected policies, indicating the start of the \textit{Recovering State}.
The system transitions to the \textit{Recovered State} when the ACR surpasses a predefined threshold for 30 consecutive iterations (i.e., Steady State length). 
If recovery is not achieved after double the Steady State length, we fix the disruptive event by turning the lights on in real-world experiments or stopping streaming dark or equalized histogram images in simulated experiments.
In the \textit{Final State}, we allow the system to transition autonomously to a \textit{Second Steady State} to understand if it forgets what it learned in the initial Steady State or not.
We stop supporting the descision-making of CORAL through our framework after five consecutive iterations (i.e., $m = 5$) of autonomous decision-making during the final steady state. In other words, when the classifier estimates a maximum probability more than the confidence threshold for one sliding window. 

\begin{figure}[ht!]
\centering
\centerline{\includegraphics[width = \textwidth]{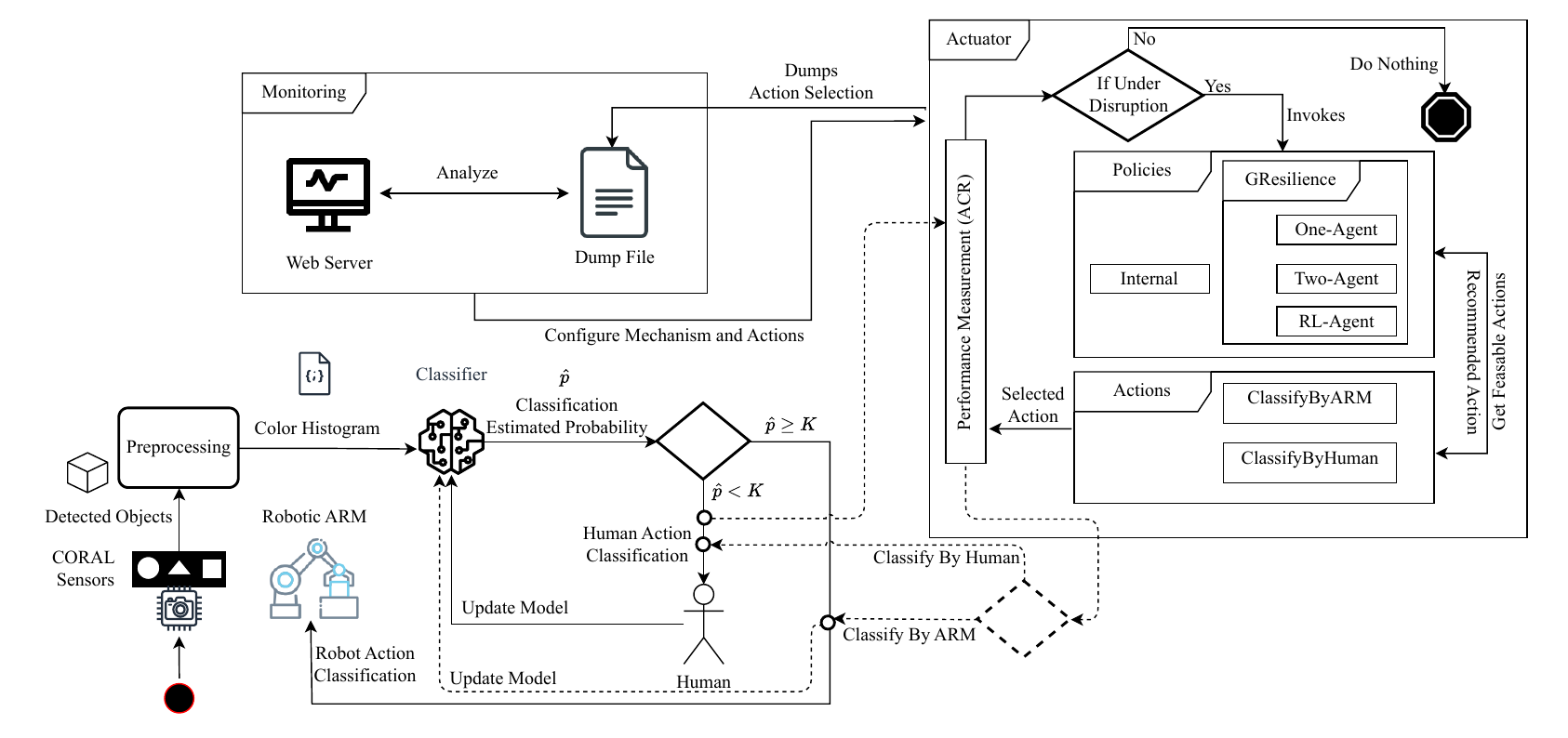}}
\caption{Integration of CORAL with CAIS-DMA in real-world experiments.}
\label{fig:coralwithcaisdma}
\end{figure}

The two main participants in the experiments are the human operator and the robotic arm. The human operator collaborates with CORAL during object classification tasks, adhering to ISO/TS 15066 safety guidelines,~\cite{noauthor_isots_2016}. 
CORAL is equipped with CAIS-DMA~\cite{kadgien_cais-dma_2024}, customized for real-time decision-making and performance monitoring.
The CAIS-DMA components used in this setup include a monitoring component that tracks the $\hat{p}$ value for performance evaluation and resilience modeling and an actuating component that executes the selected actions and updates CORAL's classifier. 
The integration of CORAL with CAIS-DMA in real-world experiments is illustrated in Fig.~\ref{fig:coralwithcaisdma}, showing the key components and their interactions.
The monitoring component configures the policies to use and the recovery actions and plots a live ACR curve of CORAL from the dump file.
The dump file contains the actions selected with their attributes (i.e., run time, CO$_2$ emission, and human interactions).
The performance measurement module in the actuator traces the ACR values to detect CORAL's current operational state. If a performance degradation is detected, it flags the start of a disruptive state and invokes the selected policies.
Then, through its APIs connected to CORAL, it actuates the recommended action.
The small circles with dashed arrows in Fig.~\ref{fig:coralwithcaisdma} show the APIs of CAIS-DMA connected to the online learning process of CORAL.

\begin{figure}[ht!]
\centering
\centerline{\includegraphics[width = \textwidth]{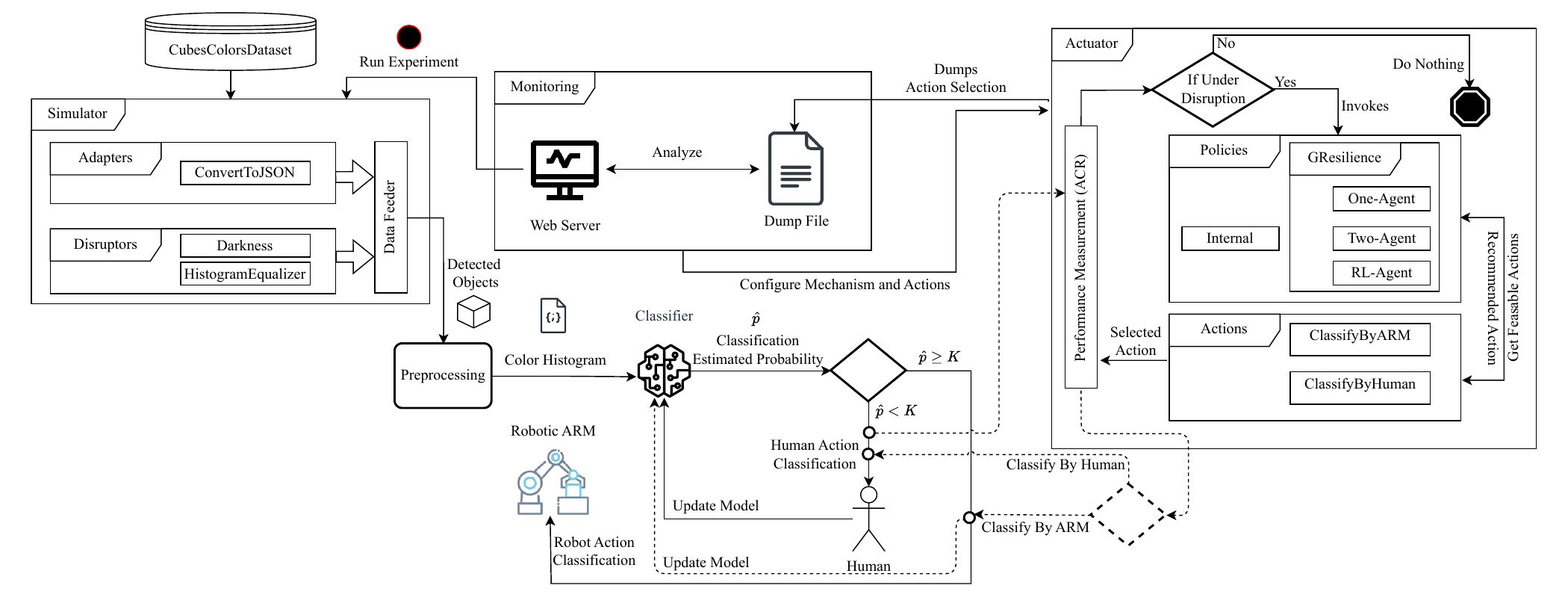}}
\caption{Integration of CORAL with CAIS-DMA in simulated experiments.}
\label{fig:coralwithcaisdmasimu}
\end{figure}

In the simulated experiments, the CAIS-DMA simulator component replaced CORAL's sensors. Where the simulator streams the objects we have collected from the real-world setup.
Through the simulator, we can control when to stream normal objects and when to stream disrupted ones.
Further, we can stress running the same experiment as many as needed  (e.g., one hundred).
Fig.~\ref{fig:coralwithcaisdmasimu} shows the simulator component streams the objects through the data feeder module, keeping the rest of the components the same as in the real-world experiments.
Additionally, the monitoring web server configures the experiment details, such as the number of iterations, when to stream disrupted objects, and what type of disruptors to use, then runs the experiment itself.

\section{Evaluating the Impact of Containerization on Greenness and Resilience}

To assess the impact of containerization on the greenness and resilience of OL-CAIS, we designed an experiment comparing the system's performance in its original multi-machinery setup (i.e., bare-metal) and its containerized deployment. 
This experiment evaluates whether transitioning to containerized components enhances resilience through improved replication mechanisms and reduces energy consumption, contributing to greener operations.

\subsection{Experimental Setup}
The experiment consists of running the system in two configurations under identical conditions:
\begin{itemize}
    \item \textit{Bare-Metal Setup:} The system operates in its original form, with separate physical machines hosting its components. The system continuously classifies objects on the conveyor belt for a fixed duration of two hours.
    \item \textit{Containerized Setup:} The system is deployed using containers, where components run as lightweight, portable instances. This setup supports dynamic resource allocation and component replication. Similar to the bare-metal configuration, the system continuously classifies objects for two hours.
\end{itemize}

In both setups, the system classifies objects as they move along the conveyor belt. Energy consumption is measured using a wattmeter throughout the operation, and the recorded data is used to evaluate differences in energy efficiency and resilience between the two configurations.

\subsection{Experimental Execution}

\begin{figure}[ht!]
\centering
\centerline{\includegraphics[width = \textwidth]{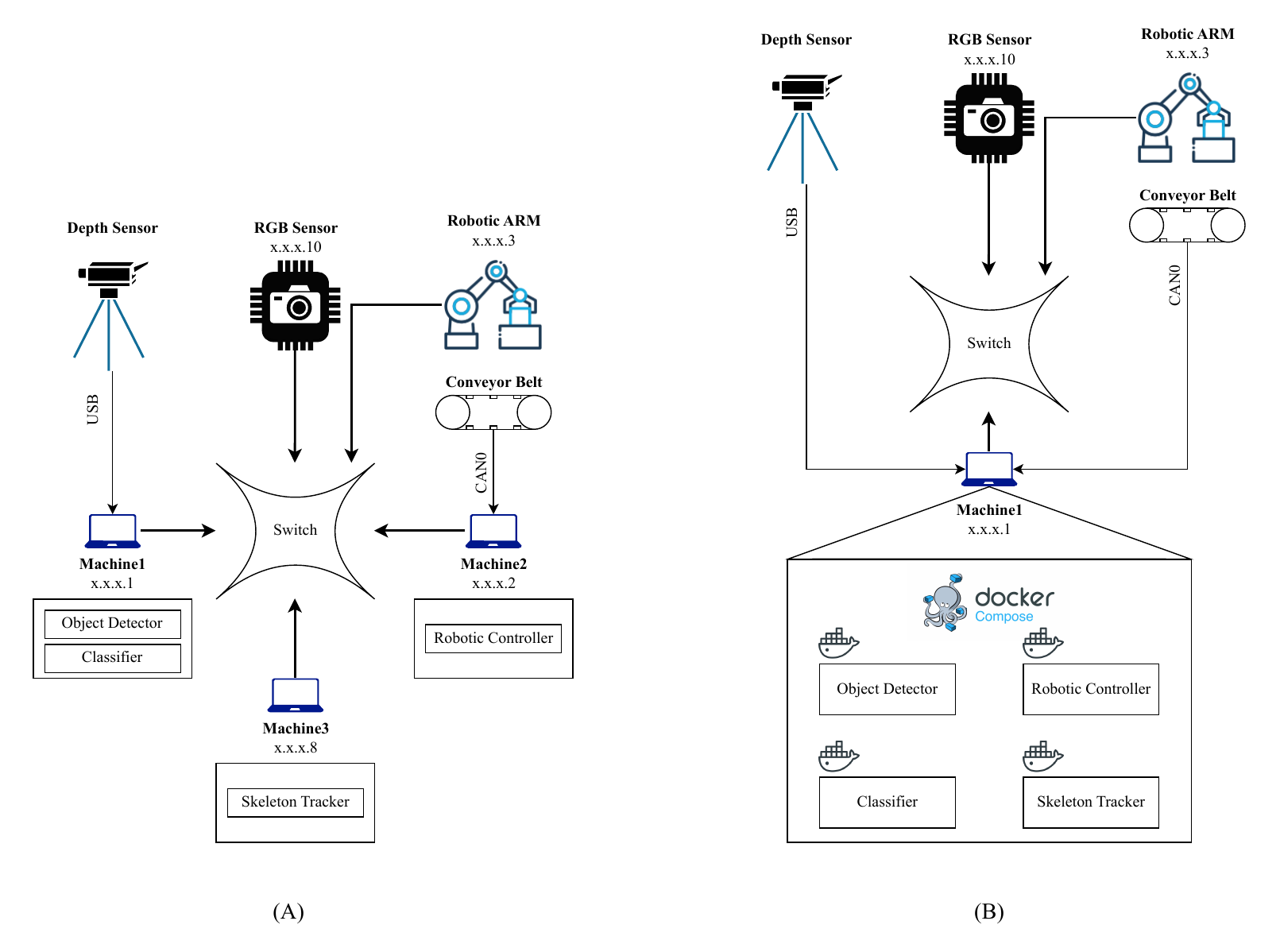}}
\caption{CORAL experimental setups for containerization evaluation: (A) Bare-metal multi-machinery configuration; (B) Containerized setup using Docker.}
\label{fig:coralbarmetal}
\end{figure}

To run this experiment, we ported all components of CORAL into docker containers and launched them using docker-compose.
Fig.~\ref{fig:coralbarmetal}~(A) shows the bare-metal multi-machinery setup, in which all the components are connected to a local network through a switch. The components are distributed across three machines: \textit{machine1} contains both the object detector and the online classifier, \textit{machine2} has the robotic controller, and \textit{machine3}  has the human skeleton tracker.
The robotic arm and RGB sensor connect to the network, while the conveyor belt and depth sensor use CAN0 and USB connections, respectively.
On the other hand, Fig.~\ref{fig:coralbarmetal}~(B) shows the containerized setup, showing all the components running from a single machine that runs and orchestrates these components using docker-compose and docker containers.

\section{Takeaways}

This chapter has highlighted the following key insights from the design and execution of our experimental study:

\begin{itemize}
    \item We introduced a structured experimental protocol that guided the setup, execution, and analysis phases, ensuring consistency across both real-world and simulated experiments.

    \item We defined a clear methodology for configuring the experimental environment, including the definition of disruptive events, recovery actions, and evaluation parameters.

    \item We implemented a repeatable process for executing decision-making experiments in OL-CAIS, which involved monitoring performance degradation, applying recovery strategies, and evaluating system behavior over time.

    \item We integrated CAIS-DMA with CORAL to support both monitoring and actuation during real-world experimentation, enabling controlled execution and data collection aligned with our framework. We extended the real-world experimental setup into a simulated environment by incorporating a data feeder and streaming mechanism, allowing us to reproduce scenarios at scale and perform controlled repetitions.

    \item We introduced an additional experiment to evaluate the impact of containerization by comparing system behavior across bare-metal and containerized deployments under identical operational conditions.

    \item We ensured that all experiments were configured to evaluate OL-CAIS performance using unified metrics for resilience and greenness, supporting the comparative analysis to be presented in the following chapter.
\end{itemize}


\chapter{Results}
\label{ch:results}
In this chapter, we present the experimental results of our empirical evaluation with respect to the research questions stated in Sec.~\ref{sec:intro:researchgoals}.
The chapter is divided into three sections according to the research goals: \textbf{RG1}: Model the resilience of OL-CAIS behavior to support the decision-makers, \textbf{RG2}: Develop agent-based policies to balance resilience and greenness in OL-CAIS upon unforeseen disruptive events, and \textbf{RG3}: Understand catastrophic forgetting in OL-CAIS in the aftermath of a disruptive event and its resolution.

\section{RG1: Modeling Resilience of OL-CAIS}
This section addresses \textbf{RG1}, which aims to model the resilience of OL-CAIS behavior to support decision-makers, specifically tackling \textbf{RC1}: Modeling performance evolution. To address this challenge, we conducted evaluations that analyze how OL-CAIS identifies and characterizes disruptive events and their impact on system performance. 
The experiments investigated the evolution of performance across steady, disruptive, and final states, providing insights into how the system transitions between these states by tracking ACR and ACR Threshold over time. 
These evaluations directly contribute to answering the associated research questions by defining performance metrics (RQ1.1), distinguishing acceptable and unacceptable performance levels (RQ1.2), and identifying the states and transitions that reflect resilience (RQ1.3).

\begin{figure}[ht!]
\centering
\centerline{\includegraphics[width = \textwidth]{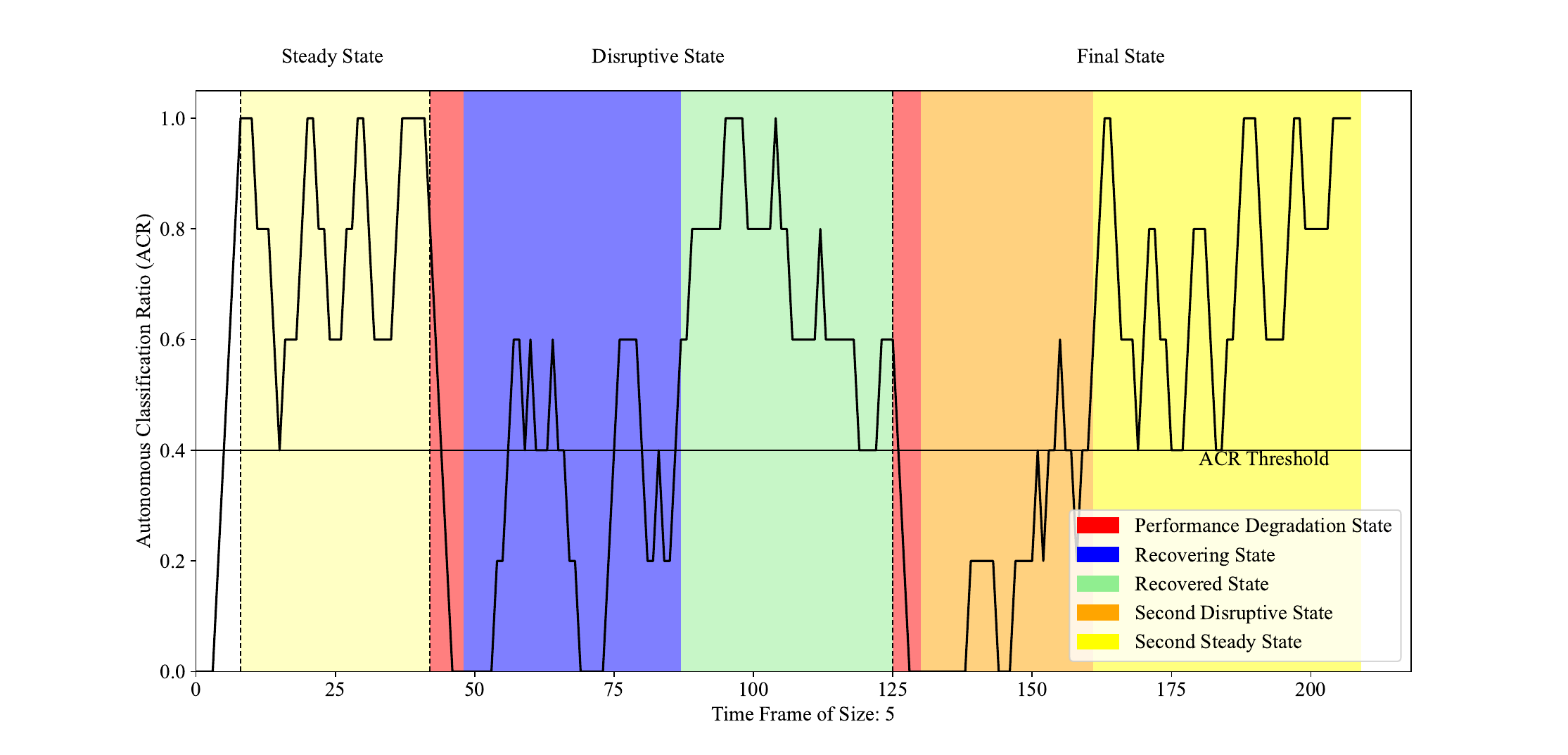}}
\caption{Resilience model of CORAL under internal policies experiment, where the system operates solely based on its internal policies without external intervention.}
\label{fig:benchmark}
\end{figure}

We have utilized the resilience model to identify the system under test performance through all our experiments.
Fig.~\ref{fig:benchmark} shows one example of the resilience model of CORAL internal policies. The resilience model shows its sensitivity in accurately tracing CORAL performance transitions.
Providing us with an empirical evidence to answer the following research questions.

\subsection{Answers to Research Questions RQ1.1 - RQ1.3}

\paragraph{Answer to RQ1.1.} \textit{What metric can measure OL-CAIS performance to render its behavior during run-time?}

One goal of automating industrial tasks is to reduce human participation in tedious tasks and automate them.
Thus, we have defined a new metric called the Autonomous Classification Ratio (ACR) that tracks autonomous classifications in OL-CAIS over a sliding window of iterations, where $ACR \in [0, 1]$.
A higher ACR value indicates autonomous operations by the system, and a low value indicates an increased reliance on human intervention.
Meanwhile, a $0.5$ ACR value equals the likelihood of autonomous and human actions.
Tracking the ACR values renders performance evolution over a series of iterations.

\paragraph{Answer to RQ1.2.} \textit{How can we define and distinguish acceptable and unacceptable performance levels in OL-CAIS from the decision-makers' perspective?}

By observing various experiments in this research, we have noticed that OL-CAIS starts by depending on human interactions to learn its tasks. During runtime, this dependency is reduced until the system starts operating autonomously.
We consider this initial state of the system to set the level for acceptable performance.
We set the ACR threshold for acceptable performance at the minimum value in this state. 
Decision-makers have the choice to change how they measurement the ACR Threshold.

\paragraph{Answer to RQ1.3.} \textit{What key states and transitions in OL-CAIS performance evolution reflect resilience during and after disruptive events?}

As shown in Fig.~\ref{fig:benchmark}, we model ACR over iterations across the different states the system goes through. We use the ACR threshold as quality gate as described in RQ1.2. 
The ACR curve shows that OL-CAIS operates in cycles of three major states.
The first state is the \textit{Steady State}, which starts with a sliding window of autonomous actions (i.e., $ACR = 1$) and ends when the ACR drops to zero, indicating a possible disruptive event.
A continuous ACR degradation to zero signals the start of the \textit{Disruptive State}. In the disruptive state, we can observe three different substrates in which the performance evolves: the Performance Degradation State, the Recovering State, the Policy Support State, and the Recovered State.
The  \textit{Performance Degradation State} precedes the \textit{Recovering State} in which the OL-CAIS resorts to internal or external policies to recover its performance to an acceptable performance level (i.e., above the ACR Threshold).
In the \textit{Recovered State}, the performance maintains an acceptable level for at least the same number of iterations as in the steady state.
After the disruptive event is fixed, the system may move to a second performance turbulence, which can cause it to move to another disruptive state.
In our empirical evaluation, we choose to fix the event after the performance is recovered. However, this might not be the case in other settings, as the event can be fixed before even starting the policies or a long time after it occurs and recovers.
The fixed event indicates the start of the third state, \textit{Final State}. The final state is similar to the disruptive state as it may start with second performance degradation, then enter a \textit{Second Disruptive State} and finally recover to a \textit{Second Steady State}.
These three states are repeated per each disruptive event that may occur.

\section{RG2: Develop agent-based policies to balance resilience and greenness}

This section focuses on \textbf{RG2}, which aims to develop intelligent policies to balance resilience and greenness in OL-CAIS during unforeseen disruptive events, addressing \textbf{RC2}: Balancing greenness and resilience upon disruptive events. 
Our evaluations investigated various decision-making policies, including internal, one-agent, two-agent, and RL-agent, to understand their impact on system recovery and energy efficiency. 
These experiments, conducted in both real-world and simulated settings, explored the trade-offs between greenness and resilience. 
Additionally, we evaluated the role of containerization as a green resource allocation technology. 
These evaluations contribute to understanding the relationship between greenness and resilience (RQ2.1), the necessary decision-making components (RQ2.2), and the effectiveness of different policies (RQ2.3, RQ2.4). Finally, the containerization experiments directly address how resource optimization enhances energy efficiency (RQ2.5).




\subsection{Real-World Experiments}
\label{subsec:realworldexp}

We conducted four real-world experiments with CORAL under different policies. These experiments focused on analyzing the collaborative robot's resilience and greenness during steady, disruptive, and final states.
\paragraph{Experiment with Internal Policies.}
The internal policies experiment aims to describe the system's performance without any intervention across the performance states. 
Fig.~\ref{fig:benchmark} shows the performance of CORAL as a function of 5-iterations sliding window. The figure shows the ACR values over 208 iterations, highlighting the Steady State, Disruptive State, and the Final State with their sub-states.
In this experiment, CORAL, with its internal policies, was able to recover both when we switched off (i.e., disruptive event) and switched back on the light (i.e., fix event). 
We noticed a second performance degradation after turning the lights back on due to CORAL's classifier memory in the disruptive state and forgetting the steady state environmental settings.

\begin{figure}[ht!]
\centering
\centerline{\includegraphics[width = \textwidth]{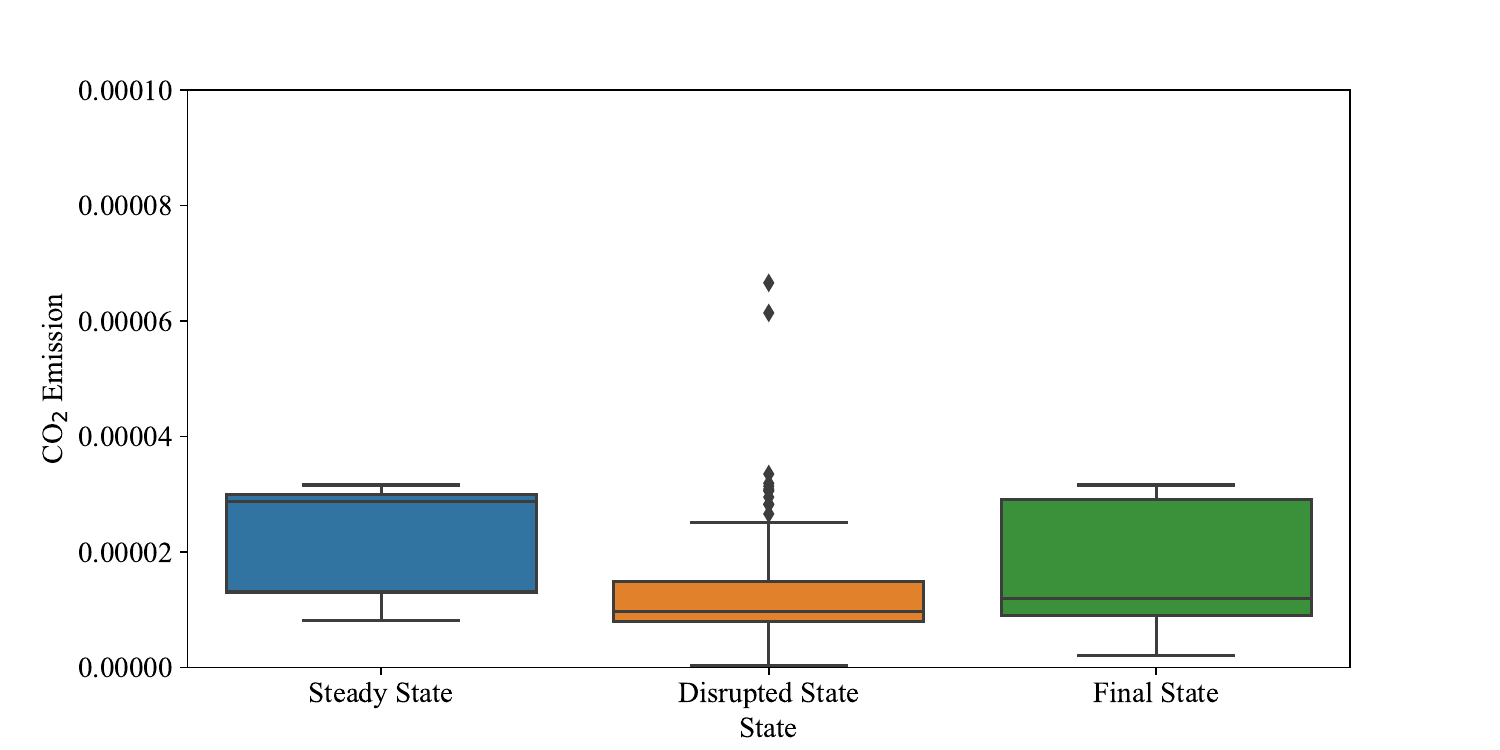}}
\caption{Box plot of CO$_2$ emission of CORAL under internal policies experiment during the three states.}
\label{fig:benchmark_co2}
\end{figure}

Fig.~\ref{fig:benchmark_co2} shows the box plot of the CO$_2$ emitted under the internal policies experiment divided over the three major states.
The mean CO$_2$ emission during the steady state is $2.1 \times 10^{-4}$, while $1.3 \times 10^{-4}$, and $1.7 \times 10^{-4}$ during the disruptive and final states, respectively.
By observing the disrupted and final states in the box plot of Fig.~\ref{fig:benchmark_co2}, we notice a close median value in both states. In both of these states, internal policies were used to recover performance by utilizing human interactions. However, in the disruptive state, we notice that the whiskers, which represent the range of the data inside the interquartile range, are noticeably reduced compared to the other two states, indicating that the final state is exhibiting some memory left from the steady state and not a total forgetting.

\begin{figure}[ht!]
\centering
\centerline{\includegraphics[width = \textwidth]{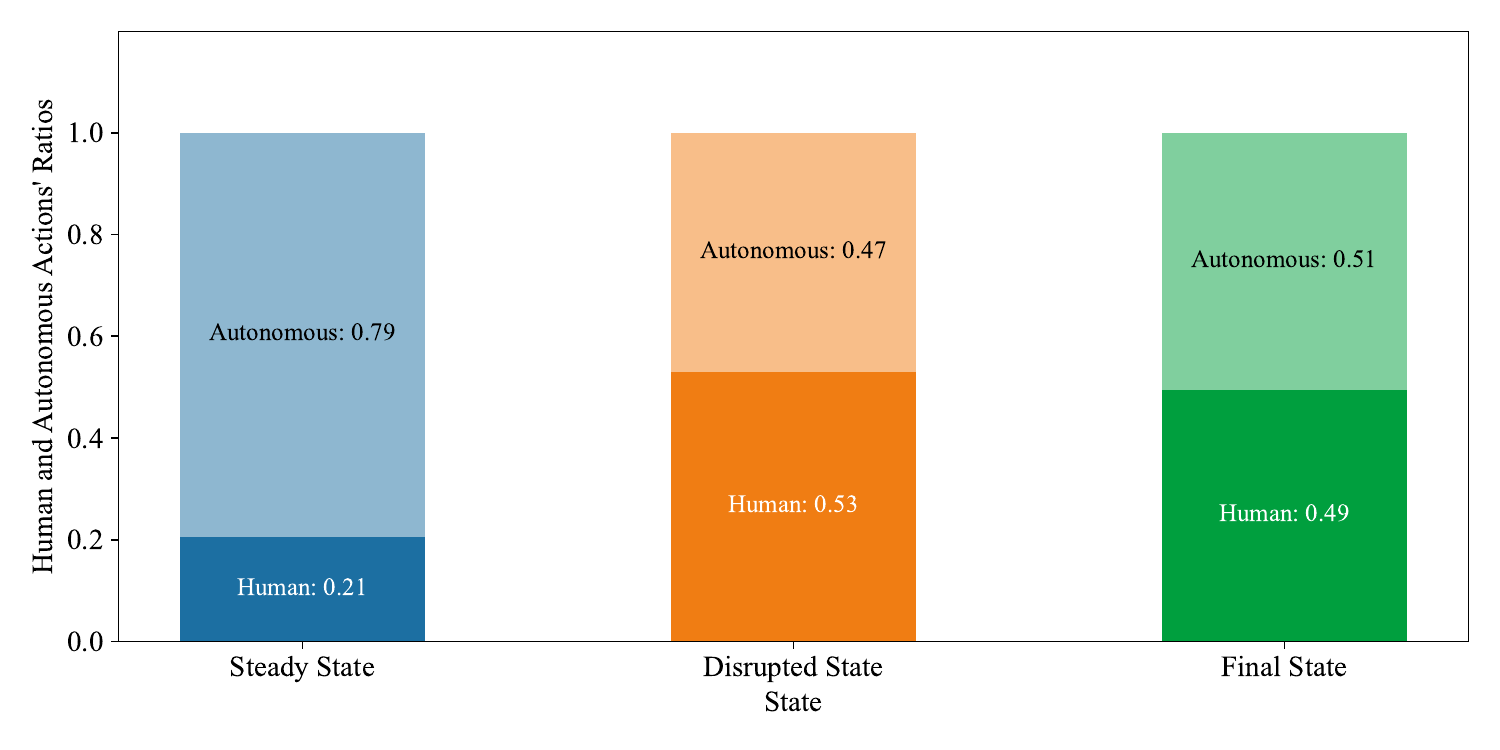}}
\caption{Bar plot comparing the Human and Autonomous actions’ ratios across the three states for the internal policies.}
\label{fig:benchmark_hi}
\end{figure}

In Fig.~\ref{fig:benchmark_hi}, we show the human and autonomous actions' ratios in each of the three states, 0.79 and 0.21 in the steady state, for autonomous actions and human actions, respectively, 0.53 and 0.47 in the disruptive state for autonomous actions and human actions respectively and 0.49 and 0.51, respectively in the final state. 
We notice that the OL-CAIS has resorted to increasing reliance on human interactions to learn its new environmental settings and restore its performance during the disruptive state. 
Moreover, similar ratios appear in the final state, indicating another change in the environmental settings (i.e., fix event).

\begin{figure}[ht!]
\centering
\centerline{\includegraphics[width = \textwidth]{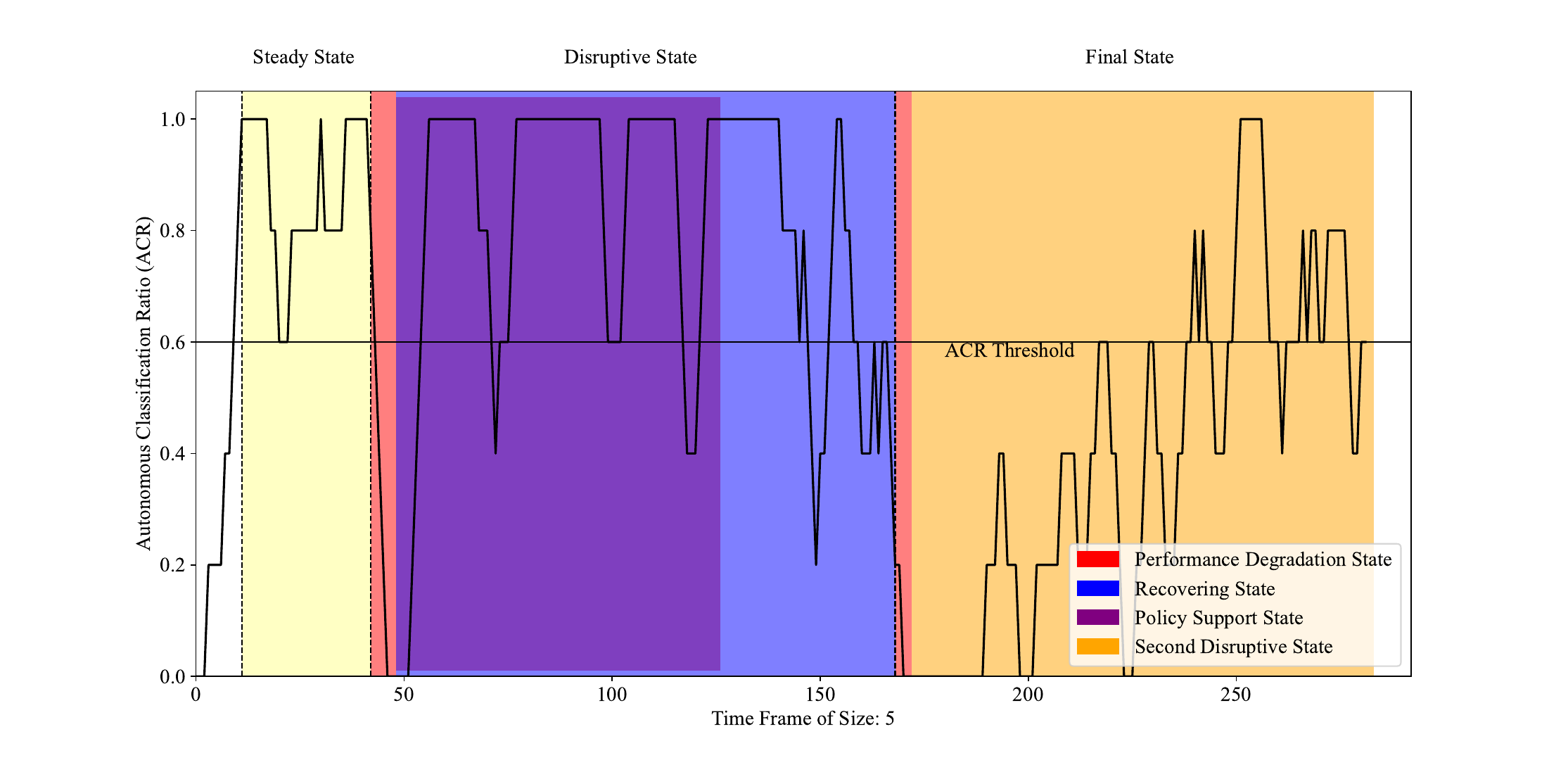}}
\caption{Resilience model of CORAL under one-agent policies experiment, where the weighted sum model supports the system's decision-making.}
\label{fig:optexpchart}
\end{figure}

\paragraph{Experiment with One-Agent Policies.}
In this experiment, we applied the action recommended by the weighted sum model after turning off the lights and entering a Performance Degradation State, Fig.~\ref{fig:optexpchart}. 
We let the recommendation run to monitor CORAL recovery. However, we could not see any recovery in either of the disruptive states, and the system has remained in a Recovering State. 
After five consecutive standalone decisions, we stopped the support from the one-agent policies and let the system operate while the disruptive event was still not fixed. 
The resilience model shows that our OL-CAIS was not able to enter a Recovered State.
The one-agent support was stuck in the recovering state.
CORAL never recovered from the disruptive event, not even when we switched back on the light and left it with the internal policies, due to the long training with disrupted data in the first disruptive state.

\begin{figure}[ht!]
\centering
\centerline{\includegraphics[width = \textwidth]{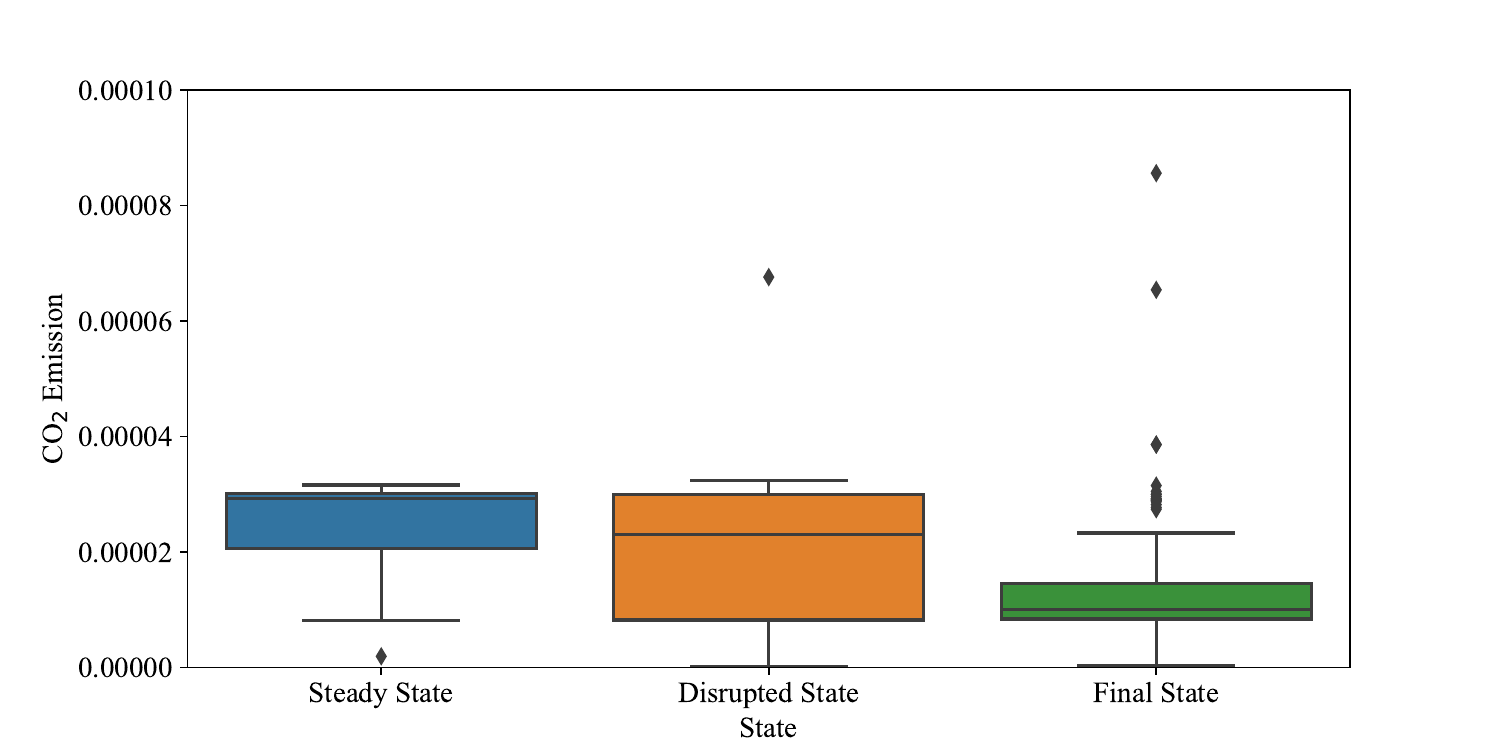}}
\caption{Box plot of CO$_2$ emission of CORAL under one-agent policies experiment during the three states.}
\label{fig:optexpchart_co2}
\end{figure}

We can notice from Fig.~\ref{fig:optexpchart_co2} that supporting CORAL with one-agent technique makes the disruptive state close to the steady state in CO$_2$ distribution. Furthermore, it changes the data ranges in the final state dramatically, showing a catastrophic forgetting from the steady state, which may be due to not being able to recover during the disruptive state.
The means of CO$_2$ emission under the one-agent technique were $2.1 \times 10^{-5}$, $1.9 \times 10^{-5}$, and $1.4 \times 10^{-5}$, during the steady, disruptive, and final states, respectively. 

\begin{figure}[ht!]
\centering
\centerline{\includegraphics[width = \textwidth]{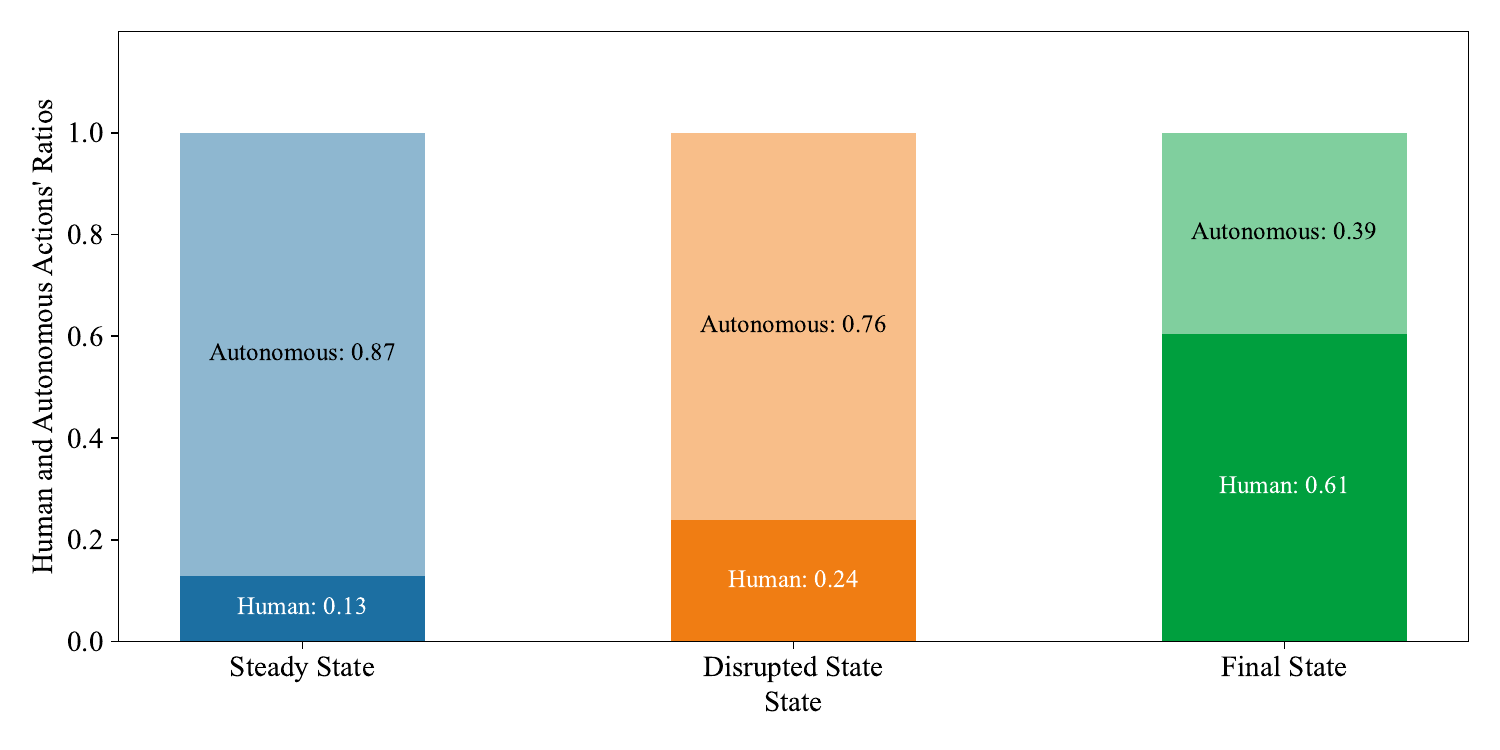}}
\caption{Bar plot comparing the Human and Autonomous actions’ ratios across the three states under one-agent policies.}
\label{fig:gropt_hi}
\end{figure}

Fig~\ref{fig:gropt_hi} shows how using one-agent technique helped our OL-CAIS to maintain similar ratios of autonomous and human actions in the disruptive state as the ratios in the steady state. The ratios were 0.87 and 0.13 for autonomous and human actions, respectively, during the steady state and 0.76 and 0.24 in the disruptive state. Applying a one-agent supporting technique for decision-making requires increased human interactions in the final state as the system attempts to restore its performance by learning more from the human, with 0.39 of autonomous actions to 0.61 of human actions.

\begin{figure}[ht!]
\centering
\centerline{\includegraphics[width = \textwidth]{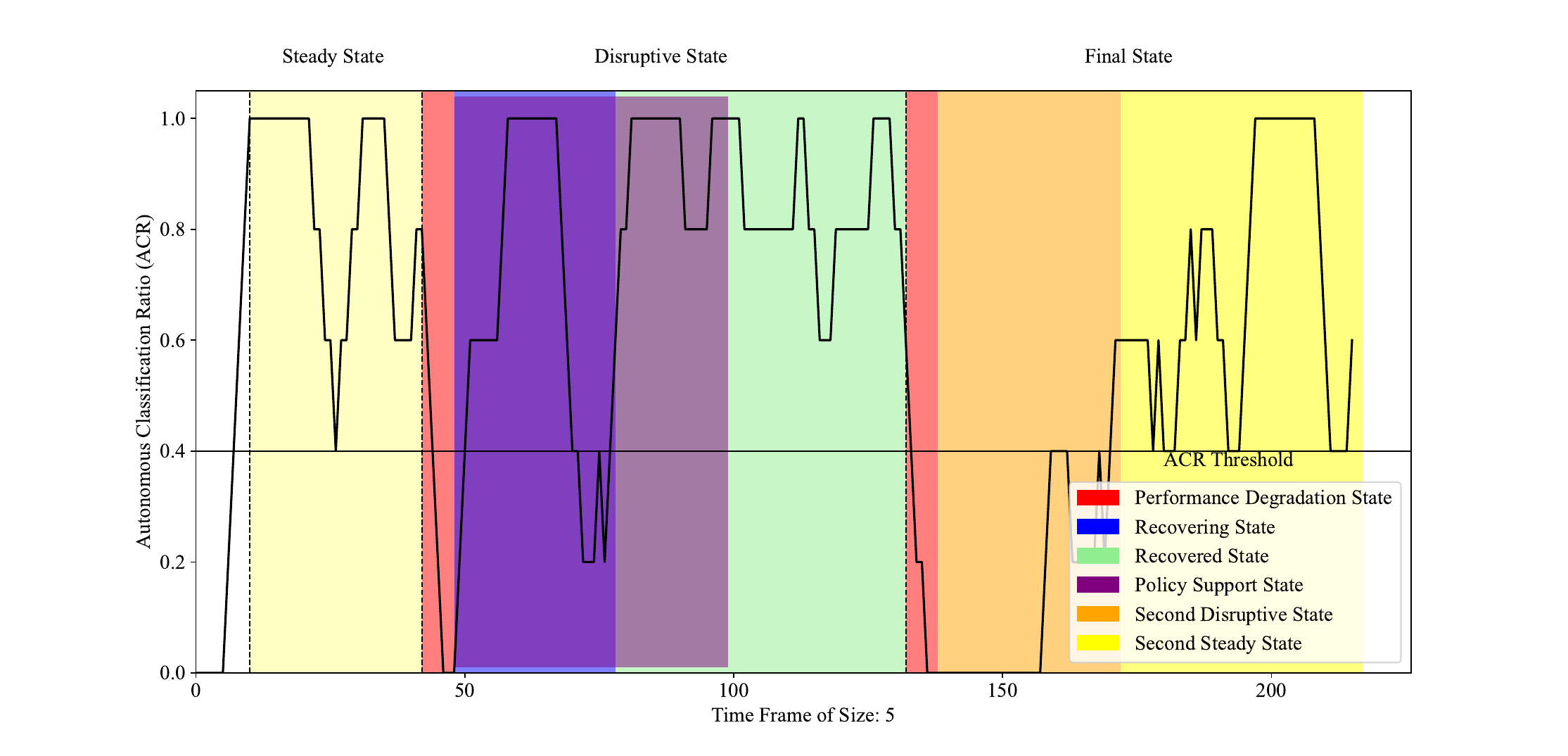}}
\caption{Resilience model of CORAL under two-agent policies experiment, where the GResilience Game supports the system's decision-making.}
\label{fig:gtexpchart}
\end{figure}

\paragraph{Experiment with Two-Agent Policies.}
In this experiment, we applied the action recommended by the GResilience Game after turning off the lights. Fig.~\ref{fig:gtexpchart} shows the ACR behavior, where the system recovered after both switching off and turning the light back on. The resilience model also shows that two-agent policies supported the decision during the Recovering State, taking it into a Recovered State soon.
Moreover, CORAL was able to learn about new environmental conditions after stopping support and maintaining a recovered state. 
At the final state, when it is recovered, the performance line shows some memory from the steady state entering into a Second Steady State.

\begin{figure}[ht!]
\centering
\centerline{\includegraphics[width = \textwidth]{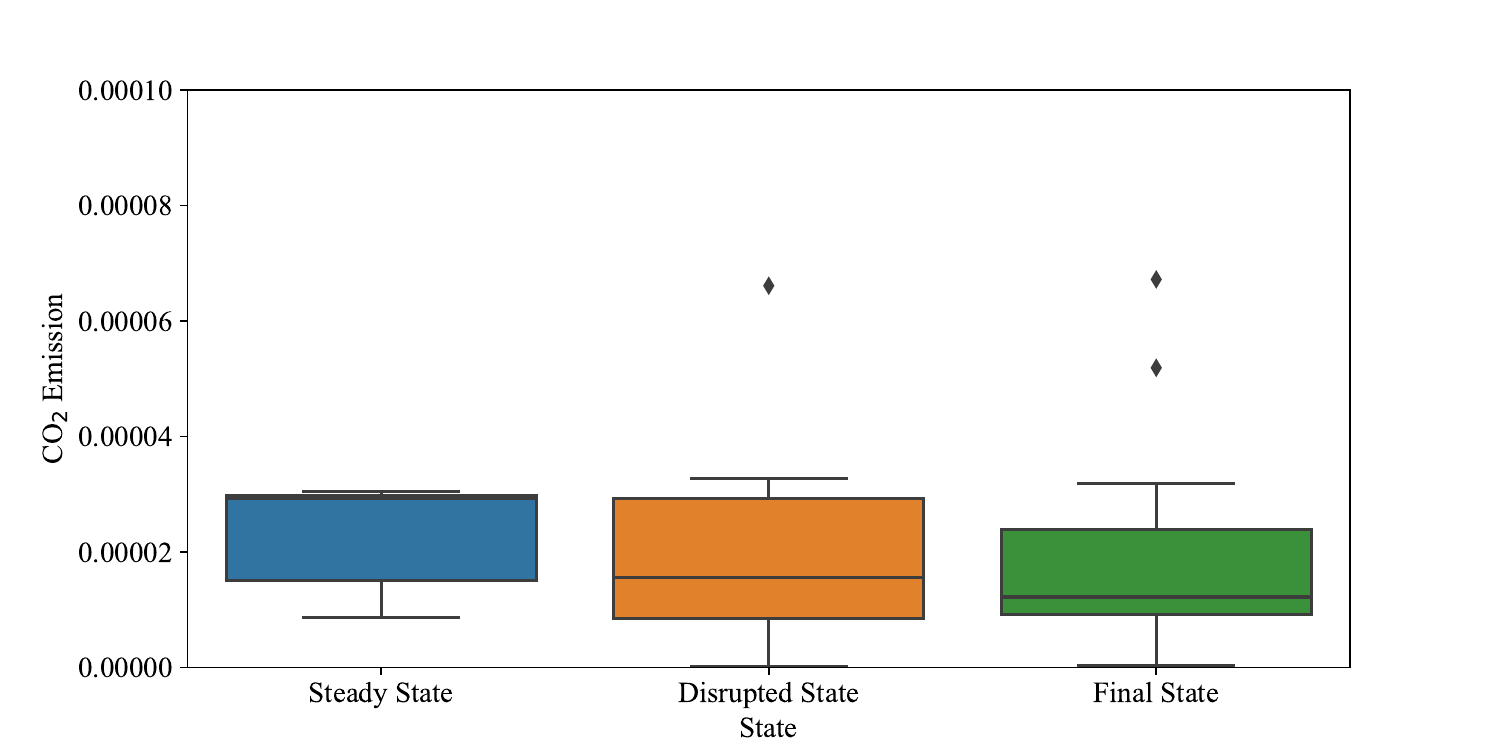}}
\caption{Box plot of CO$_2$ emission of CORAL under two-agent policies experiment during the three states.}
\label{fig:gtexpchart_co2}
\end{figure}

Fig.~\ref{fig:gtexpchart_co2} shows a coherent behavior of CO$_2$ emissions among the three states. The means were $2.1 \times 10^{-5}$, and $1.8 \times 10^{-5}$ during the steady and disruptive states, respectively, and $1.6 \times 10^{-5}$ during the final state. From these results presented in the three experiments, we can notice a similar distribution of CO$_2$ emissions among three steady states; however, different behavior in the other states is influenced by the applied policies to support decision-making.

\begin{figure}[ht!]
\centering
\centerline{\includegraphics[width = \textwidth]{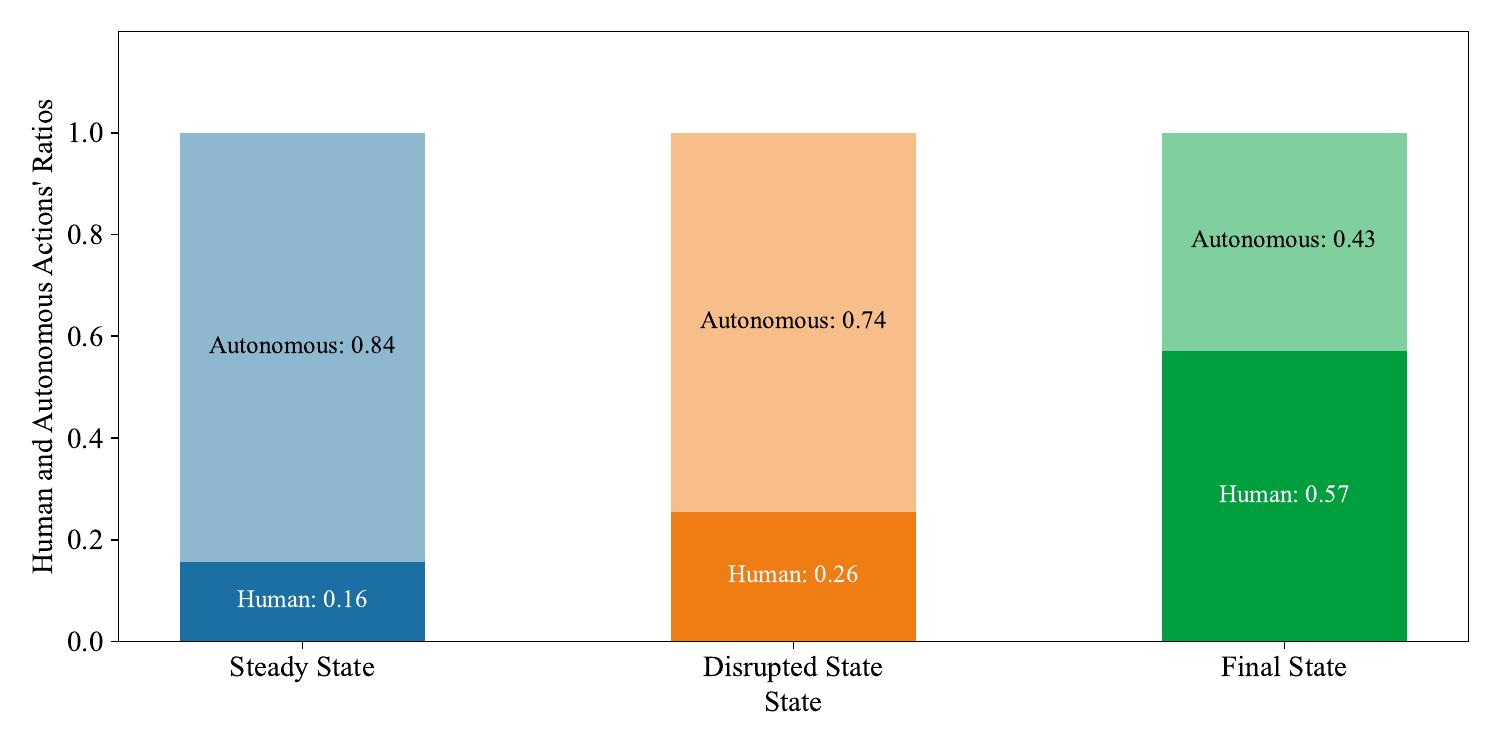}}
\caption{Bar plot comparing the Human and Autonomous actions’ ratios across the three states under two-agent policies.}
\label{fig:grgame_hi}
\end{figure}

In Fig.~\ref{fig:grgame_hi}, the two-agent policies have helped the system to maintain similar ratios of autonomous and human actions in the steady state 0.84 and 0.16, and 0.74 and 0.26, in the disruptive state.
In the final state, leaving the system with its internal policies, it shows an increased reliance on human interactions to learn the environmental changes in the final state, with 0.43 of autonomous actions and 0.57 of human actions.

\begin{figure}[ht!]
\centering
\centerline{\includegraphics[width = \textwidth]{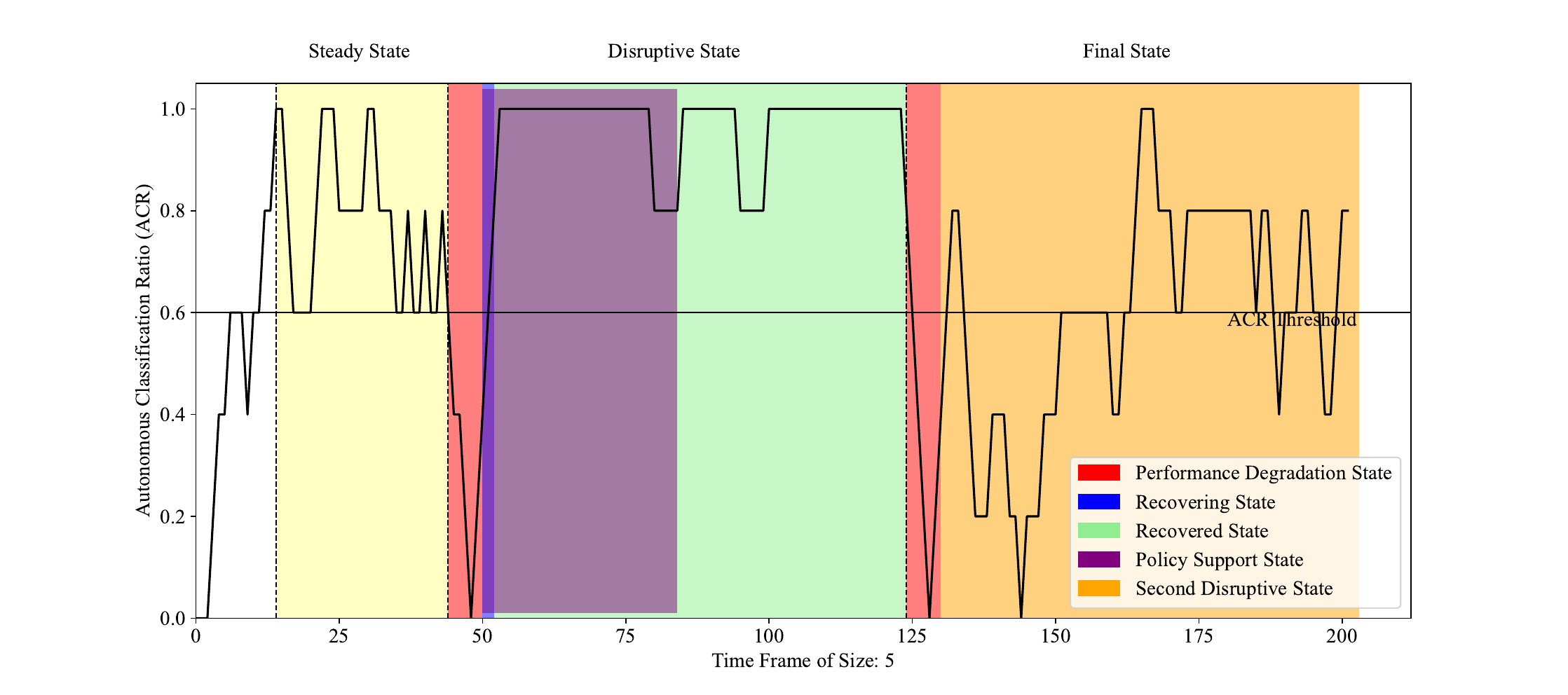}}
\caption{Resilience model of CORAL under RL-agent policies experiment, where the reinforcement learning agent supports the system's decision-making.}
\label{fig:rlagentacr}
\end{figure}

\paragraph{Experiment with RL-Agent Policies.}
We applied the action recommended by the reinforcement learning agent after turning off the lights. Fig.~\ref{fig:rlagentacr} shows the ACR curve entering into the Recovered State at the time the RL-agent started. Even after stopping the support, our OL-CAIS managed to maintain a recovered state. 
In the final state, the resilience models show that the system has forgotten what it learned in the steady state, leading to a second Performance Degradation State, which the system didn't manage to transfer into a Second Steady State after.

\begin{figure}[ht!]
\centering
\centerline{\includegraphics[width = \textwidth]{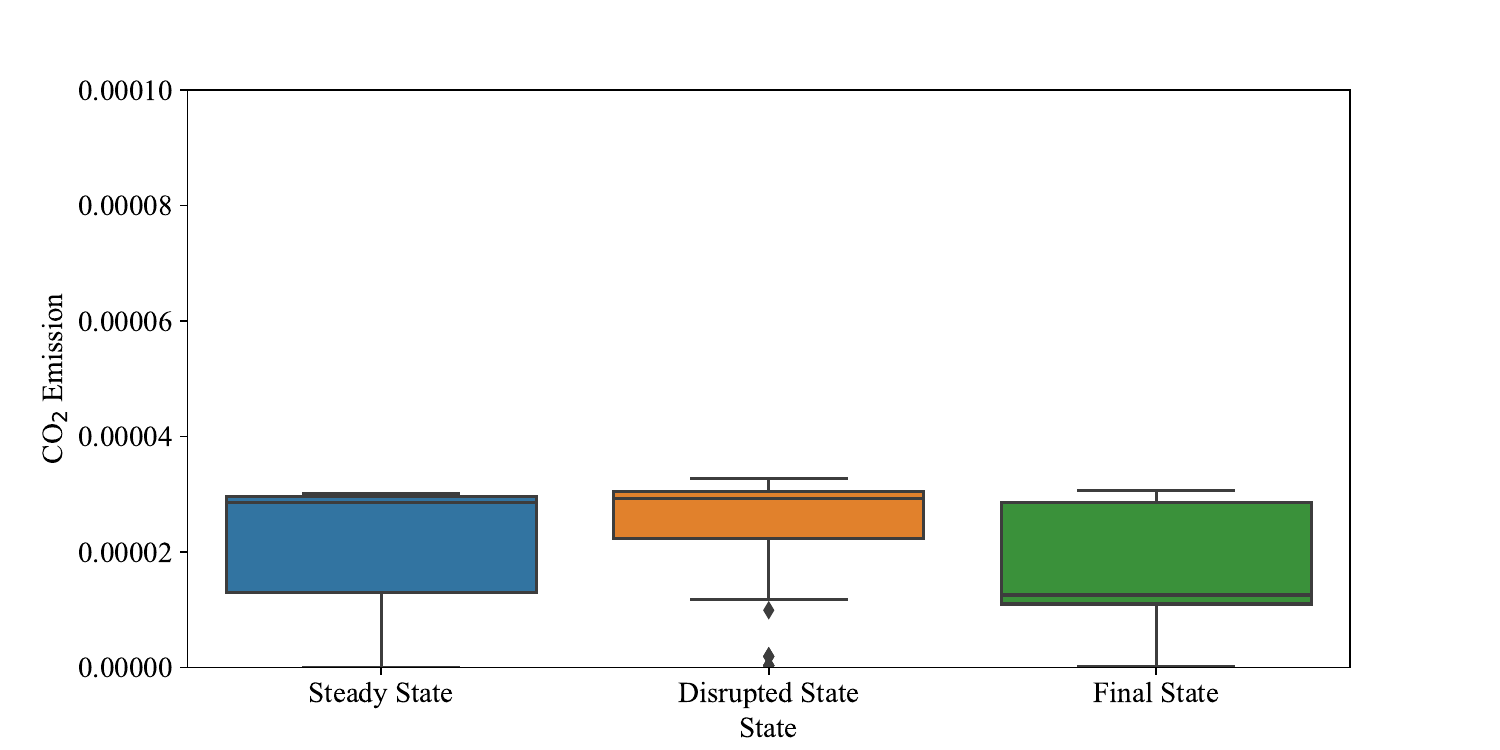}}
\caption{Box plot of CO$_2$ emission of CORAL under RL-agent policies experiment during the three states.}
\label{fig:rlagentco2}
\end{figure}

The box plot in Fig.~\ref{fig:rlagentco2} shows higher emission in the disruptive state than the steady and final states.
This is expected as the RL-agent keeps monitoring the environment and learns the best policy, which consumes additional computation energy.
The means were $2.1 \times 10^{-5}$, and $3.5 \times 10^{-5}$ during the steady and disruptive states, respectively, and $1.7 \times 10^{-5}$ during the final state. 
From these results presented in the four experiments, we notice a similar distribution of CO$_2$ emissions among the four steady states as we intentionally wanted to have a similar steady state for experiments. However, this distribution was different in the other states as it is influenced by the applied policies to support decision-making.

\begin{figure}[ht!]
\centering
\centerline{\includegraphics[width = \textwidth]{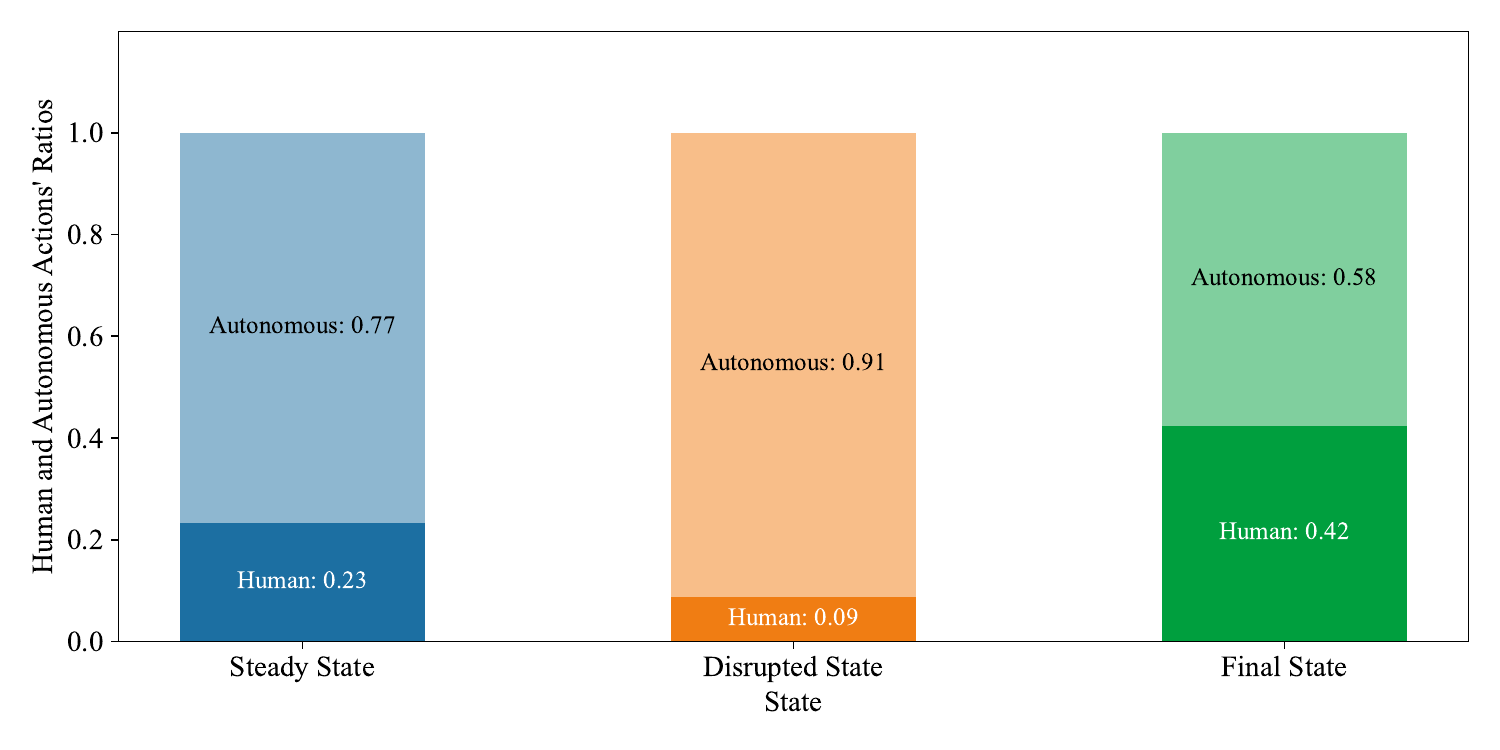}}
\caption{Bar plot comparing the Human and Autonomous actions’ ratios across the three states under RL-agent policies.}
\label{fig:rlagenthi}
\end{figure}

Finally, Fig.~\ref{fig:rlagenthi} shows the ratios between autonomous and human actions during three states. We notice that the RL-agent policies achieved the highest autonomous ratios in both the disruptive and final states, although our OL-CAIS could not recover in the final state.
The autonomous ratios were 0.77 in the steady state and then increased to 0.91 under the RL-agent support. Even without support in the final state, the autonomous ratio was still higher than that of the other three experiments, which was 0.58.
The ratio in the final state decreased from the disruptive state, indicating a higher dependency on human interactions to learn back what it forgot during the disruptive state.

\subsection{Simulated Experiments}
\begin{figure}[ht!]
\centering
\centerline{\includegraphics[width = \textwidth]{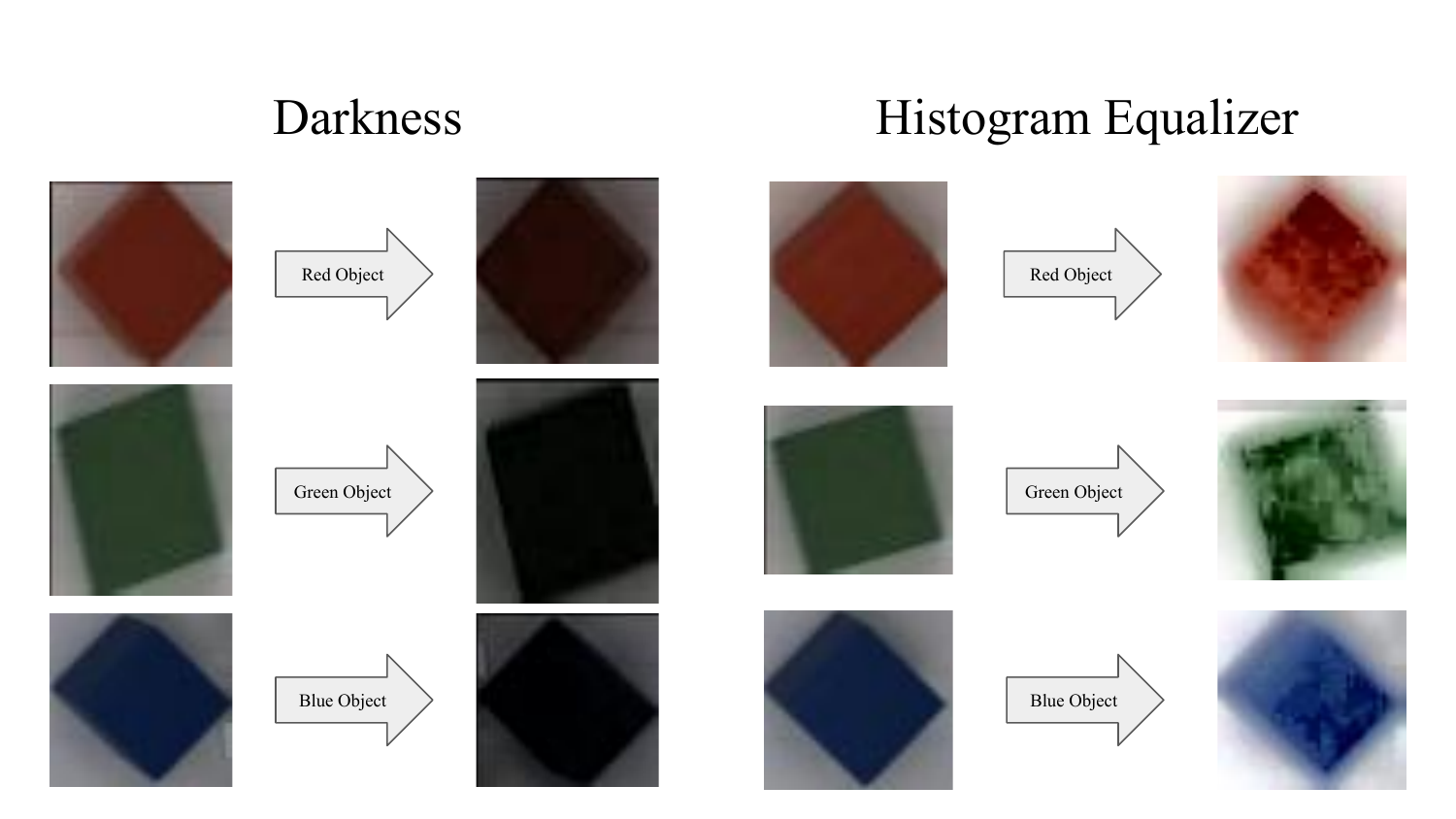}}
\caption{Object instances from CORAL with darkness and histogram equalizer filters.}
\label{fig:darkandattack}
\end{figure}

In the simulated experiments, we ran the same experiments, once simulating turning off the lights with a darkness filter and second simulating the adversarial attack using a histogram equalizer.
Fig.~\ref{fig:darkandattack} shows some examples\footnote{The full dataset is available at our CAIS-DMA Github Repository, and further available on Robo Flow at: \url{https://universe.roboflow.com/diaeddin-rimawi/objects-shapes-and-colors}} before and after applying the darkness and histogram equalizer filters to create the two disruptive events.
We present both results categorized based on the simulated disruptive event.

\begin{figure}[ht!]
\centering
\centerline{\includegraphics[width = \textwidth]{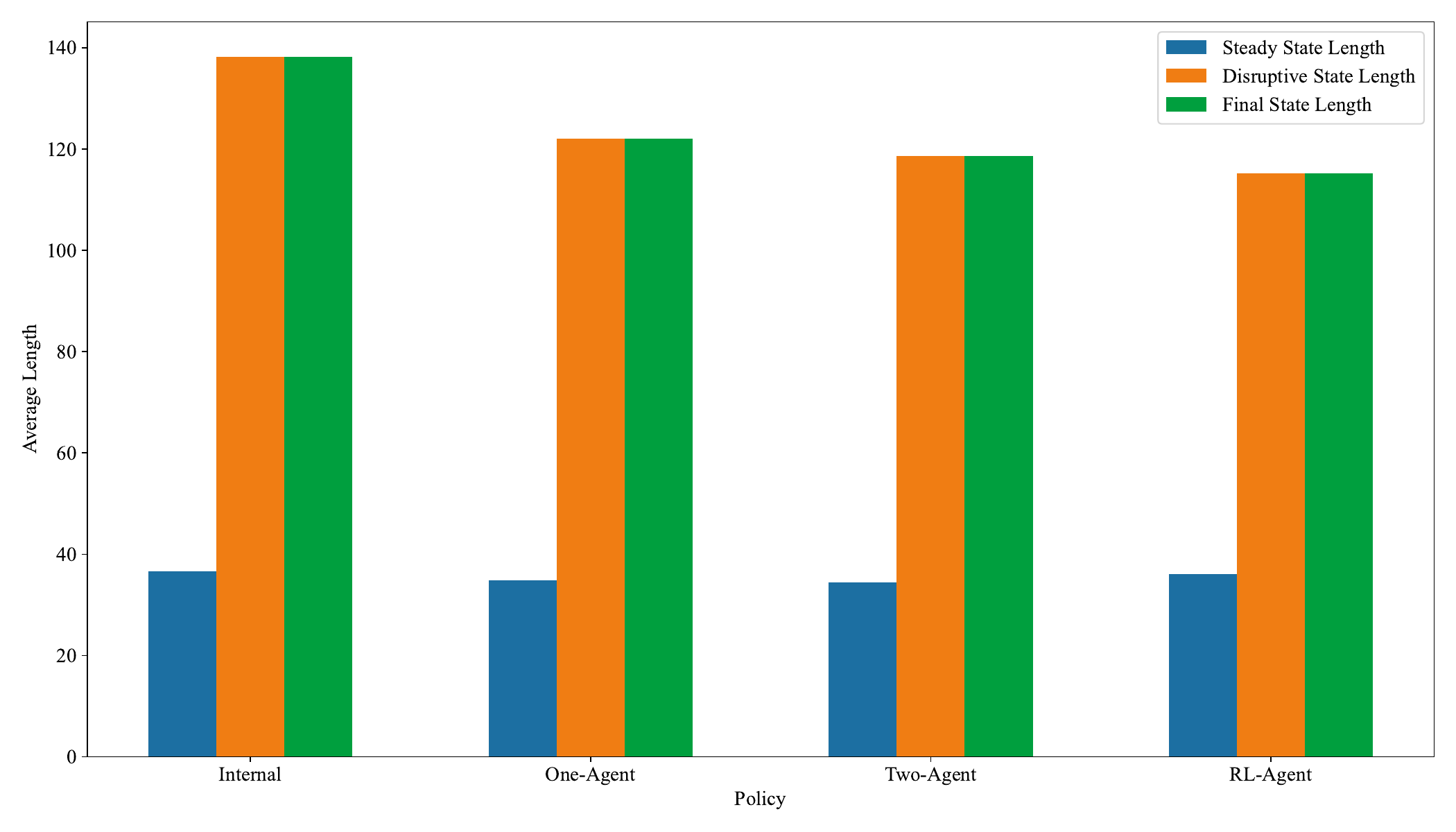}}
\caption{Bar Plot comparing the states' lengths across the four policies in experiments simulating darkness as a disruptive event.}
\label{fig:darknessstateslengths}
\end{figure}

\paragraph{Darkness filter.}
For each decision-making policy, we run one hundred replications of the same real-world experiments using our simulator in CAIS-DMA.
Fig.\ref{fig:darknessstateslengths} shows the average lengths for each of the three states across the four experiments.
We kept the same steady state lengths across the four hundred experiments, and we stopped the disruptive state when the performance recovered. The final state runs the same length as the disruptive state.
Fig.\ref{fig:darknessstateslengths} gives an indication of the policies that have led to faster recovery of the system. We can notice this information by observing the policies with the minimum disruptive state length, that is, RL-agent policies with an average length of 115 iterations, then two-agent, one-agent, and internal policies, with 118, 122, and 138, respectively.

\begin{figure}[ht!]
\centering
\centerline{\includegraphics[width = \textwidth]{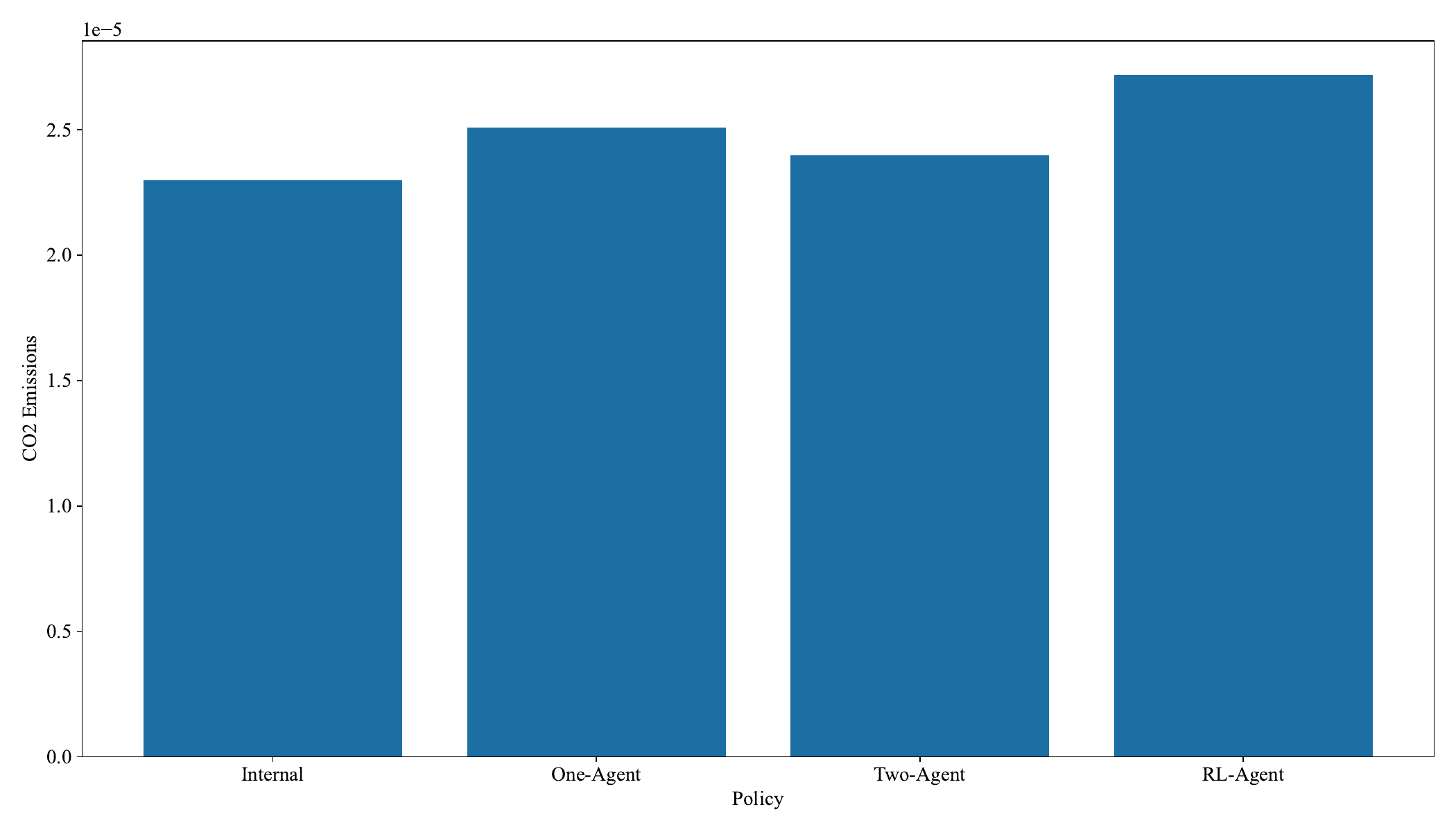}}
\caption{Bar Plot comparing the CO$_2$ emission across the four policies in experiments simulating darkness as a disruptive event.}
\label{fig:darknessco2}
\end{figure}

In Fig.~\ref{fig:darknessco2}, we show the average CO$_2$ emission in the one hundred experiments per policy.
The internal policies scored the lowest emission, followed by two-agent, one-agent, and RL-agent, aligning with our findings in real-world experiments.

\begin{figure}[ht!]
\centering
\centerline{\includegraphics[width = \textwidth]{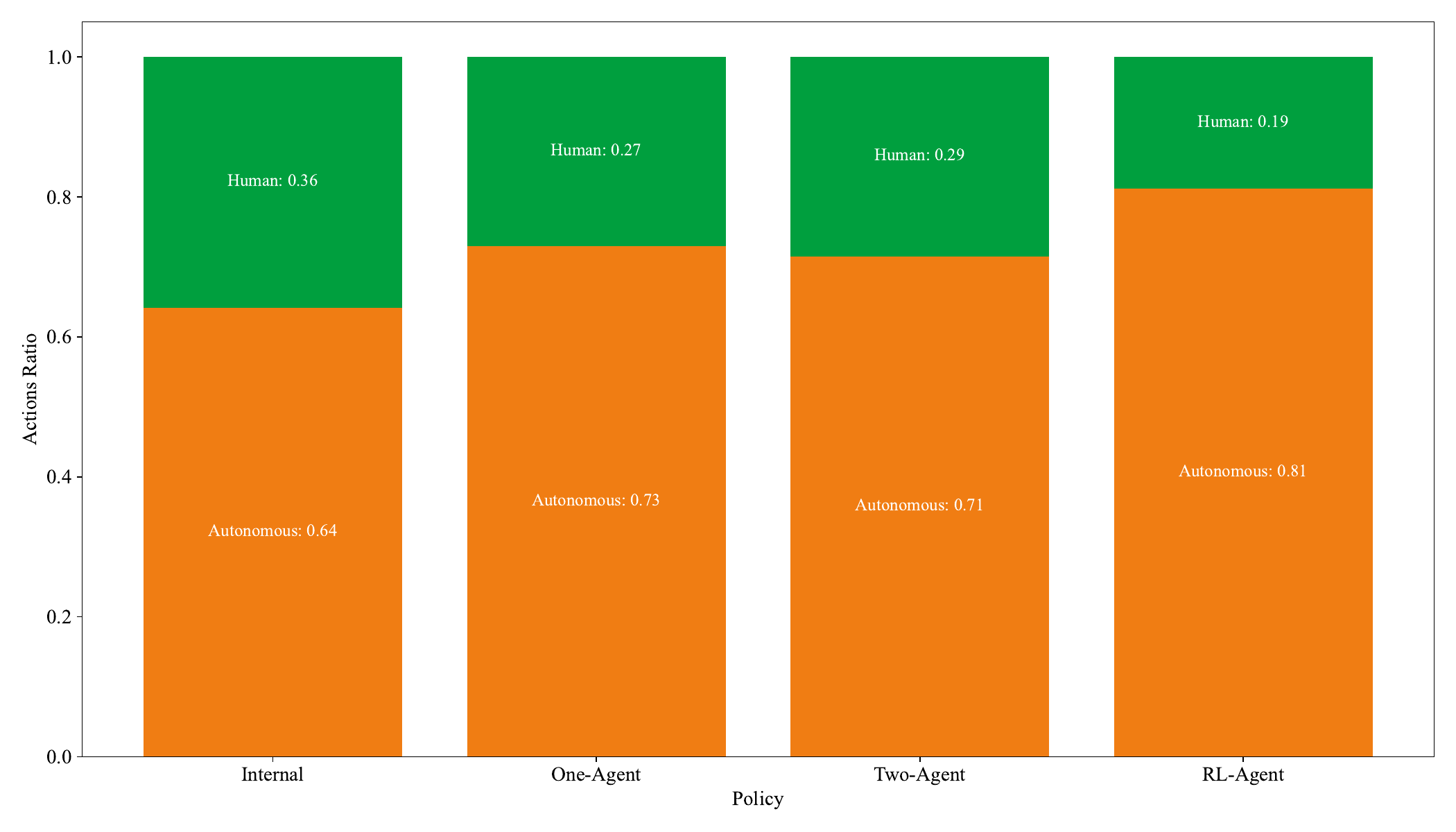}}
\caption{Bar plot comparing the autonomous/human actions ratios across the four policies in experiments simulating darkness as a disruptive event.}
\label{fig:darknessactions}
\end{figure}

We further compared the policies in terms of the autonomous and human action ratios. Fig.~\ref{fig:darknessactions} shows the ratios per policies in our experiments. 
In industrial settings, one of the main goals is increasing the system's ability to perform autonomous actions.
All our decision-making policies have improved the system's autonomy. 
The RL-agent achieved the highest autonomous actions ratio among the four policies, followed by the one-agent, two-agent, and finally, the internal policies. 

\begin{figure}[ht!]
\centering
\centerline{\includegraphics[width = \textwidth]{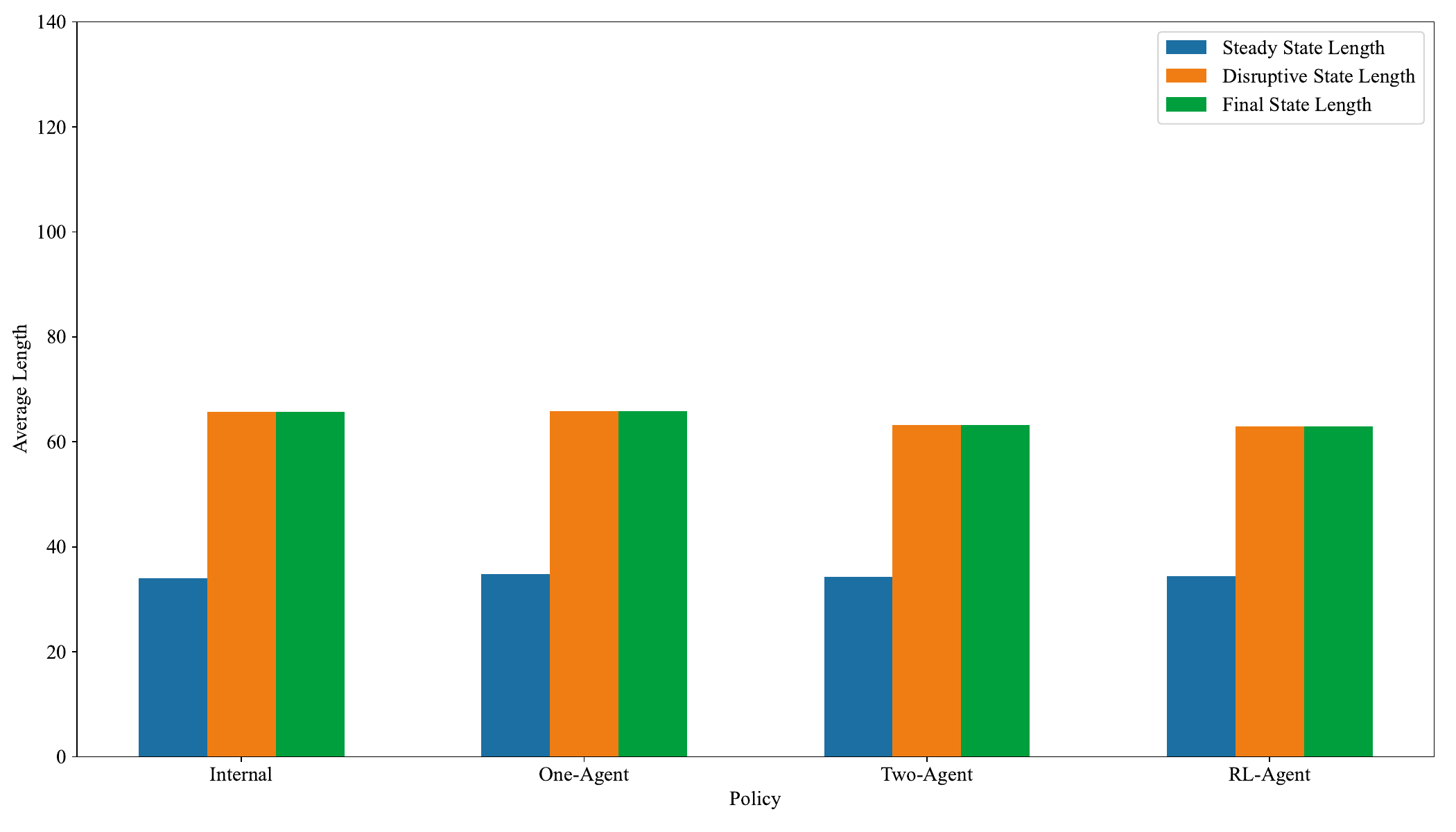}}
\caption{Bar plot comparing the states' lengths across the four policies in experiments simulating histogram equalizer attack as a disruptive event.}
\label{fig:adversarialstateslengths}
\end{figure}

\paragraph{Adversarial attack.}
Like the darkness filter, we run one hundred replications of the same experimental protocol using our simulator in CAIS-DMA, but simulating an adversarial attack as a disruptive event.
Fig.\ref{fig:adversarialstateslengths} shows the average lengths for each of the three states across the four experiments.
Keeping the same steady state lengths, stopping the disruptive state when the performance recovered, and setting the final state the same length as the disruptive state.
Fig.\ref{fig:adversarialstateslengths} gives an indication of the policies that have led to faster recovery of the system. 
The lengths of the disruptive state were close, ranging between 62 and 65 iterations. 62 and 63 iterations are the lengths achieved by RL-agent and two-agent experiments, respectively, and 65 is the length scored by both internal and one-agent policies experiments.
The policies' responses to different disruptive events yielded different disruptive state lengths. However, the order of the policies that achieve faster recovery still stands.

\begin{figure}[ht!]
\centering
\centerline{\includegraphics[width = \textwidth]{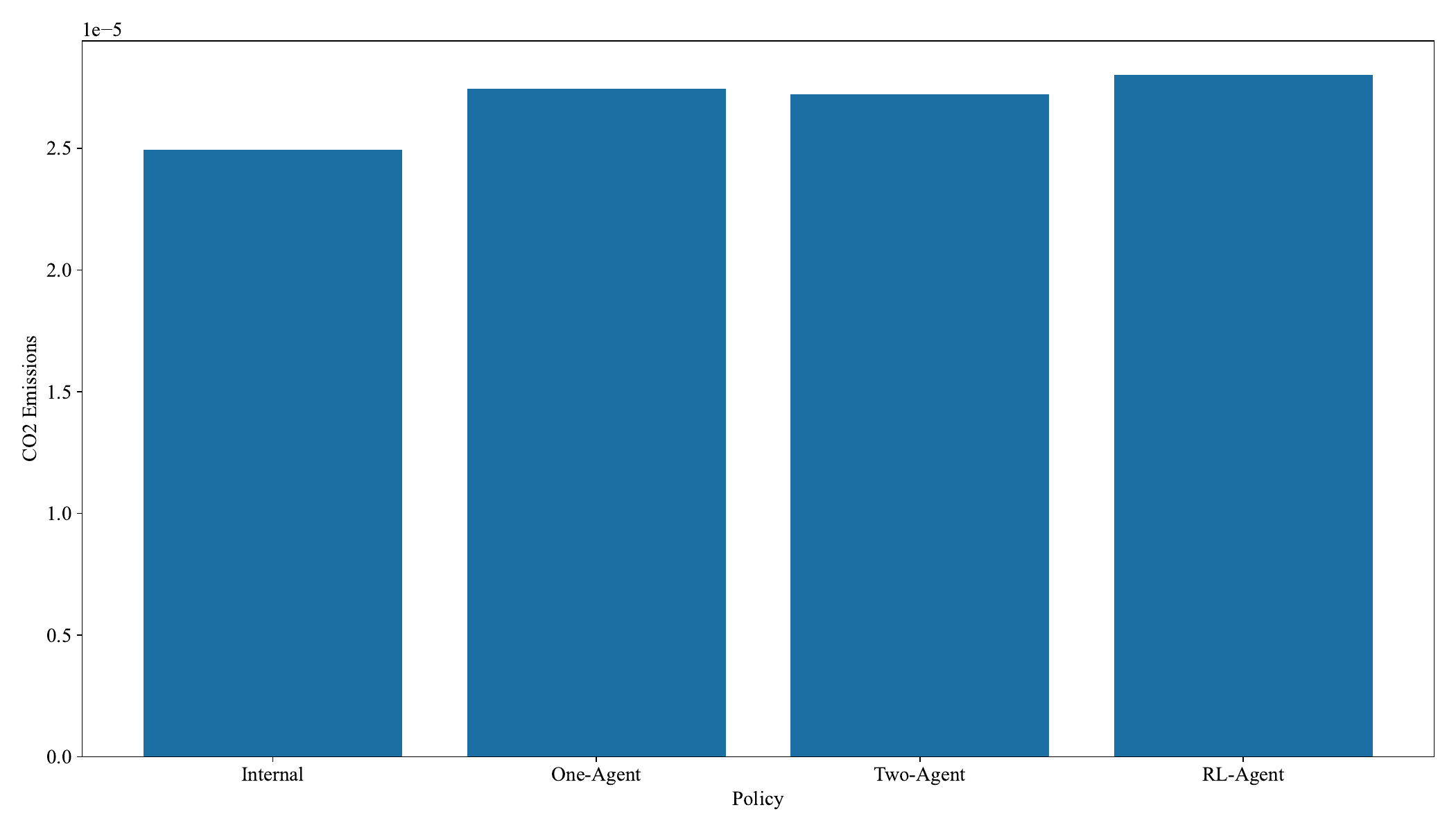}}
\caption{Bar plot comparing the CO$_2$ emission across the four policies in experiments simulating histogram equalizer attack as a disruptive event.}
\label{fig:adversarialco2}
\end{figure}

In Fig.~\ref{fig:adversarialco2}, we show the average CO$_2$ emission in the one hundred experiments per policy.
Aligning with the results we gained from real-world experiments and simulation with a darkness filter, the CO$_2$ emissions in adversarial attack experiments were in the same order.
The internal policies scored the lowest emission, followed by two-agent, one-agent, and RL-agent.

\begin{figure}[ht!]
\centering
\centerline{\includegraphics[width = \textwidth]{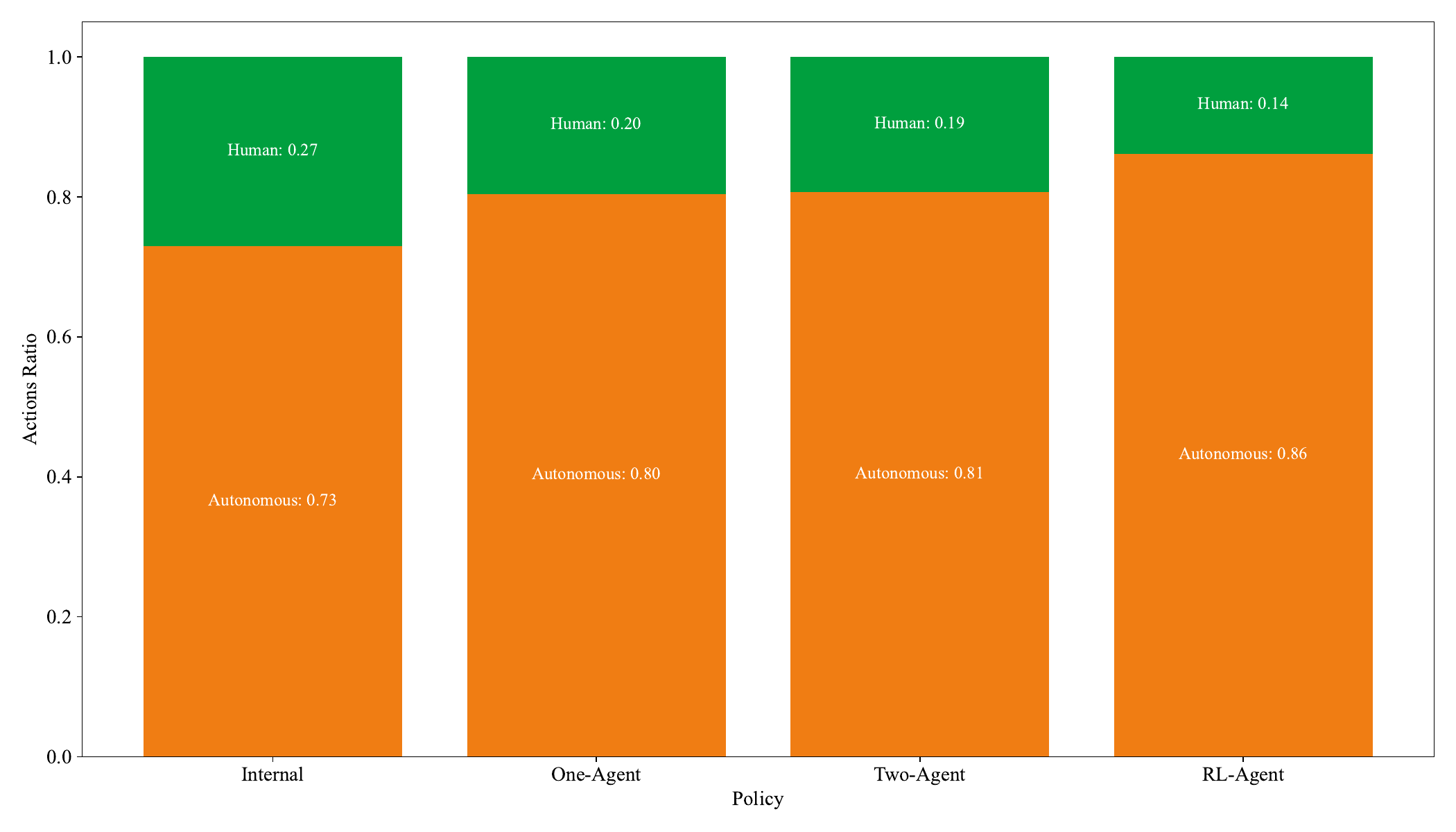}}
\caption{Bar plot comparing the autonomous/human actions ratios across the four policies in experiments simulating histogram equalizer attack as a disruptive event.}
\label{fig:adversarialactions}
\end{figure}

When we compared the policies in terms of the autonomous and human action ratios in the adversarial attack experiments, the results again aligned with the previous findings. Fig.~\ref{fig:adversarialactions} shows the ratios per policies in our experiments. 
All our decision-making policies have improved the system's autonomy. 
The RL-agent achieved the highest autonomous actions ratio among the four policies, followed by the one-agent, two-agent, and finally, the internal policies.

\subsection{Measurements Framework Evaluations}
\label{subsec:measurementframework}
\begin{figure}[ht!]
\centering
\centerline{\includegraphics[width = \textwidth]{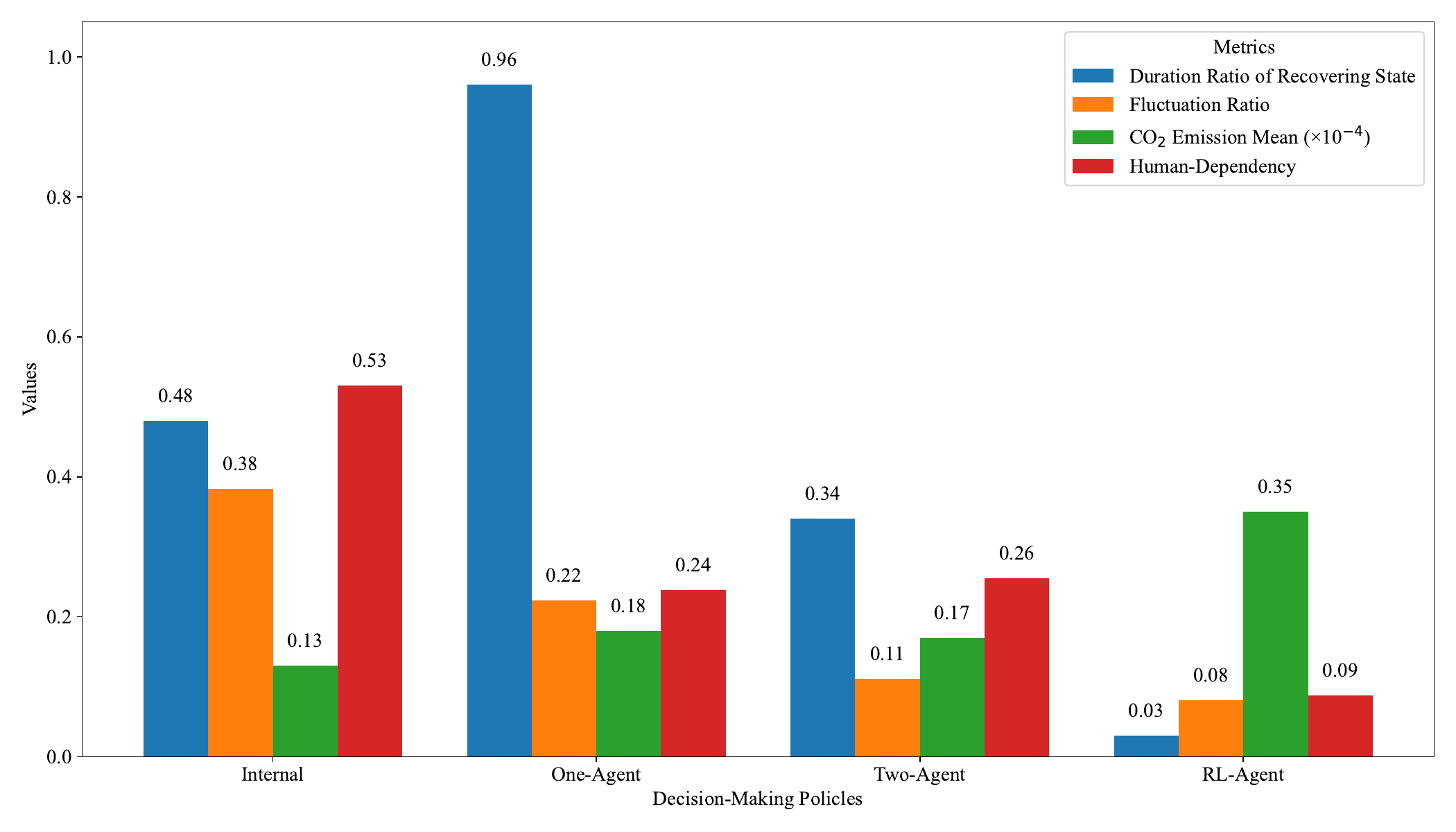}}
\caption{Measurement framework results from the real-world experiments.}
\label{fig:measurerealworld}
\end{figure}

The measurements framework compares the four agent-based policies by measuring resilience and greenness concepts: recovery speed, performance steadiness, green efficiency, and autonomy.
We present the measurements grouped by the decision-making policies per each experiment group.
Fig.~\ref{fig:measurerealworld} shows the values for the duration ratio of recovering state, which quantifies the recovery speed. The fluctuation ration to quantify performance steadiness. The CO$_2$ emission means quantifies green efficiency, noting that this value is scaled to $10^{-4}$, and finally, the human-dependency that quantities autonomy.
The results show that one-agent policies exhibited a more steady performance than internal policies although our OL-CAIS did not recover. It has reduced the human interactions from both internal and two-agent policies experiments, which indicates an autonomous attempt to restore performance. 
The two-agent policies show intermediate results between one-agent and RL-agent. RL-agent shows superior results compared to one-agent and two-agent in terms of rapid recovery and higher autonomous ratio. However, it comes at a higher CO$_2$ emission cost as it requires more computational power than the other agent-based policies.

\begin{figure}[ht!]
\centering
\centerline{\includegraphics[width = \textwidth]{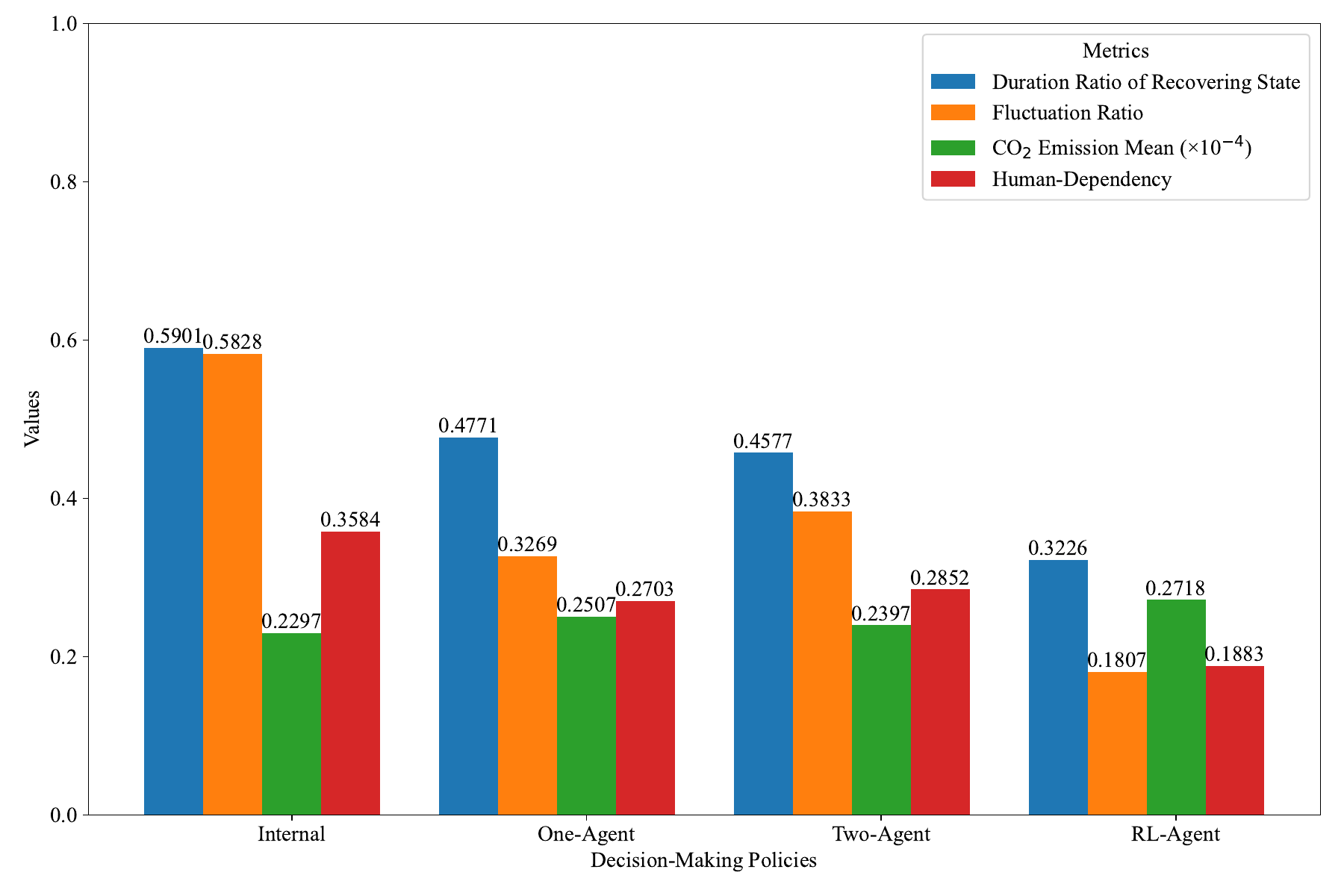}}
\caption{Measurement framework results from the simulated experiments with darkness as a disruptive event.}
\label{fig:measuredarkness}
\end{figure}

Fig.~\ref{fig:measuredarkness} shows the results from the simulated experiments with darkness filter as a disruptive event. The results align with the real-world experiments, emphasizing  RL-agent as the best in achieving fast recovery, maintaining steady performance, and increasing autonomy, however, with slightly extra CO$_2$ emission than other policies. Two-agent shows intermediate results between one-agent and RL-agent in terms of autonomy, green efficiency, and recovery speed, but with slightly less performance stability.
The results also emphasize that internal policies are still poor and require extra support through our policies.

\begin{figure}[ht!]
\centering
\centerline{\includegraphics[width = \textwidth]{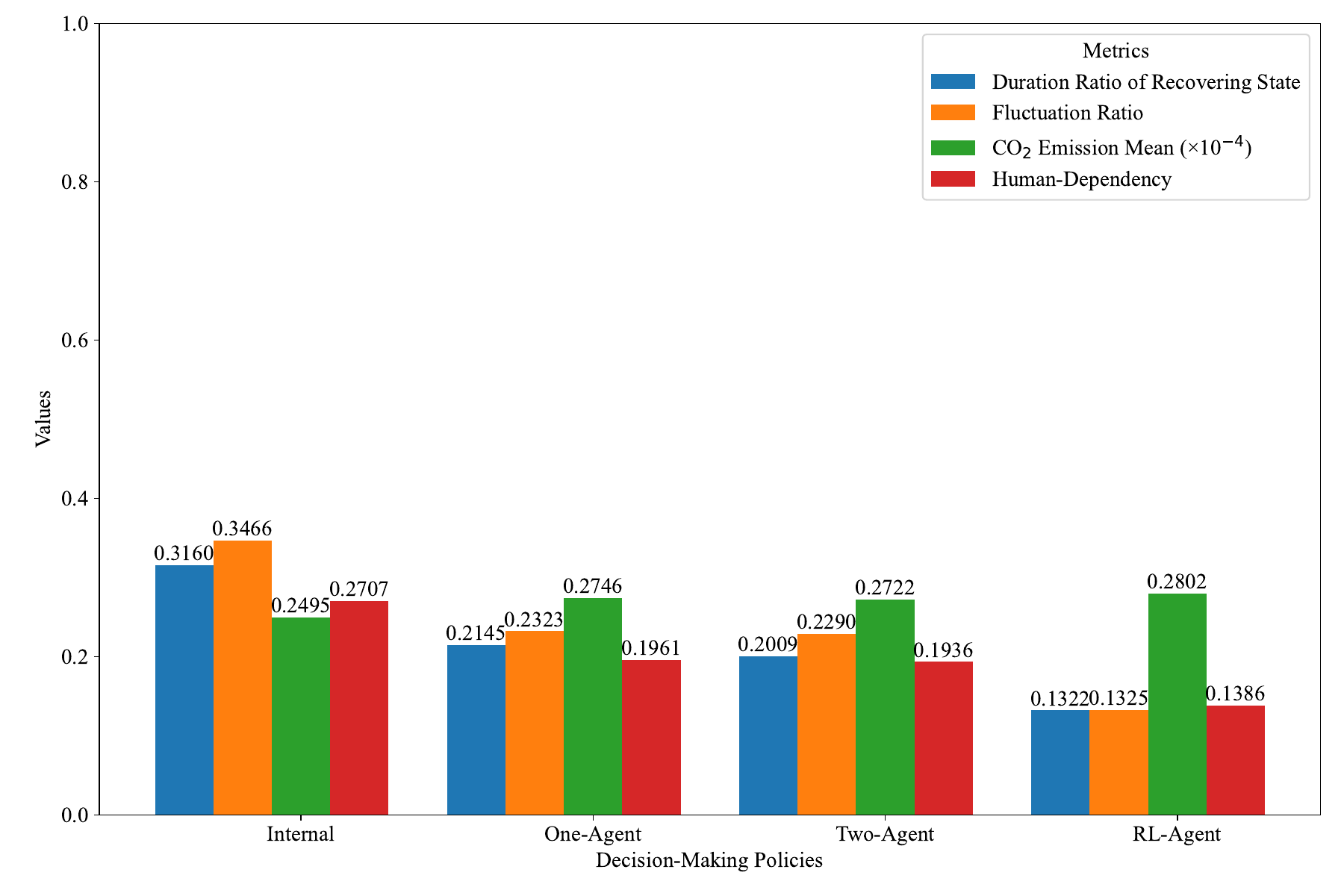}}
\caption{Measurement framework results from the simulated experiments with histogram equalizer as a disruptive event.}
\label{fig:measuradvers}
\end{figure}

The measurements coming from the adversarial attack simulated experiments shown in Fig.~\ref{fig:measuradvers} align with the previous results in terms of improving the greenness and resilience of the internal policies. However, as the histogram equalized was not as disruptive as the darkness, we noticed less disruption than the other experiments.
The experiment results emphasize RL-agent effectiveness over the other policies. Two-agent as an intermediate solution is slightly better than one-agent policies.

\subsection{Containerization Experiments}
We have evaluated the impact of containerization by presenting CORAL's energy consumption in a bare-metal setup and a containerized setup. 

\paragraph{Bare-Metal Setup}
We ran our collaborative robot to classify objects for two hours. During these two hours, we ran a total of $266$ of actions, $247$ of which were autonomous actions and $19$ human actions.
In the first hour, the system consumed a total energy of$ 0.3$ kilowatts per hour(kwh), which doubled to $0.6$ kwh in the next hour. The CO$_2$ emission during the two hours reached $0.198$ CO$_2$eq.

\paragraph{Containerized Setup}
Similar to the bare-metal setup, we ran our containerized collaborative robot to classify objects for two hours. 
We notice our ability to classify more objects, which may be affected by the local communication between the system components.
During the two hours, we ran a total of $275$ of actions, $253$ of which were autonomous actions and $22$ human actions.
We notice a drop in the system energy consumption to $0.1$ kwh in the first hour, which increased to $0.3$ kwh in the next hour. The CO$_2$ emission during the two hours reached $0.099$ CO$_2$eq, which is half the bare-metal setup.

\subsection{Answers to Research Questions RQ2.1 - RQ2.5}

\paragraph{Answer to RQ2.1.} \textit{What are the different metrics required to measure resilience and greenness in OL-CAIS?}

In our GResilience framework, we define three metrics to quantify resilience and greenness for recovery actions.
The action running time quantifies its contribution to the resilience attribute.
On the other hand, we quantify greenness by combining the action's CO$_2$ emission and the number of human interactions it consumes.

\paragraph{Answer to RQ2.2.} \textit{What are the necessary components to automatically monitor OL-CAIS behavior, and support decision-making?}

We address this research question by defining three components: A \textit{monitoring component} that gives the decision-makers of OL-CAIS a preview of the performance evolution and allows them to select the decision-making policies.
An \textit{actuator component} that connects to the OL-CAIS classifier and tracks the ACR values to identify the system's state.
The actuator invokes the selected decision-making policies to recommend the action to execute and then actuate it on OL-CAIS.
The last component is a \textit{simulator component}, which allows decision-makers to simulate their system's running environment to explore different disrupted events and decision-making policies in a protected environment to facilitate their decisions.
We equip the three components into an extendable decision-making assistant to continuously monitor the evolution of the OL-CAIS performance, detect performance degradation, invoke internal or external decision-making policies, recommend recovery action, and actuate it.

\paragraph{Answer to RQ2.3.} \textit{Are the agent-based policies valuable in supporting OL-CAIS to restore its services while balancing greenness and resilience upon disruptions?}

We have developed one-agent policies utilizing a multi-objective optimization model, a two-agent that leverages the game of Battle of Sexes payoff matrix, and an RL-agent that maximizes a reward based on the system's greenness-resilience states.
The empirical evaluation of these policies shows a clear improvement for green recovery in all policies over the internal policies of the system.
This improvement shows the value behind these agent-based policies in balancing greenness and resilience upon disruptions.

\paragraph{Answer to RQ2.4.} \textit{What are the major differences between the agent policies in terms of resilience and greenness?}

To address this research question, we have defined a measurement framework that addresses resilience and greenness qualities through two concepts each: recovery speed and performance steadiness indicate the effectiveness of decision-making policies toward resilience, while green efficiency and autonomy indicate the effectiveness of decision-making policies toward greenness.
We define the duration ratio of the recovering state as a metric to measure recovery speed, the performance fluctuation metric measures performance steadiness, the CO$_2$ emission mean metric measures green efficiency, and the human-dependency metric measures autonomy. Minimizing each of these metrics indicates a higher level of greenness and resilience.
As discussed in Sec.~\ref{subsec:measurementframework}, the results emphasized that RL-agent has a superior improvement toward greenness and resilience over two-agent and one-agent, sequentially. 

\paragraph{Answer to RQ2.5.} \textit{How can we optimize resource allocation using containerization to ensure efficient energy consumption?}

We have defined a new methodology to wrap the multi-machinery bare-metal setup with containers in our attempt to save energy consumption. By applying our methodology to our collaborative robot CORAL, we have reduced the energy consumption and thus its adverse effect (i.e., CO$_2$ emission) to half the bare-metal setup. 
Furthermore, by wrapping the OL-CAIS components in docker containers, we were able to use docker-compose to orchestrate these containers, creating replicas of the components, which increased our system resilience toward component failure and reduced the response time as everything was running on a localhost network.

\section{RG3: Understand catastrophic forgetting in OL-CAIS}

This section addresses \textbf{RG3}, which aims to understand catastrophic forgetting in OL-CAIS following disruptive events and their resolution, tackling \textbf{RC3}: Understanding catastrophic forgetting. 
Our evaluations examined whether the system retains memory of steady-state settings after resolving disruptions, highlighting the potential for catastrophic forgetting. 
These evaluations also investigated mitigation strategies, such as using intelligent policies, to maintain consistent performance over time. 
These experiments analyze patterns in resilience models and assess transitions between steady, disruptive, and final states. They provide evidence for detecting catastrophic forgetting (RQ3.1) and suggest strategies to ensure steady performance (RQ3.2).

\subsection{Catastrophic Forgetting Detection}
\begin{figure}[ht!]
\centering
\centerline{\includegraphics[width = \textwidth]{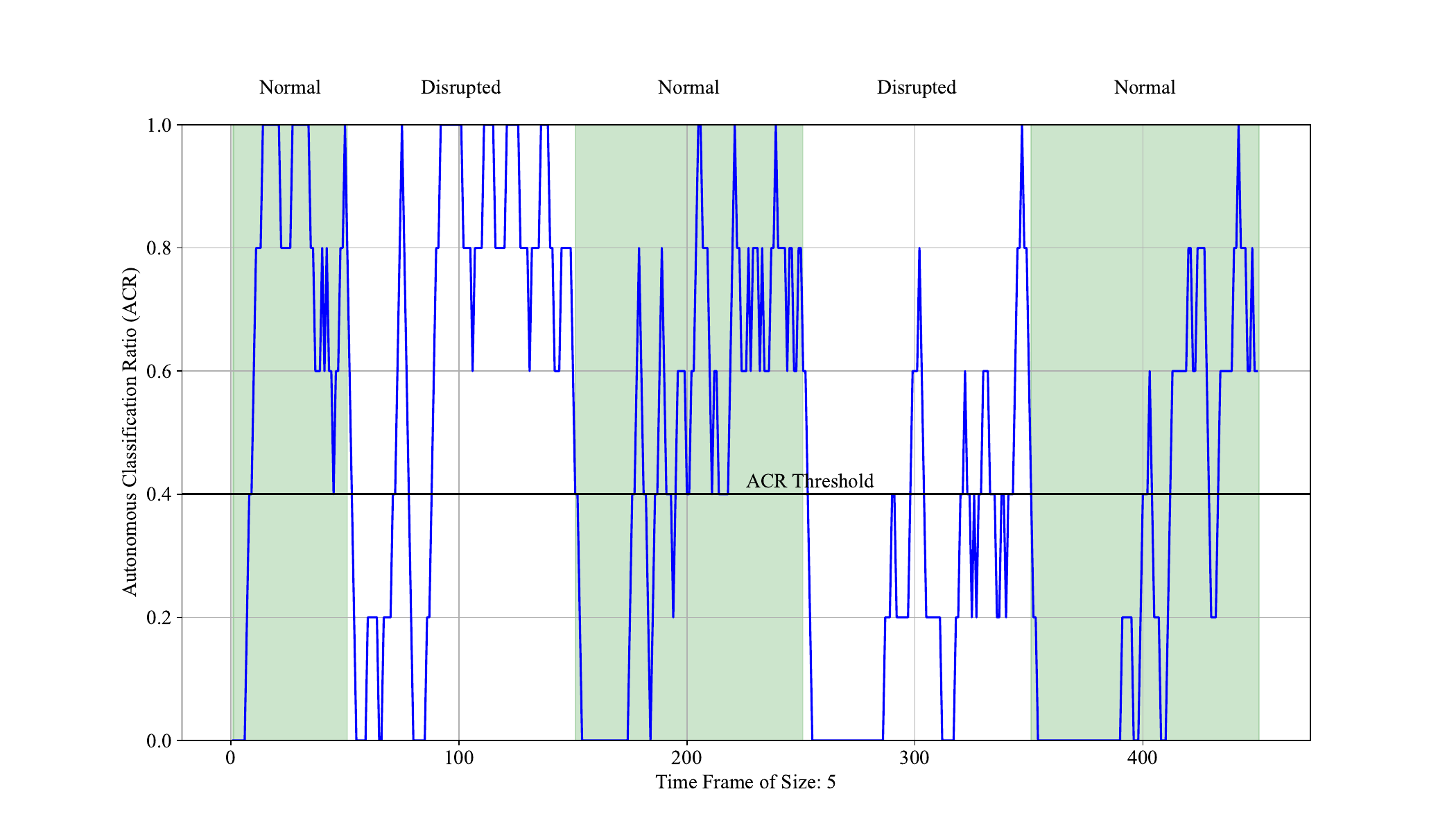}}
\caption{Observing performance degradation upon disruption and resolution of events due to catastrophic forgetting.}
\label{fig:catasinternal}
\end{figure}

Catastrophic forgetting implications extend to online learning scenarios, due to the continuous acquisition of new features, classes, or tasks can also contribute to memory performance degradation,~\cite{lesort2020continual}.
In this experiment, we run our OL-CAIS to experience multiple disruptions and fixes.
Fig.~\ref{fig:catasinternal} illustrates that at each transfer between disruption and normal data, the system enters another disruptive state, and as we go, the system becomes even more disrupted with the change of its environment. This is due to the learning of the different environmental settings, which always hamper the learning memory.

\begin{figure}[ht!]
\centering
\centerline{\includegraphics[width = \textwidth]{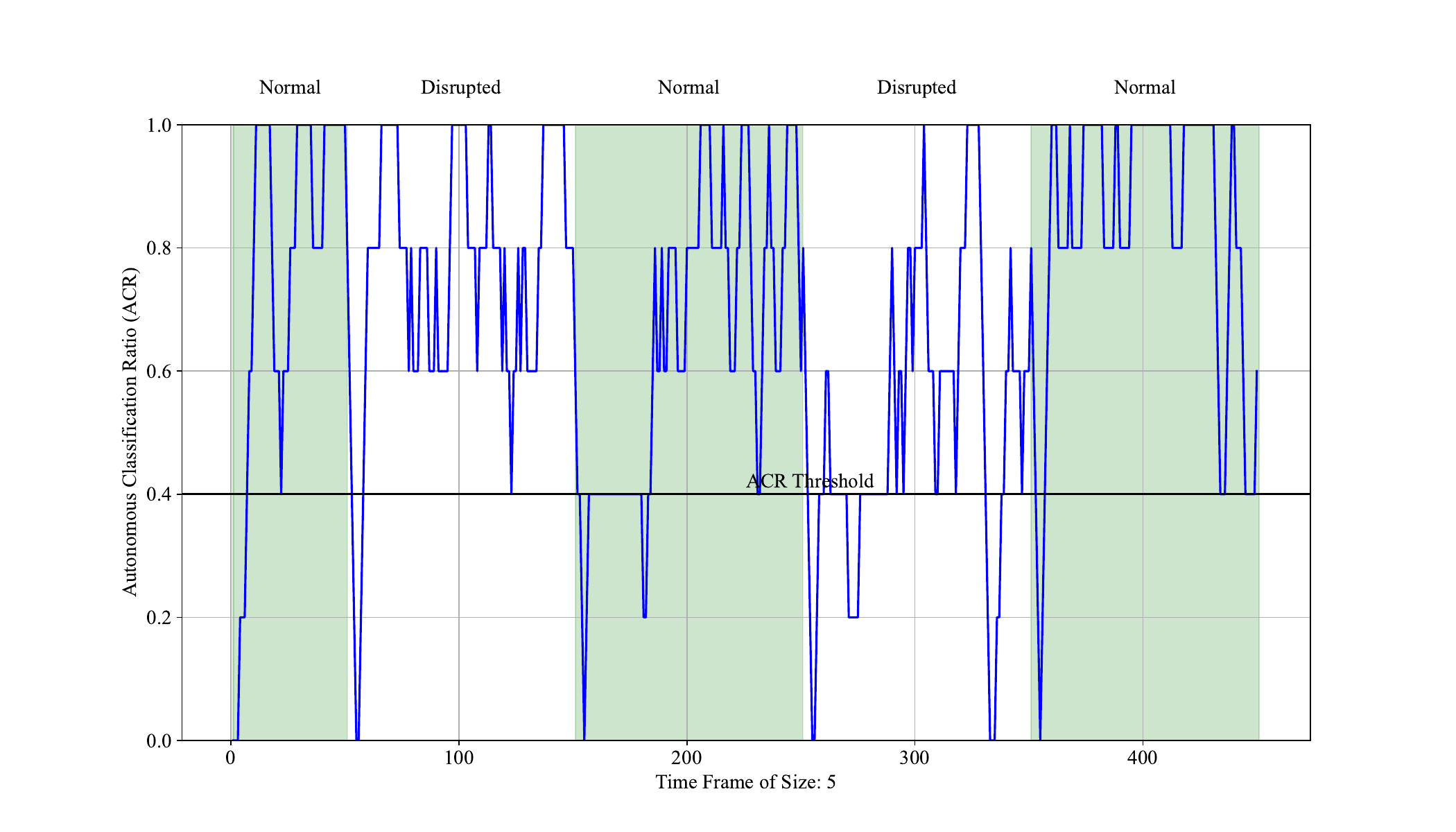}}
\caption{Runing RL-agent to support decision-making upon multiple disruptive events.}
\label{fig:catasall}
\end{figure}

To address this issue, we propose allowing policies different from internal policy to support decision-making. For this work, we have selected the RL-agent to support the decision, leaving the comparison between different policies upon multiple disruptions for future work.
Fig.~\ref{fig:catasall} shows how RL-agent has managed to achieve faster recovery and consistent performance across the different states.
In the first disruptive state, our policies managed to reduce the fluctuation ratio from 0.35 without our support to 0.04 under RL-agent support.
Then ratio has changed in the next state from 0.36 to 0.05, and then from 1.6 to 0.17. Finally, from 1.4 to 0.03.

\subsection{Answers to Research Questions RQ3.1 and RQ3.2}
\paragraph{Answer to RQ3.1.} \textit{What are the patterns in the resilience model that indicate catastrophic forgetting in OL-CAIS?}

In all our real-world or simulated experiments, we have noticed that the OL-CAIS classifier maintains a memory from the disruptive state. This means when we fixed the disruptive event and restored the original environment settings as they were in the steady state, the classifier forgot the original environment and experienced a second performance degradation. 
In the final state, we observed that the system started learning again from human interactions as this was the first time to see these objects. 
Thus, a second performance degradation state is a clear indication of catastrophic forgetting in OL-CAIS going out from a disruptive state.

\paragraph{Answer to RQ3.2.} \textit{How can we ensure that OL-CAIS maintains steady performance over time, even when facing frequent disruptions?}

Our current solution considers the fixed event another disruptive event. In other words, any performance degradation is a cause of some disruptive event.
Thus, we always put the system under the GResilience decision-making policies to ensure consistent, steady performance and fast recovery.

\myPart{Reflections and Conclusions}
\huge{T}\normalsize{he final part synthesizes the insights gained from the research while reflecting on its limitations and proposing directions for future work. Then conclude this work by discussing the key contributions and takeaways}


\chapter{Limitations and Future Work}
\label{ch:discussion}
This chapter reflects on the limitations of the research presented in this thesis and explores potential directions for future work. First, we identify and discuss the threats to validity, considering the scope and assumptions of the proposed methodologies and experiments. Following this, we outline opportunities for future research, emphasizing how the insights gained from this work can guide advancements in the field of green and resilient OL-CAIS.

\section{Threats to Validity} 
Threats to validity define validity as the degree of support for a fallible inference,~\cite{wieringa_six_2015}.
\textcite{verdecchia_threats_2023} address the shortcomings in how researchers approach, document, and evaluate threats to validity. 
They stress the importance of defining threats to validity as trade-offs between different threats to design a better study that aligns with its goals. Thus, we accepted some threats during our study to achieve a better design.
Our study defines holistic theoretical frameworks to achieve green recovery in an OL-CAIS. Then, using these frameworks, we performed empirical evaluations by conducting several real-world experiments and simulation experiments to validate our theories. 
The following summarizes the threats we identified during the construction and validation process.

\paragraph{Internal validity}
Threats in this category are related to internal factors that could have influenced the results. 
One potential threat for real-world experiments is the presence of confounding variables that may influence the observed relationship between the disruptive event and our robot's decision-making. To mitigate this, we invested more effort in maintaining and controlling the same environmental settings (such as the same object colors, lab lights, and conveyor belt speed) across the four experiments.

\paragraph{Construct validity}
Threats to construct validity refer to the extent to which the experiment setting actually reflects the construct under study.
The accurate and reliable measurements, particularly the measurements of the color histograms, threats construct validity. 
To mitigate this, we implemented calibration procedures (such as white balance and exposure adjustments) to ensure the consistency of the data collection.

In the RL-agent experiments, we rounded the states for the Q-table to construct a finite number of states for the second decimal point.
This might lead to loose information about the greenness and resilience state.
However, this solution is acceptable due to the state's simplicity and the need to converge soon.
Further solutions, such as splitting the state into a tuple of greenness and resilience states and using approximators such as deep learners, can be explored in future work.

Another threat to construct validity in the containerization experiment is that the wattmeter reads to the closest decimal point, which leads to information loss.
To mitigate this threat, we ran the experiment for two hours, making sure that the two-hour readings were double the first hour.

\paragraph{Conclusion validity}
Threats in this category  
deal with the degree to which conclusions are
reasonably related to the observed outcome.
In the experiments we conducted, there were cases of performance degradation that the resilience model did not detect.
This raises the question, if the continues degradation of the performance to zero is good enough or this value can be refined.
Although these cases can threaten the validity of our conclusion, they are rare and outliers in our experiments. However, this limitation is worth further investigation in future work.

\paragraph{External validity} 
Threats to external validity concern the generalizability of the results. 
Our policies are mainly exploratory, and to consolidate our claims, generalization beyond our collaborative robot case study is needed. 
Thus, we consider our policies to be generalized as a lab-to-lab study for OL-CAISs, specifically, level 4 of collaborative robots.
However, our study equips OL-CAIS decision-makers with general and extendable tools to investigate their system's performance evolution and green recovery.

\section{Future Work}

While this thesis provides foundational insights and practical methodologies for balancing greenness and resilience in OL-CAIS, several dimensions for future work remain to be explored.

One immediate direction involves enhancing our current work by investigating key variables and their impact on system performance and learning. For example, future research could explore the effects of varying the sliding window size. 
Exploring edge cases such as when the steady-state threshold approaches zero, will we still be able to detect degradation or everything will be accepted, and maybe explore different technique to find this threshold. 
Additionally, stress-testing the lengths of operation states, such as simulating very short steady states or extremely brief disruptive events, could reveal insights into their effects on catastrophic forgetting. These investigations would deepen our understanding of system dynamics and inform the development of more robust and adaptive methodologies.

Building on this, a promising direction involves designing a state machine framework that incorporates multiple machine learning models. This framework would include a primary model responsible for classifying states as either anomaly states, new states, or one of the existing predefined states. The primary model would then moderate transitions between these states while preserving the learning rates of the individual models. By mitigating the risks of catastrophic forgetting across states, this approach could significantly enhance the adaptability and resilience of OL-CAIS, especially in dynamic and uncertain environments.

Finally, achieving a truly comprehensive OL-CAIS requires considering greenness and resilience not only in system design but also during runtime. To this end, future work will focus on integrating the decision-making toolbox developed in this thesis as an intermediate container. This container would monitor interactions between system containers, dynamically redirecting decisions to maintain an optimal balance between greenness and resilience. Such an integration would ensure a runtime-aware OL-CAIS capable of sustainable and adaptive operation in real-world scenarios.

\chapter{Discussion and Conclusion}
\label{ch:conclusion}
This chapter conclude the thesis work by synthesizing  the findings of this research to address the overarching goals outlined at the start of this thesis.
We highlight the contribution of this work to advancing the understanding and development of OL-CAIS, particularly in balancing greenness and resilience. 
This discussion bridges theoretical results with practical applications and outlines the implications for decision-makers, researchers, and industrial practitioners.

\section{Modeling Resilience in OL-CAIS}
The first research goal explores the resilience model in OL-CAIS to support decision-makers in monitoring their system's performance evolution, detecting performance degradation, and identifying performance levels and states.

Our main contribution is the introduction of the Autonomous Classification Ratio (ACR). 
A practical metric for monitoring OL-CAIS performance during run-time.
Furthermore, by defining the ACR Threshold, decision-makers can understand whether their system operates at acceptable or unacceptable levels and identify the current performance state (i.e., steady, disruptive, and final state). 
Finally, through our empirical evaluation, the experiments demonstrated how these states evolve and interact during disruptions, offering insights into system resilience.
The following summarizes our main contributions to the first research goal.

\begin{itemize}
    \item \textit{ACR as a Performance Metric:} The ACR metric, defined as the ratio of autonomous classifications within a sliding window, captures the system’s reliance on human intervention. 
    The metric evolves dynamically, enabling run-time monitoring of performance levels. Decision-makers can interpret the ACR curve to identify performance degradation and recovery patterns, ensuring suitable interventions during disruptions.
    \item \textit{Defining Performance Levels:} By analyzing the minimum ACR during steady state, we established the ACR Threshold for acceptable performance. This threshold, representing the lowest ACR value under normal operating conditions, serves as a reference for identifying unacceptable performance during disruptions. 
    Decision-makers can adapt this threshold using statistical measures, such as the mean or median, to suit specific operational needs.
    \item \textit{Resilience Model and State Transitions:} The resilience model defined three major states: i) Steady State, ii) Disruptive State, and iii) Final State and their transitions.
    Further, we identify sub-states like performance degradation, recovery, and recovered states within the disruptive state.
    This provides a detailed understanding of how OL-CAIS navigates disruptions. The model also highlighted cyclical patterns in performance, with repeated transitions between states as disruptive events occurred and resolved.
\end{itemize}

These metrics equip decision-makers with the ability to monitor system performance evolution in runtime and understand state transitions.
By detecting performance degradation, decision-makers can anticipate potential disruptive events and the time they occur, which helps them effectively mitigate these events and recover performance from degradation to an acceptable level.
This work highlights the importance of maintaining steady and consistent operations by framing resilience as a dynamic process rather than a static attribute.

\section{Balancing Resilience and Greenness}
The second research goal addressed the development of agent-based policies to balance resilience and greenness in OL-CAIS upon disruptive events. 
We introduced three decision-making policies to support OL-CAIS during disruption: one-agent, two-agent, and RL-agent policies.
Our evaluations explored the impact of decision-making policies on recovery speed, performance steadiness, green efficiency, and service delivery autonomy. 
In addition, we introduced containerization as a green resource allocation strategy, and we demonstrated its potential to enhance both greenness and resilience.
The following summarizes our contribution to the second research goal.

\begin{itemize}
    \item \textit{One Extendable Assistant:} Through CAIS-DMA, we equip OL-CAIS decision-makers with all the toolboxes required to simulate, actuate, and monitor their system.
    Through the simulator, they can create a safe environment to explore how their system will behave upon different disruptions.
    Investigate different decision-making policies and their effectiveness toward green recovery.
    Finally, running an endless number of configurations, such as state lengths, combinations of disruptions, and more.
    \item \textit{Metrics for Evaluation:} To measure resilience, measurable concepts such as recovery speed and performance steadiness were introduced, while greenness was evaluated using green efficiency and autonomy concepts. The results showed that the trade-offs between these metrics vary depending on the chosen policy and the nature of the disruptive event.
    \item \textit{Agent-Based Policy Evaluation:} The comparative analysis of one-agent, two-agent, and RL-agent policies revealed significant differences in their impact on greenness and resilience. 
    RL-agent policies are superior to those of the other two agents. RL-agent policies achieve faster recovery, steady performance, and higher autonomy, which requires more computational power. This results in higher CO$_2$ emissions and, therefore, lower green efficiency.
    Conversely, two-agent policies offered intermediate effectiveness between one-agent and RL-agent in terms of recovery speed and green efficiency; however, slightly less in performance steadiness and autonomy than one-agent policies.
    These findings highlight the need for context-specific policy selection, depending on whether resilience or greenness is prioritized.
    \item \textit{Role of Containerization:} By wrapping OL-CAIS components in Docker containers, we reduced energy consumption by half compared to bare-metal setups. 
    Additionally, container orchestration improved system resilience by enabling component replication and localized recovery during failures. This dual benefit positions containerization as an effective strategy for improving greenness and resilience.
\end{itemize}

The findings offer actionable insights for decision-makers. By leveraging the measurements framework and the resilience model, they can evaluate decision-making policies and select those that align with their operational priorities. 
Finally, including containerization as a green strategy further empowers decision-makers to optimize energy consumption and resilience.

\section{Understanding Catastrophic Forgetting}
The third research goal focused on understanding catastrophic forgetting in OL-CAIS and developing strategies to maintain consistent, steady performance.
Through experimental evaluations, we identified patterns in OL-CAIS resilience models that signal catastrophic forgetting and proposed strategies to maintain steady performance over time.
The following highlights our contribution to the third research goal.

\begin{itemize}
    \item \textit{Patterns of Catastrophic Forgetting:} Observations revealed that OL-CAIS retained a memory of disruptive states, leading to second performance degradation when returning to original settings. 
    This phenomenon, observed consistently across experiments, underscores the challenge of preserving steady-state knowledge while addressing disruptive conditions.
    \item \textit{Maintaining Steady Performance:} Our approach treated restored steady state conditions as new disruptive events. 
    By continuously applying GResilience decision-making policies, we ensured that OL-CAIS maintained consistent performance, even in the face of frequent disruptions. 
    This proactive strategy reduces the likelihood of performance fluctuation and performance degradation and ensures rapid recovery.
\end{itemize}

These findings highlight the importance of decision-making frameworks in addressing catastrophic forgetting. 
Decision-makers can leverage these insights to design OL-CAIS that maintains steady performance over time, ensuring resilience in dynamic environments.

\section{Takeaways}
\begin{itemize}
    \item \textit{Metrics and Frameworks:} This research introduced novel metrics and resilience models that provide actionable tools for decision-makers to monitor and manage OL-CAIS performance.
    \item \textit{Policy Evaluation:} The comparative analysis of agent-based policies highlighted the trade-offs between greenness and resilience, offering stakeholders flexibility in addressing operational needs.
    \item \textit{Behavioral Insights:} Identifying performance states and transitions, coupled with understanding catastrophic forgetting, provides a foundation for designing OL-CAIS that remains resilient under dynamic conditions.
    \item \textit{Resource Optimization:} Containerization emerged as an important strategy for improving resilience and greenness, showcasing its potential to reduce energy adverse effects significantly.
\end{itemize}

In conclusion, this thesis advances the understanding and development of OL-CAIS by providing theoretical insights, practical tools, and actionable strategies for decision-makers. 
The introduced frameworks and policies equip stakeholders to navigate the trade-offs between resilience and greenness, ensuring green recovery of system performance. 
These contributions represent a step forward in designing intelligent, adaptive, and environmentally responsible CPS that can thrive in the face of disruptive challenges.








\printbibliography[heading=bibintoc]

\appendix
\chapter{Study Characteristics Tables}
\label{app:Study_Characteristics_Tables}
\begin{table}[hbt!]
\centering
\caption{Included studies that address green problems in cyber-physical systems context.}\label{tab:green}
\begin{tabular}{|p{0.05\linewidth}|p{0.13\linewidth}|p{0.13\linewidth}|p{0.15\linewidth}|p{0.18\linewidth}|p{0.12\linewidth}|p{0.08\linewidth}|}
\hline
\textbf{ID} & \textbf{Cite} & \textbf{Date} & \textbf{Published in} & \textbf{Learning Method Used} & \textbf{Addressed Topic} & \textbf{Number of Studies} \\
\hline
S1 & \cite{hou_cyber-physical_2021, sargolzaei_machine_2017, xu_adaptive_2021, andronie_artificial_2021, valaskova_deep_2021, pandey_greentpu_2019, pandey_greentpu_2020, deniz_reconfigurable_2020, kovacova_sustainable_2021} & 2017, 2019 - 2021 & Conference / Journal  & Deep Learning & Green & 9 \\\hline
S2 & \cite{airlangga_initial_2019} & 2019 & Conference  & Classification Algorithm & Green & 1 \\\hline
S3 & \cite{hou_cyber-physical_2021, xu_adaptive_2021} & 2021 & Conference / Journal  & Reinforcement Learning & Green & 2 \\\hline
S4 & \cite{xu_adaptive_2021} & 2021 & Journal  & Simisupervised Learning & Green & 1 \\\hline
S5 & \cite{pinzone_framework_2020, yan_stochastic_2019, li_sustainable_2019, mohandes_advancing_2018, guo_agricultural_2018, konstantinou_chaos_2021, kharchenko_concepts_2017, zhu_cyber-physical_2020, lu_developing_2019, mohammed_eco-gresilient_2018, estrela_emergency_2017, munoz_energy-aware_2019, pavlenko_estimating_2019, rodriguez_green_2020, atat_green_2019, hu_guest_2018, zhang_guest_2020, zhang_impact_2021, gurdur_interoperable_2017, wang_jamming_2018, hussin_sensor_2019, pavlenko_sustainability_2018, fritzson_openmodelica_2021, xu_thermal_2020} & 2017 - 2021 & Conference / Journal  / Book  & No Learning-based solution & Green & 24 \\\hline
\end{tabular}
\end{table}

\begin{table}[hbt!]
\centering
\caption{Included studies that address resilience problems in cyber-physical systems context.}\label{tab:resilience}
\begin{tabular}{|p{0.05\linewidth}|p{0.13\linewidth}|p{0.13\linewidth}|p{0.15\linewidth}|p{0.18\linewidth}|p{0.12\linewidth}|p{0.08\linewidth}|}
\hline
\textbf{ID} & \textbf{Cite} & \textbf{Date} & \textbf{Published in} & \textbf{Learning Method Used} & \textbf{Addressed Topic} & \textbf{Number of Studies} \\
\hline
S6 & \cite{zarandi_detection_2020, eke_detection_2020, pandey_greentpu_2019, pandey_greentpu_2020, lee_integration_2020, veith_adversarial_2020} & 2019, 2020 & Conference / Journal  & Deep Learning & Resilience & 6 \\\hline
S7 & \cite{hopkins_foundations_2020} & 2020 & Conference  & Classification Algorithm & Resilience & 1 \\\hline
S8 & \cite{veith_adversarial_2020, vanderhaegen_towards_2017} & 2017, 2020 & Journal  & Reinforcement Learning & Resilience & 2 \\\hline
S9 & \cite{olowononi_security_2019} & 2019 & Conference  & Simisupervised Learning & Resilience & 1 \\\hline
S10 & \cite{januario_distributed_2019, liu_distributionally_2022, dimase_holistic_2020, bakirtzis_ontological_2020, kouicem_artificial_2019, xu_automatic_2019, konstantinou_chaos_2021, chekole_cima_2020, alcaraz_cloud-assisted_2018, bagozi_context-based_2021, haque_contract-based_2018, zhu_cyber-physical_2020, kong_cyber-physical_2018, moura_cyber-physical_2019, segovia_cyber-resilience_2020, sahu_design_2021, tomic_design_2018, liu_dynamic_2017, mohammed_eco-gresilient_2018, kouicem_emotional_2021, salazar_enhancing_2019, rodriguez_green_2020, zhang_guest_2020, balogh_learning_2019, barbeau_metrics_2020, bennaceur_modelling_2019, bi_novel_2019, lakshminarayana_performance_2019, horowitz_policy_2018, mouelhi_predictive_2019, ramasubramanian_privacy-preserving_2020, zhang_real-time_2020, cheng_reputation-based_2020, fauser_resilience_2021, ti_resilience_2022, dubey_resilience_2017, januario_resilience_2018, barbeau_resilience_2021, bellini_resilience_2021, fauser_resilience_2020, murino_resilience_2019, wu_resilience-based_2021, lin_resilience-oriented_2021, yong_switching_2018, kouicem_towards_2020, loeve_towards_2020, hasan_towards_2019, jackson_towards_2017, severson_trust-based_2018, mohamed_understanding_2020} & 2017 - 2022 & Conference / Journal  & No Learning-based solution & Resilience & 50 \\\hline
\end{tabular}
\index{tables}
\end{table}

\begin{table}[hbt!]
\centering
\caption{Included studies that address performance problems in cyber-physical systems context without mentioning resilience explicitly.}\label{tab:performance}
\begin{tabular}{|p{0.05\linewidth}|p{0.13\linewidth}|p{0.13\linewidth}|p{0.15\linewidth}|p{0.18\linewidth}|p{0.12\linewidth}|p{0.08\linewidth}|}
\hline
\textbf{ID} & \textbf{Cite} & \textbf{Date} & \textbf{Published in} & \textbf{Learning Method Used} & \textbf{Addressed Topic} & \textbf{Number of Studies} \\
\hline
S11 & \cite{al-sharman_sensorless_2021, chiu_integrative_2020, li_conaml_2021, xi_data-correlation-aware_2022, luo_deep_2020, sharmeen_identifying_2019, ding_krakenbox_2021, patel_low-latency_2020, ding_-line_2019, pauli_optimal_2019, song_physical_2019, taormina_real-time_2017, abokifa_real-time_2019, rajawat_reliability_2022, jindal_sedative_2018, zhou_temperature-constrained_2021} & 2017 - 2022 & Conference / Journal  & Deep Learning & Performance & 16 \\\hline
S12 & \cite{an_modeling_2019, meyer_analytics-based_2020, tsai_apply_2019, nandy_carbon_2018, arman_cyber_2018, xi_data-correlation-aware_2022, li_ensemble_2021, patel_low-latency_2020, jan_machine_2020, perrone_machine_2021, gartziandia_microservice-based_2021, taormina_real-time_2017} & 2017 - 2022 & Journal  / Book  /Conference  & Classification Algorithm & Performance & 12 \\\hline
S13 & \cite{fei_learn--recover_2020} & 2020 & Conference  & Reinforcement Learning & Performance & 1 \\\hline
S14 & \cite{meyer_analytics-based_2020, kafle_automation_2020, puri_interpretable_2020, taormina_real-time_2017} & 2017, 2020 & Book  /Conference  & Linear Regression & Performance & 4 \\\hline
S15 & \cite{xi_data-correlation-aware_2022, luo_deep_2020, wickramasinghe_explainable_2021, puri_interpretable_2020, abokifa_real-time_2019, vatanparvar_self-secured_2019} & 2019 - 2022 & Conference / Journal  & Unsupervised Learning & Performance & 6 \\\hline
S16 & \cite{xi_data-correlation-aware_2022} & 2022 & Journal  & Simisupervised Learning & Performance & 1 \\\hline
S17 & \cite{pagliari_case_2017, parto_novel_2020, rosales_actor-oriented_2018, zhang_eat-ml_2021, alrimawi_incidents_2019, chaudhary_machine_2019, bures_performance_2018, pinciroli_qn-based_2021, smith_software_2020, pagliari_what_2019} & 2017 - 2022 & Conference / Journal  / Book  & No Learning-based solution & Performance & 10 \\\hline
\end{tabular}
\index{tables}
\end{table}

\begin{table}[hbt!]
\centering
\caption{Included studies that address security problems in cyber-physical systems context without mentioning resilience explicitly.}\label{tab:Security}
\begin{tabular}{|p{0.05\linewidth}|p{0.13\linewidth}|p{0.13\linewidth}|p{0.15\linewidth}|p{0.18\linewidth}|p{0.12\linewidth}|p{0.08\linewidth}|}
\hline
\textbf{ID} & \textbf{Cite} & \textbf{Date} & \textbf{Published in} & \textbf{Learning Method Used} & \textbf{Addressed Topic} & \textbf{Number of Studies} \\
\hline
S18 & \cite{chen_deep_2018, yang_chapter_2021, li_conaml_2021, xi_data-correlation-aware_2022, luo_deep_2020, hussain_deep_2021, taormina_deep-learning_2018, ma_deep-learning-based_2021, li_deepfed_2021, sharmeen_identifying_2019, song_physical_2019, bakalos_protecting_2019, taormina_real-time_2017, abokifa_real-time_2019, akowuah_recovery-by-learning_2021, rajawat_reliability_2022, liu_secure_2021, dhir_study_2020, zhou_temperature-constrained_2021} & 2017 - 2022 & Conference / Journal  & Deep Learning & Security & 19 \\\hline
S19 & \cite{tertytchny_classifying_2020, arman_cyber_2018, xi_data-correlation-aware_2022, ma_deep-learning-based_2021, wu_detecting_2019, wang_detecting_2017, li_ensemble_2021, hao_hybrid_2021, dhiman_machine_2021, perrone_machine_2021, clark_machine_2018, oyekanlu_osmotic_2018, taormina_real-time_2017, khan_rigorous_2020, saha_sharks_2021, dhir_study_2020} & 2017 - 2022 & Conference / Journal  & Classification Algorithm & Security & 16 \\\hline
S20 & \cite{hao_hybrid_2021, fei_learn--recover_2020} & 2020, 2021 & Conference / Journal  & Reinforcement Learning & Security & 2 \\\hline
S21 & \cite{puri_interpretable_2020, taormina_real-time_2017} & 2017 2020 & Conference  & Linear Regression & Security & 2 \\\hline
S22 & \cite{chen_deep_2018, xi_data-correlation-aware_2022, luo_deep_2020, wu_detecting_2019, wickramasinghe_explainable_2021, hao_hybrid_2021, puri_interpretable_2020, abokifa_real-time_2019, vatanparvar_self-secured_2019} & 2018 - 2022 & Conference / Journal  & Unsupervised Learning & Security & 9 \\\hline
S23 & \cite{xi_data-correlation-aware_2022, bakalos_protecting_2019, kriebel_robustness_2018} & 2018, 2019, 2022 & Conference / Journal  & Simisupervised Learning & Security & 3 \\\hline
S24 & \cite{odonovan_fog_2018, alrimawi_incidents_2019, chaudhary_machine_2019, zhang_real-time_2021} & 2018, 2019, 2021 & Conference / Journal  & No Learning-based solution & Security & 4 \\\hline
\end{tabular}
\index{tables}
\end{table}

\begin{table}[hbt!]
\centering
\caption{Included studies that address different properties in cyber-physical systems context, other than the one mentioned before.}\label{tab:none}
\begin{tabular}{|p{0.05\linewidth}|p{0.13\linewidth}|p{0.13\linewidth}|p{0.15\linewidth}|p{0.18\linewidth}|p{0.12\linewidth}|p{0.08\linewidth}|}
\hline
\textbf{ID} & \textbf{Cite} & \textbf{Date} & \textbf{Published in} & \textbf{Learning Method Used} & \textbf{Addressed Topic} & \textbf{Number of Studies} \\
\hline
S25 & \cite{kumar_machine_2021, wu_weighted_2018, monedero_cyber-physical_2021, hao_deep_2021, leong_deep_2018, gati_differentially_2021, yamagata_falsification_2021, vos_incorporating_2020, snijders_machine_2020, yong_multi_2019, petrenko_multi-agent_2021, benini_plenty_2017, mekala_resource_2020} & 2017 - 2021 & Conference / Journal  & Deep Learning & Other properties & 13 \\\hline
S26 & \cite{monedero_cyber-physical_2021, shah_machine_2019, bunte_model-based_2019, yong_multi_2019} & 2019, 2021 & Conference / Journal  & Classification Algorithm & Other properties & 4 \\\hline
S27 & \cite{hao_deep_2021, leong_deep_2018, yamagata_falsification_2021, petrenko_multi-agent_2021, mekala_resource_2020} & 2018, 2020, 2021 & Conference / Journal  & Reinforcement Learning & Other properties & 5 \\\hline
S28 & \cite{li_machine_2019} & 2019 & Conference  & Linear Regression & Other properties & 1 \\\hline
S29 & \cite{vos_incorporating_2020} & 2020 & Conference  & Unsupervised Learning & Other properties & 1 \\\hline
S30 & \cite{ren_interactive_2022, snijders_machine_2020, putnik_machine_2021} & 2020, 2021, 2022 & Conference / Journal  & Simisupervised Learning & Other properties & 3 \\\hline
S31 & \cite{oumimoun_solo-checkpointing_2021, iv_feasibility_2018, pavlov_preferences_2021} & 2018, 2021 & Conference  / Book  & No Learning-based solution & Other properties & 3 \\\hline
\end{tabular}
\index{tables}
\end{table}






\clearpage
\newpage
\includepdf[pages=-,pagecommand={},width=\paperwidth]{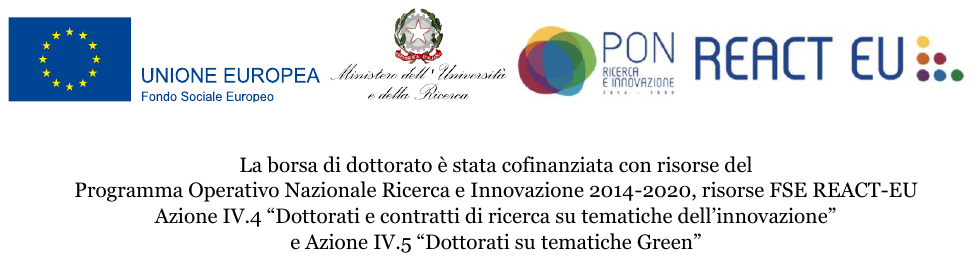}

\end{document}